\documentclass[twocolumn]{aastex63}

\usepackage{amsmath}

\submitjournal{AJ}
\shorttitle{Limb-darkening and TESS}
\shortauthors{Patel \& Espinoza}

\begin{document}

\title{Empirical limb-darkening coefficients \& transit parameters of known exoplanets from \textit{TESS}}

\correspondingauthor{Jayshil A. Patel}
\email{jayshil.patel@astro.su.se}

\author[0000-0001-5644-6624]{Jayshil A. Patel}
\affiliation{Department of Astronomy, Stockholm University, AlbaNova University Center, 10691 Stockholm, Sweden}

\author[0000-0001-9513-1449]{N\'estor Espinoza}
\affiliation{Space Telescope Science Institute, 3700 San Martin Drive, Baltimore, MD 21218, USA}
\affiliation{Department of Physics \& Astronomy, Johns Hopkins University, Baltimore, MD 21218, USA}

\received{}
\revised{}





\begin{abstract}

Although the main goal of the Transiting Exoplanet Survey Satellite (\textit{TESS}) is to search for new transiting exoplanets, its data can also be used to study in further detail already known {{systems}}. The \textit{TESS} bandpass is particularly interesting to study the limb-darkening effect of the stellar host which is imprinted in transit lightcurves, as the widely used \textsc{phoenix} and \textsc{atlas} stellar models predict different limb-darkening profiles. Here we study this effect by fitting the transit lightcurves of  {{176 known}} exoplanetary systems observed by \textit{TESS}{{, which allows us to extract 
empirical limb-darkening coefficients (LDCs) for the widely used quadratic law, but also updated transit parameters (including ephemerides refinements) as 
a byproduct. Comparing our empirically obtained LDCs}} with theoretical predictions, we find significant offsets {{when using tabulated \textit{TESS} LDCs}}. Specifically, the $u_2$ coefficients obtained using \textsc{phoenix} models show the largest discrepancies depending on the method used to derive them, with offsets that can reach up to {{$\Delta u_2 \approx 0.2$ on average. Most of those average offsets disappear, however, if one 
uses the SPAM algorithm introduced by \cite{2011MNRAS.418.1165H} to calculate the LDCs instead. Our results suggest, however, that for 
stars cooler than about 5000 K, no methodology is good enough to explain 
the limb-darkening effect: we observe a sharp deviation between measured and 
predicted LDCs on both quadratic LDCs of order $\Delta u_1, \Delta u_2 \approx 0.2$ 
for those cool stars. We recommend caution when assuming limb-darkening coefficients 
as perfectly known thus, in particular for these cooler stars when analyzing \textit{TESS} 
transit lightcurves.}}

\end{abstract}



\section{Introduction}\label{sec:intro}
The phenomenon of limb-darkening on stars --- the observed intensity decrease 
towards their limb --- has for long been recognized as a natural consequence 
of temperature gradients in stellar atmospheres \citep{Sch:1906}. Although a complex 
effect in nature, parametric formulations of it have been shown to be very practical to 
use when the effect needs to be modelled. These so-called ``limb-darkening laws" have been 
mainly motivated from an observational perspective and have been formulated with certain use-cases 
in mind \citep[see, for example][]{2000A&A...363.1081C, 1992A&A...259..227D, 2009AA...505..891S, 1970AJ.....75..175K, 2003AA...412..241C}. {{In exoplanetary science, in particular, the most popular laws to date are the linear law \citep{Sch:1906}, the quadratic law \citep{1950HarCi.454....1K} and the non-linear law \citep{2000A&A...363.1081C}. Their functional forms are,}}

\begin{eqnarray*}
        \frac{I(\mu)}{I(1)} &=& \ 1 - a(1 - \mu);\ \textnormal{(the linear law),} \\
        \frac{I(\mu)}{I(1)} &=& \ 1 - u_1(1 - \mu) - u_2{(1 - \mu)}^2;\ \textnormal{(the quadratic law),} \\
        \frac{I(\mu)}{I(1)} &=& \ 1 - \sum_{n=1}^{4}c_n(1 - \mu^{n/2});\ \textnormal{(the non-linear law)}.
\end{eqnarray*}

Here $a$, $(u_1, u_2)$ and $(c_1, c_2, c_3, c_4)$ are the so-called limb darkening coefficients (LDCs) of these respective laws and 
$\mu$ represents the cosine of the angle $\theta$ - the angle between the line of sight and the normal to the surface. The reason 
for the popularity of these laws is most likely related with the fact that these give rise to very fast lightcurve 
computations \citep{2002ApJ...580L.171M}, which in turn makes them the most practical to use. Among them, 
the quadratic law is by far the most widely used in the exoplanet literature mainly due to the fact that 
it gives rise to {{fast and efficient computation of lightcurves}} in some of the most popular lightcurve computing methods/algorithms \citep[e.g., \texttt{batman};][]{2015PASP..127.1161K}. 
The simplicity of this law also makes it an attractive one to use for 
researchers: it is defined by only two parameters which can even be fitted along 
with the rest of the other transit parameters if stellar model atmospheres are 
suspected to not be appropriate for a given use-case. {{While it might not always be 
the most accurate law to use \citep[see, e.g.,][]{2008MNRAS.386.1644S, 2012AA...539A.102H, 2016MNRAS.457.3573E}, its usage is nonetheless widespread across transiting exoplanet studies 
because of the aforementioned reasons}}.

The details of the limb-darkening effect become important when one wants to fit an exoplanet transit lightcurve because 
its shape is strongly modulated by the effect \citep{2002ApJ...580L.171M, 2003ApJ...585.1038S}. Incorrect 
assumptions about the effect, thus, have the potential to lead to biases on the retrieved transit parameters. Indeed, simulations 
have studied in detail the impact of either fitting or fixing the LDCs 
\citep[see, e.g., ][]{2013AA...549A...9C, 2015MNRAS.450.1879E, 2017ApJ...845...65N, 2017AJ....154..111M} and even the impact the selection of a given limb-darkening law has on the retrieval of the transit parameters \citep{2016MNRAS.457.3573E}. Most of 
these studies conclude that the most conservative assumption is to fit for the coefficients in the transit lightcurve 
fitting procedure, taking special care on the selection of the limb-darkening law to use for very precise (better than about 
200 ppm) transit lightcurves. This suggestion, however, is impractical in cases on which the lightcurve is poorly sampled 
and/or dominated by systematic noise such as in, e.g., \textit{Hubble Space Telescope} (HST) spectro-photometric measurements 
\citep[see, e.g.,][]{2018haex.bookE.100K}. It is also not clear from the literature how far off empirical LDCs are 
from stellar model atmospheres predictions, which of all the stellar models available in the literature are the best to use, and which exact method/table one should use if one needed a set of LDCs to either fix or use as priors in the analysis 
\citep[see, e.g., ][]{2012AA...539A.102H, 2015MNRAS.450.1879E, 2017AJ....154..228S}. {{Studies such as that of, e.g., \citet{2012AA...539A.102H}, \citet{2013A&A...560A.112M}, \citet{2015MNRAS.450.1879E} and \citet{2018A&A...616A..39M}, have approached the problem by comparing LDCs empirically determined from transit 
lightcurves to that of predictions from stellar model atmospheres. This intercompraison between empirical 
LDCs and those obtained from actual data are critical}} not only to inform stellar modellers, but also 
the community on best practices and uses of stellar model atmospheres for exoplanet transit lightcurves. 
Such a comparison has the power to aid us in the search for better models, and empirical LDCs 
could provide good priors especially useful when confronted with low precision lightcurves.

The \textit{Transiting Exoplanet Survey Satellite} \citep[\textit{TESS}; ][]{tess} mission does not only provide exquisite, 
almost all-sky photometry with which to look for new transiting exoplanets, but also provides a unique dataset of already 
\textit{known} exoplanets which is ideal to perform tests on stellar model atmospheres through stellar limb-darkening. Given that most known transiting 
exoplanetary systems have been thoroughly characterized with ground-based instruments, precise spectroscopic parameters 
are available for their host stars and thus limb-darkening predictions can be extracted from stellar model atmospheres. 
The precise photometry being collected by the \textit{TESS} mission, in turn, should allow us to tightly constrain the 
limb-darkening effect through the LDCs, thus providing a rich dataset to compare against theoretical predictions. 
Interestingly (and as it will be shown in this work), LDCs from the most widely used stellar model atmospheres to model limb-darkening in exoplanet transit lightcurves, 
namely, the \textsc{atlas} \citep{2003IAUS..210P.A20C} and \textsc{phoenix} \citep{2013AA...553A...6H} model atmospheres, have very different values in the \textit{TESS} bandpass depending on which methods one uses to obtain them, which provides a 
perfect opportunity to test these two stellar model atmospheres (and the methods used 
to obtain LDCs) against empirical data. In this work, we use recently 
released \textit{TESS} transit lightcurve data in order to extract limb darkening coefficients for a subset of the known 
systems observed by the mission in order to perform these tests.



We have organized this work as follows. In Section \ref{sec:data}, we describe how we select our targets and download the data for them, along with a description of 
the modelling procedure used to analyze the lightcurves and to obtain theoretical 
predictions for our LDCs. We present our findings in Section \ref{sec:results}, which are followed by a Discussion in Section \ref{sec:discussion} and Conclusions and Future Work in Section 
\ref{sec:conclusion}.

\section{Data, sample selection and modelling}\label{sec:data}

\subsection{Selection of targets}
Because the main objective of the present work was to characterize the stellar hosts, we chose \textit{TESS} targets that had already known and confirmed transiting exoplanets, and that had precise follow-up observations that allowed for 
precise stellar (e.g., stellar effective temperatures, gravity, metallicity) and planetary (e.g., via radial-velocity variations) characterization using high-resolution spectroscopy. In addition, only exoplanets observed in 2-minute cadence 
by \textit{TESS} were selected in order to minimize lightcurve distortions due to the binning of the 30-minute cadence 
lightcurves \citep{binningsinning}.

\begin{figure*}
	\includegraphics[width=\textwidth]{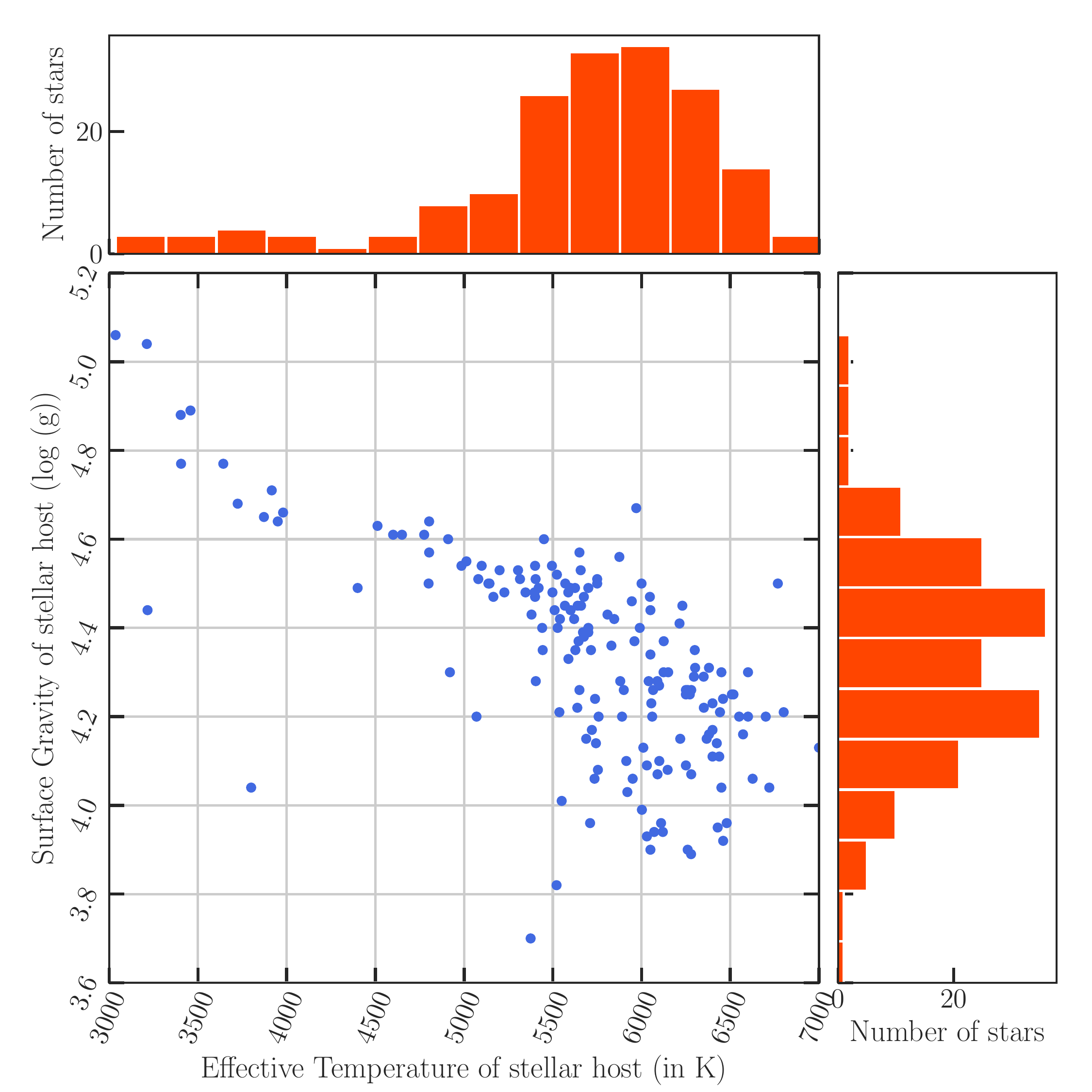}
    \caption{Range and distribution of Effective Temperature ($T_{eff}$) and Surface Gravity ($\log g$) of target exoplanet host stars.}
    \label{fig:fig2}
\end{figure*}

According to the NASA Exoplanet Archive\footnote{\url{https://exoplanetarchive.ipac.caltech.edu/cgi-bin/TblView/nph-tblView?app=ExoTbls&config=planets}}, when we started our analysis, more than 4000 exoplanets were known. Among them, we chose transiting exoplanets which were going to be or already were observed by \textit{TESS}. We filtered these planets using the Web TESS Viewing tool (WTV)\footnote{\url{https://heasarc.gsfc.nasa.gov/cgi-bin/tess/webtess/wtv.py}}. At the time of our analysis, the \textit{TESS} mission had observed targets up to sector 32, so we selected targets only up to this sector. There were a total of 1745 of them. Among these systems, we removed from 
our sample:

\begin{itemize}
    \item Systems that did not have follow-up high-resolution spectroscopic 
    observations that allowed us to confirm their orbits in an independent 
    way to the transits.
    \item Multi-planetary systems, which would involve extra complications in the analyses (TTVs, joint fitting of some properties, checks 
    for planet-planet transits).
    \item Systems that had known physically associated 
    stellar companions, as they could significantly dilute the 
    exoplanet transit lightcurves.
\end{itemize}

When we analysed these filtered systems, we found that some of them had 
very low signal-to-noise ratio transit lightcurves. We decided to 
measure the signal-to-noise ratio of the transits through their 
transit depths, and decided that systems for which the signal-to-noise 
ratio of the transit depths were less than 5 would also be 
removed from our study. After this, a total of 176 targets remained 
in our sample. Figure \ref{fig:fig2} shows the range and the distribution of effective temperature ($T_{eff}$) and surface gravity ($\log(g)$ - in cgs units) of the exoplanet host stars of our target systems. Table \ref{tab:stellar_properties} shows the values of various stellar 
properties of each of our host target stars.

\subsection{Data and Modelling}

We used \texttt{juliet} \citep[][]{juliet} to download and fit 
the 2-minute cadence lightcurve data directly from the Mikulski 
Archive for Space Telescopes (MAST\footnote{\url{https://mast.stsci.edu/portal/Mashup/Clients/Mast/Portal.html}}) 
portal in an automated fashion. Times, fluxes and errors were extracted from 
the \texttt{PDC}-corrected lightcurve products.

The transit lightcurves were modelled within \texttt{juliet} using the 
\texttt{batman} \citep{2015PASP..127.1161K} model. In addition to it, 
each lightcurve fit also included a Gaussian Process (GP) in order to marginalize either astrophysical or instrumental systematic trends present in the \textit{TESS} lightcurves. In particular, for each lightcurve, we fitted a GP using two kernels: an Exponential-Mat\`ern kernel and a Quasi-periodic 
kernel. The first one is a result of multiplying an exponential kernel, which typically samples smooth functions, and 
a Mat\`ern 3/2 kernel, which allows to model abrupt changes in the lightcurves, making the resultant multiplied kernel 
a very flexible one to model instrumental systematics or non-periodic phenomena. The Quasi-periodic kernel we opted 
to use, on the other hand, was the one introduced in \cite{celerite}, and of the form, 
\begin{equation}
k(x_l,x_m) = \frac{B}{2+C}e^{-\tau/L}\left[\cos\left(\frac{2\pi\tau}{P_{\textnormal{rot}}}\right) + (1+C)\right],
\end{equation}

where $B$, $C$, $L$ and $P_{\textnormal{rot}}$ are the hyperparameters of the kernel. This is a very useful 
kernel when it comes to modelling quasi-periodic phenomena such as the one observed in, e.g., starspot rotational modulations. These two kernels 
are implemented within \texttt{juliet} using \texttt{celerite} \citep{celerite}. 

In practice, our fits were done in a two-step fashion for each target. 
First, the out-of-transit data for each sector was analyzed separately, 
and fitted with a GP, plus a jitter term and a flux normalization 
factor. We then compared the bayesian evidence on each 
sector for each of the kernels outlined above, and the one that 
was preferred for most of the sectors was defined as \textit{the} 
kernel to use for that target. Then, the posteriors on each of these 
parameters for each sector were used as priors for a multi-sector joint 
fit of the in-transit data, to which we added the transit model. {{This two-step process 
for fitting these transit lightcurves had two main advantages. The first advantage is that this 
provides a much more efficient sampling of the parameter space given we use nested sampling algorithms 
to explore it (which are much better behaved in low dimensional spaces). The second advantage, which is a byproduct of the first one, is that it allows for a much 
faster convergence of the algorithms. We follow the recommendation in \cite{juliet}, and use importance 
nested sampling with the MultiNest algorithm \citep{MultiNest} via the PyMultinest library 
\citep{PyMultiNest} to perform fits with up to about 20 free parameters, and use dynamic nested sampling 
via the \texttt{dynesty} library \citep{Dynesty} to explore higher dimensional parameter spaces. In practice, 
this means that we use MultiNest for multi-sector fits involving less than 4 sectors, and \texttt{dynesty} 
for multi-sector fits involving 4 or more sectors.}} Our automated routines to perform this two-step fitting procedure 
is available at Github\footnote{\url{https://github.com/nespinoza/tess-limb-darkening/}}.

\begin{deluxetable}{lcc}
 \tablecaption{{Summary of priors used in our analysis. $\mathcal{U} (a,b)$ and $\mathcal{J} (a,b)$ show the uniform and log-uniform distribution between $a$ and $b$, respectively; $\mathcal{N}(\mu,\sigma^2)$ gives the normal distribution with mean $\mu$ and the standard deviation $\sigma$. The ``$lit$" subscript to symbols means that the value of that parameter is taken from the literature. A single value indicates that the parameter was fixed.}}
 \label{tab:prior}
 \tablehead{ \colhead{Parameter} & \colhead{Unit} & \colhead{Prior}}
 \startdata
    \multicolumn{3}{c}{\text{Planetary Parameters}}\\ \hline
    $P$ & days & $\mathcal{N}(\mu_{lit}, \sigma^2_{lit})$\\
    $T_0$ & BJD & $\mathcal{N}(\mu_{lit}, 0.1)$\\
    $R_p/R_*$ & & $\mathcal{U}(0., 1.)$\\
    $b$ & & $\mathcal{U}(0., 1.)$\\
    $a/R_*$ & & $\mathcal{J}(1., 100.)$\\
    eccentricity & & $ecc_{lit}$\\
    $\omega$ & deg & $\omega_{lit}$\\\hline
    \multicolumn{3}{c}{\text{Instrumental Parameters}}\\ \hline
    $q_1$ (transformed LDC) & & $\mathcal{U}(0.,1.)$\\
    $q_2$ (transformed LDC) & & $\mathcal{U}(0.,1.)$\\
    $m_{flux}$ & ppm & $\mathcal{N}(0., 10^5)$\\
    $m_{dilution}$ & & 1\\
    $\sigma_w$ (jitter) & ppm & $\mathcal{J}(0.1, 10000)$ \\ \hline
    \multicolumn{3}{c}{\text{Exponential-Mat\`ern kernel}}\\ \hline
    Amplitude of the GP & ppm & $\mathcal{J}(10^{-5}, 10000)$ \\
    Timescale (Exp part) & days & $\mathcal{J}(0.001, 100)$ \\
    Timescale (Mat\`ern part) & days & $\mathcal{J}(0.001, 100)$ \\ \hline
    \multicolumn{3}{c}{\text{Quasi-Periodic kernel}}\\ \hline
    B & & $\mathcal{J}(10^{-5}, 10^3)$\\
    C & & $\mathcal{J}(10^{-5}, 10^4)$\\
    L & days & $\mathcal{J}(10^{-3}, 10^3)$\\
    $P_{rot}$ & days & $\mathcal{J}(1., 100.)$
\enddata
\end{deluxetable}

The priors used in our fits were, in general, wide enough to explore a vast part of the parameter space. The 
only exception was the period of the orbits, whose priors were normal distributions 
with means and standard deviations taken from previous works in the literature --- our posteriors on this parameter, 
then, can be seen effectively as an update over those previous estimates. The time-of-transit center also had a more or less constrained prior; we also defined a normal distribution as a 
prior on this parameter based on previous works, but the standard deviation of this distribution was set to 0.1 days (i.e., about 2.4 hours). To parametrize the limb-darkening we decided 
to use the widely used quadratic law via the uninformative sampling scheme proposed by \citet{2013MNRAS.435.2152K}. 
In this setup, instead of fitting for the limb-darkening coefficients directly, we fit for the transformed 
parameters $q_1$ and $q_2$, each of which we define to have a uniform prior distribution between 0 and 1. We defined wide priors 
between 0 and 1 for the planet-to-star radius ratio and the impact parameter of the orbits. For $a/R_*$, the scaled semi-major axis, we decided to use a wide {{log-}}uniform prior between 1 and 100. In order to account for possible out-of-transit offsets, we also fitted for a mean out-of-transit flux in our transit fits, $m_\textnormal{flux}$, which normalizes the lightcurve via $1/(1 + m_\textnormal{flux})$ \citep[see][for details]{juliet}; we defined a normal prior for this parameter with zero-mean and a standard deviation of $10^5$ ppm. A dilution factor was set to 1 for all of our fits 
(i.e., assuming no dilution by nearby sources), as most of our targets are bright and 
the \texttt{PDC} algorithm is supposed to correct for the dilution of nearby objects 
anyways. Finally, for our GP kernels, we also defined wide priors. For the Exponential-Mat\`ern kernel, we used a log-uniform prior 
on the amplitude of the GP from $10^{-5}$ to 10,000 ppm, and time-scales for both the exponential and Matern parts of the kernel between 0.001 and 100 days. For the quasi-periodic kernel, the B, C and L parameters had log-uniform priors 
between $10^{-5}$ to $10^3$, $10^{-5}$ to $10^4$, and $10^{-3}$ to $10^3$. The rotation period of the kernel had a {{log-}}uniform prior between 1 and 100 days. All lightcurve models also included a white-gaussian zero-mean noise component, whose standard deviation was also fitted 
in our procedure. The prior for that ``jitter" parameter was defined to be log-uniform between 0.1 and 10,000 ppm. The eccentricity and argument of 
periastron on our fits was fixed to literature values.

 We also note here that although our target list consist of already known exoplanets observed by TESS up to Sector 32, we downloaded \textit{all} available data for these targets, which may include, for some of the systems, data from the most recent data release of Sector 34. 

\subsection{Theoretical calculation of Limb Darkening Coefficients}
\label{sec:theo_LDC}
{{One of the key parts of our work involves computing theoretical LDCs in order to compare them with our empirical estimates obtained from \textit{TESS} lightcurve fits. In this work, we used two main methods to compute these. The first, and the most popular method used by the community, is to use various tables/codes published in the literature to directly retrieve those coefficients. These coefficients, in turn, are obtained by fitting the intensity 
profiles of model stellar atmospheres, weighted by the instrumental bandpasses 
\citep[see, e.g.,][]{Claret2012, 2015MNRAS.450.1879E}. The second method we use is the one suggested by \cite{2011MNRAS.418.1165H}: 
the Synthetic-Photometry/Atmosphere-Modelling (or, SPAM) technique. This is a technique explicitly designed to work with exoplanetary transit lightcurves, and involves generating synthetic transit lightcurves with limb-darkening coefficients estimated from the same tables/codes as the ones described above, which are then fitted with a limb-darkening law 
of choice to retrieve the LDCs. We detail how we explicitly obtain the LDCs we use in 
this work with these two methods below.}}

\subsubsection{LDCs from Tables/Codes}\label{sec:LDC_tables}
In order to perform theoretical predictions for LDCs obtained from our lightcurve fits, we made use of limb-darkening 
tables and/or algorithms present in the literature that use both \textsc{atlas} and \textsc{phoenix} model stellar atmospheres as inputs. 
To put all those predictions on equal footing and to maximize the amount of information extracted from those tables, 
instead of extracting tabulated quadratic limb-darkening coefficients we followed the approach of \citet{2015MNRAS.450.1879E}, in which ``limiting'' limb darkening coefficients are derived using the non-linear 
law. This approach guards against the fact that different models and methods to fit for the intensity profiles of 
model stellar atmospheres give rise to slightly different results depending on the number of points sampled from 
those profiles. In essence, the method simply performs $\chi^2$ - minimization between the non-linear law and the 
quadratic law, to find a relationship that gives the quadratic limb-darkening coefficients as a function of the 
non-linear limb-darkening coefficients {{\citep[see][for a derivation]{2015MNRAS.450.1879E}}}, which is given by: 
\begin{equation}\label{eq2}
    \begin{split}
        u_1 &= \frac{12}{35}c_1 + c_2 + \frac{164}{105}c_3 + 2c_4 \\
        u_2 &= \frac{10}{21}c_1 - \frac{34}{63}c_3 - c_4.
    \end{split}
\end{equation}

Here $c_1$, $c_2$, $c_3$ and $c_4$ are the LDCs of the non-linear law, and $u_1$ and $u_2$ are the resultant ``limiting" 
LDCs for the quadratic law. We use these latter ones to compare our retrieved limb-darkening coefficients from 
our transit fits.

We retrieve theoretical LDC calculations on the \textit{TESS} bandpass for our targets using two 
different sources/methods. The first set is obtained using the \texttt{limb-darkening}\footnote{\url{https://github.com/nespinoza/limb-darkening}} code outlined in \citet{2015MNRAS.450.1879E}. The second are the tables published by \citet{2017AA...600A..30C}, which are 
arguably very popular among researchers that analyze \textit{TESS} data. 

There are several details and/or assumptions one has to be careful with when dealing with those sources of 
LDCs, which we here pay attention to when comparing them against our empirically determined LDCs using \textit{TESS} 
lightcurves. The first is that, as put forward by \cite{2004AA...413..711W}, \textsc{atlas} and \textsc{phoenix} model 
stellar atmospheres disagree about where the stellar atmosphere ends, which in turn implies that different results 
between both are practically guaranteed if not accounted for. Different authors have approached this problem in different 
ways. On the one hand, \citet{2015MNRAS.450.1879E} follow \cite{2004AA...413..711W} and look for an inflection point in the 
intensity profile of spherically symmetric models (i.e., where the derivative of the intensity as a function of 
$r = \sqrt{1 - \mu^2}$ reaches its maximum), redefining that point to be at $r = 1$ (i.e., $\mu = 0$). 
\citet{2017AA...600A..30C}, on the other hand, suggests that a ``better" approach --- 
where ``better" is defined as attaining a lower residual sum of squares with respect 
to the fitted intensity profile (see their Figure 1) --- is to 
simply not use values where $\mu<0.1$, a method they refer to as the ``quasi-spherical" method. Those two methods are fundamentally different approaches, and it is 
furthermore unclear which one of those is better (or if they can be even distinguished 
with actual data). In order to attempt at testing their efficacy at predicting 
empirically determined LDCs, here we extract LDCs from the tables 
of \citet{2017AA...600A..30C} using both methods. Following that same study, in this 
work we refer to the former as the \textit{r-method}, and to the latter as 
the \textit{q-method}.

The second important detail has to do with the fact that \citet{2015MNRAS.450.1879E} 
used the ``vanilla" \texttt{PHOENIX - COND} models, available from the \textsc{phoenix} website\footnote{\url{ftp://phoenix.astro.physik.uni-goettingen.de/SpecIntFITS/}}. \citet{2017AA...600A..30C}, however, suggested that their work uses slightly different 
versions of those models \citep[the ones from][]{Claret2012}, and that furthermore, discrepancies between the works of \cite{Claret2012} and \cite{2015MNRAS.450.1879E} for 
stars with effective temperatures lower than 3000 K could be explained due to \cite{Claret2012} using \texttt{PHOENIX-DRIFT} models. While the latter cannot be tested with 
the data obtained in this work (as none of our stars has effective temperatures smaller 
than 3000 K), the impact of the slightly different \texttt{PHOENIX-COND} models used by these 
authors \textit{can} be tested, by paying attention to the prediction error between 
the tables of \citet{2017AA...600A..30C} using the \textit{q-method} and the predictions made 
using the \texttt{limb-darkening} code of \citet{2015MNRAS.450.1879E}, which for 
\textsc{phoenix} use the \textit{r-method} by default. Here, thus, we also 
take the opportunity to test which of these versions of the \textsc{phoenix} models 
is actually preferred to minimize mismatches with actual empirical LDCs obtained 
from \textit{TESS} lightcurves.

\subsubsection{Synthetic-Photometry/Atmosphere-Model LDCs}\label{sec:LDC_SPAM}

{{Obtaining LDCs from fits to model intensity profiles is an 
inherently different optimization problem to that of obtaining LDCs from fits to 
transiting exoplanet lightcurves. They have different optimization functions as 
they are optimizing different supposed observables. As shown by \cite{2011MNRAS.418.1165H}, {{this fact may give rise to completely different LDCs, depending on the adopted fitting method}}, even if the same coefficients are used to generate both 
the intensity profile and the transiting exoplanet lightcurve. In order to take 
this into account when comparing theoretical to empirically obtained LDCs 
through transit lightcurve fitting, \cite{2011MNRAS.418.1165H} proposed an alternative way of computing theoretical LDCs: the Synthetic-Photometry/Atmosphere-Model (SPAM).}}

{{The first step of the SPAM algorithm is to calculate a ``synthetic'' 
transit lightcurve with an accurate representation of the stellar intensity profile 
(in the case of the present work, described by the non-linear law) assuming full knowledge 
of planetary and stellar properties \citep[although this can be relaxed; see][]{2015MNRAS.450.1879E}. This synthetic transit lightcurve is then fitted by a lightcurve model with fixed planetary properties and any limb-darkening law of interest 
with free LDCs. The retrieved LDCs through this procedure are the SPAM LDCs. Here, we 
follow this latter procedure by fitting the synthetic lightcurves with a 
\textit{quadratic} limb-darkening law.}} {{We use the \texttt{ExoCTK}\footnote{\url{https://github.com/ExoCTK/exoctk} \\ see also, \url{https://github.com/Jayshil/ld-spam/blob/main/p2.py} for our implementation} \citep[][]{matthew_bourque_2021_4556063} package to compute the SPAM LDCs using the non-linear LDCs derived from the various stellar model atmospheres discussed in Section \ref{sec:LDC_tables}.}}


\section{Results}\label{sec:results}
{{Having fitted the \textit{TESS} transit lightcurves of the 176 transiting 
exoplanets in our sample, we are now in a position to compare the empirically 
obtained LDCs with the theoretical ones described in the previous section.}} In what follows, we present these results in two parts. In the first part, we compare the retrieved planetary parameters obtained from our \texttt{juliet} fits to the \textit{TESS} data with their corresponding literature values in order to validate our modelling procedure. Then, we compare the retrieved LDCs from those transit fits with the ones calculated theoretically from model stellar atmospheres obtained as discussed in Section \ref{sec:theo_LDC}. The codes which we used to produce all the results in this section are publicly available on Github\footnote{\url{https://github.com/nespinoza/tess-limb-darkening},\\ \url{https://github.com/Jayshil/ld-project-updated}, and,\\ \url{https://github.com/Jayshil/my_thesis}}.


\subsection{Comparison with literature values}\label{sec:comp_lit}
Our \texttt{juliet} fits allow us to estimate various planetary parameters along with the limb darkening coefficients. The former, in turn, provide us with an excellent dataset which is useful not only to refine these planetary parameters, but also to validate our procedures by comparing them to the ones already estimated in the literature. {{While such a validation in principle assumes these literature values are unbiased, we believe the risk of this not being the case 
is somewhat mitigated by the wide range of instruments, analyses and assumptions made by different authors that analyzed these 
systems in the past. To perform this comparison}}, we chose three planetary parameters to compare in this work, which are the 
ones for which we obtain the most precise constraints: the scaled semi-major axis ($a/R_*$), the 
planet-to-star radius ratio ($R_p/R_*$) and the time of transit center ($t_c$). 

\begin{figure*}
	\includegraphics[width=2.14\columnwidth]{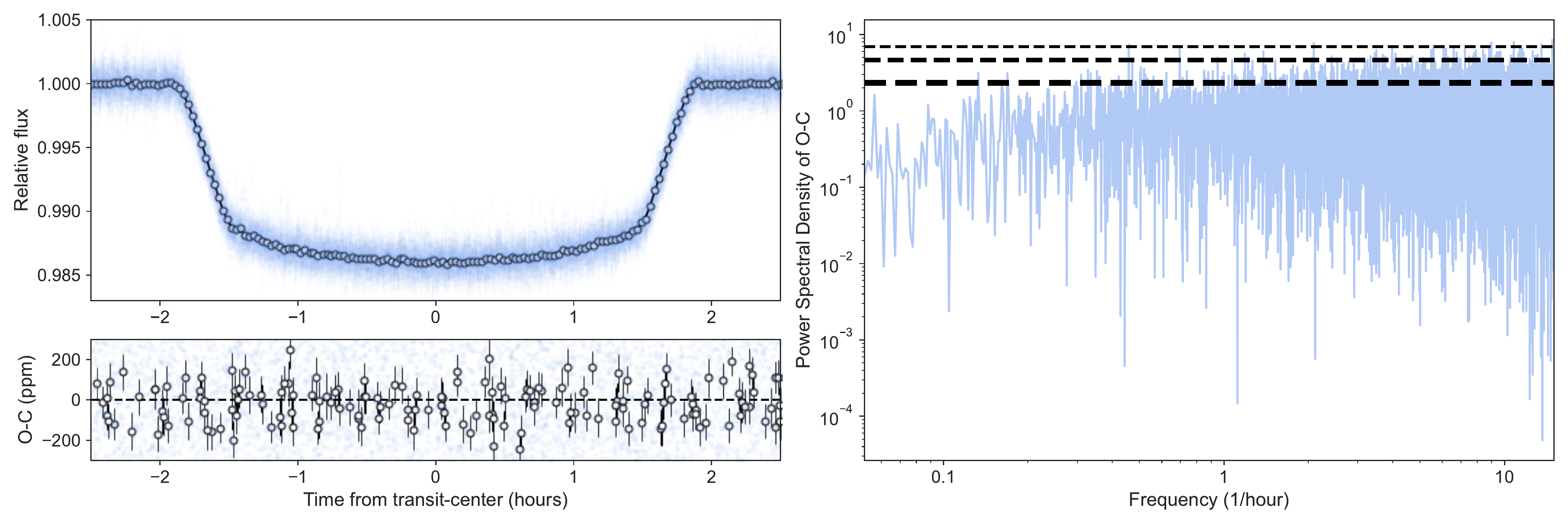}
    \caption{ {{(\textit{Left}) Phase-folded lightcurve for the systematics-removed WASP-62b system with data taken by TESS combining the observations of all sectors (top left plot; blue datapoints), and the residuals from 
    the best-fit model (bottom left plot). The white circles with black errorbars are binned datapoints in 2-minute intervals in this phased lightcurve. (\textit{Right})  Power spectral density (PSD) of the residuals of the 
    best fit, which doesn't show any significant signal on the frequencies relevant to the transit event. 
    Dashed lines from top to bottom show where 99.9\%, 99\% and 90\% of the PSD of simulated gaussian noise with the same properties at those of the errorbars on our data falls; the observed PSD 
    of the residuals falling below these lines suggests it is consistent with that of white gaussian noise)}}.}
    \label{fig:fig4}
\end{figure*}

\begin{figure*}
	\includegraphics[width=\textwidth]{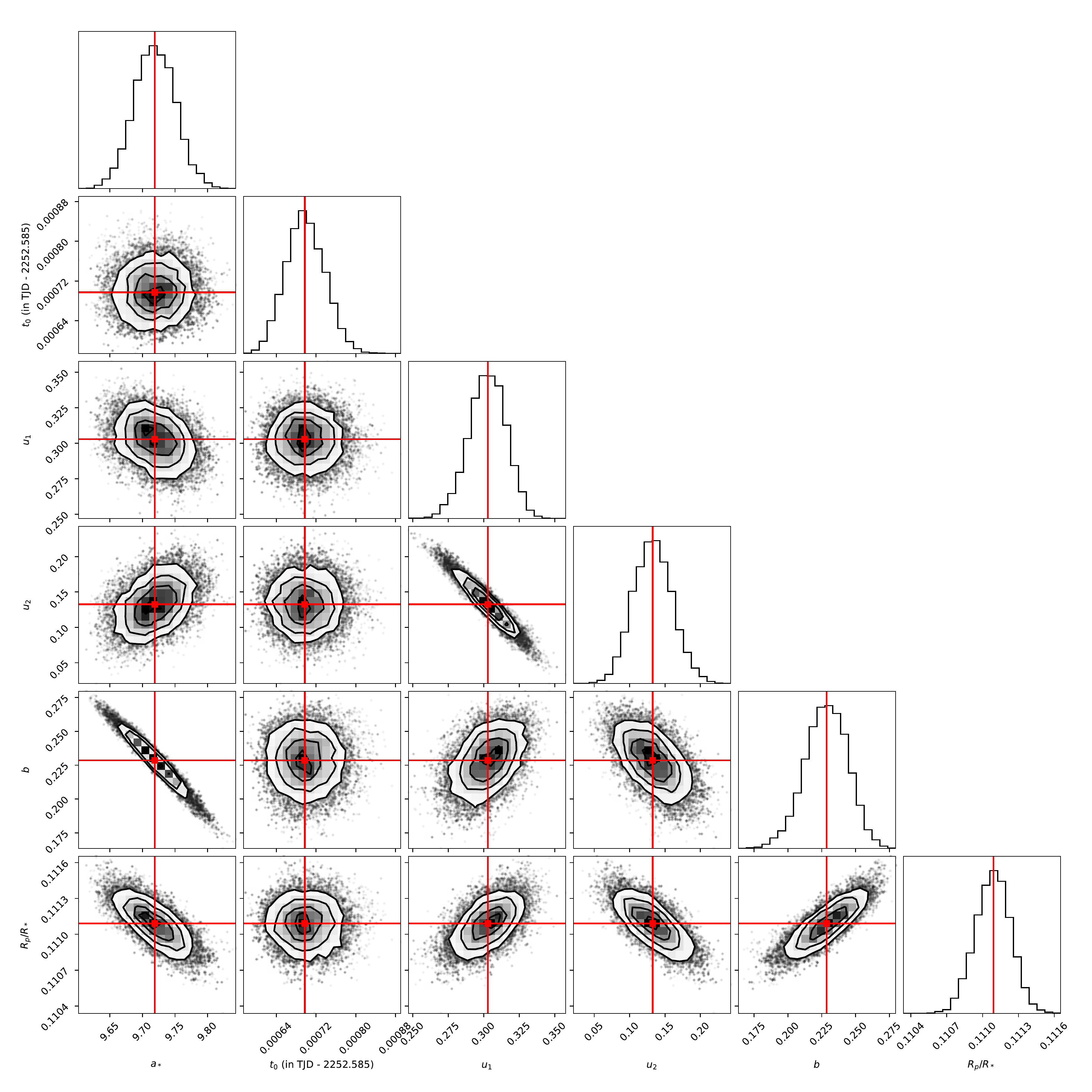}
    \caption{Corner plot of various retrieved parameters for WASP-62b system}
    \label{fig:corner}
\end{figure*}

Before presenting the main results of our study for all the planets in our sample, we 
first present a representative fit for the WASP-62b system as observed by \textit{TESS} 
{{(including all sectors on which this exoplanet was observed). The summary 
of this fit is presented in Figure \ref{fig:fig4}. As can be seen, 
the phase-folded transit lightcurve has exquisite precision, and doesn't give rise to any 
noticeable leftover signal in the residuals. Indeed, a power spectral density analysis reveals 
no significant signals are present in the residual time-series, suggesting our Exponential Matern 
kernel did take care of any correlated noise structure in the data. In terms of achieving the 
photon-noise level of these observations, for some sectors a significant jitter term of a couple of 
hundreds of ppm is found. This is the case for Sector 1, 4, 7, 8 and Sector 27 through 
34. This probably points to the fact that the PDC algorithm has some extra uncorrected systematics 
in those sectors, but the amplitude seems to be overall small and, judging from our power spectral 
density analysis, it is unimportant in terms of correlated noise in the time-scales of interest for 
our analyses ($\sim$ hours).}} The posterior distribution of the transit lightcurve parameters for this target 
are presented in Figure \ref{fig:corner}, where we have transformed the $q_1$ and 
$q_2$ limb-darkening parameters back to the $u_1$ and $u_2$ plane using the 
transformations in \cite{2013MNRAS.435.2152K}. These posterior parameters, 
in turn, {{are in very good} agreement} with previous values in the literature. For instance, the planet-to-star radius ratio ($R_p/R_*$) from our fits is 
$0.1111 \pm 0.0001$, which is within $3-\sigma$ from the value found by \citet{2020AcA....70..181M} of $0.1105 \pm 0.0003$. A similar agreement is found for 
the rest of the transit parameters: our estimated scaled semi-major axis ($a/R_*$) 
is $9.72\pm 0.03$, which is within 1-sigma from the value found by \citet{2017AJ....153..136S} of $9.55 \pm 0.41$. Finally, the difference between the 
transit center predicted by that same work on the time-span of the \textit{TESS} 
observations and the one estimated by our transit fits are also in {{very good} agreement} agreement --- 
they show a difference of $28.757$ seconds, which is consistent with zero within 1-sigma. The 
predicted ephemerides from the work of \cite{2017A&A...602A.107B} has 
considerably deteriorated since its publication, showing a bias of around 12 minutes for 
this target. Our \textit{TESS} updated ephemerides, however, significantly improve 
its precision to only 3 seconds.


\begin{figure}
	\includegraphics[width=\columnwidth]{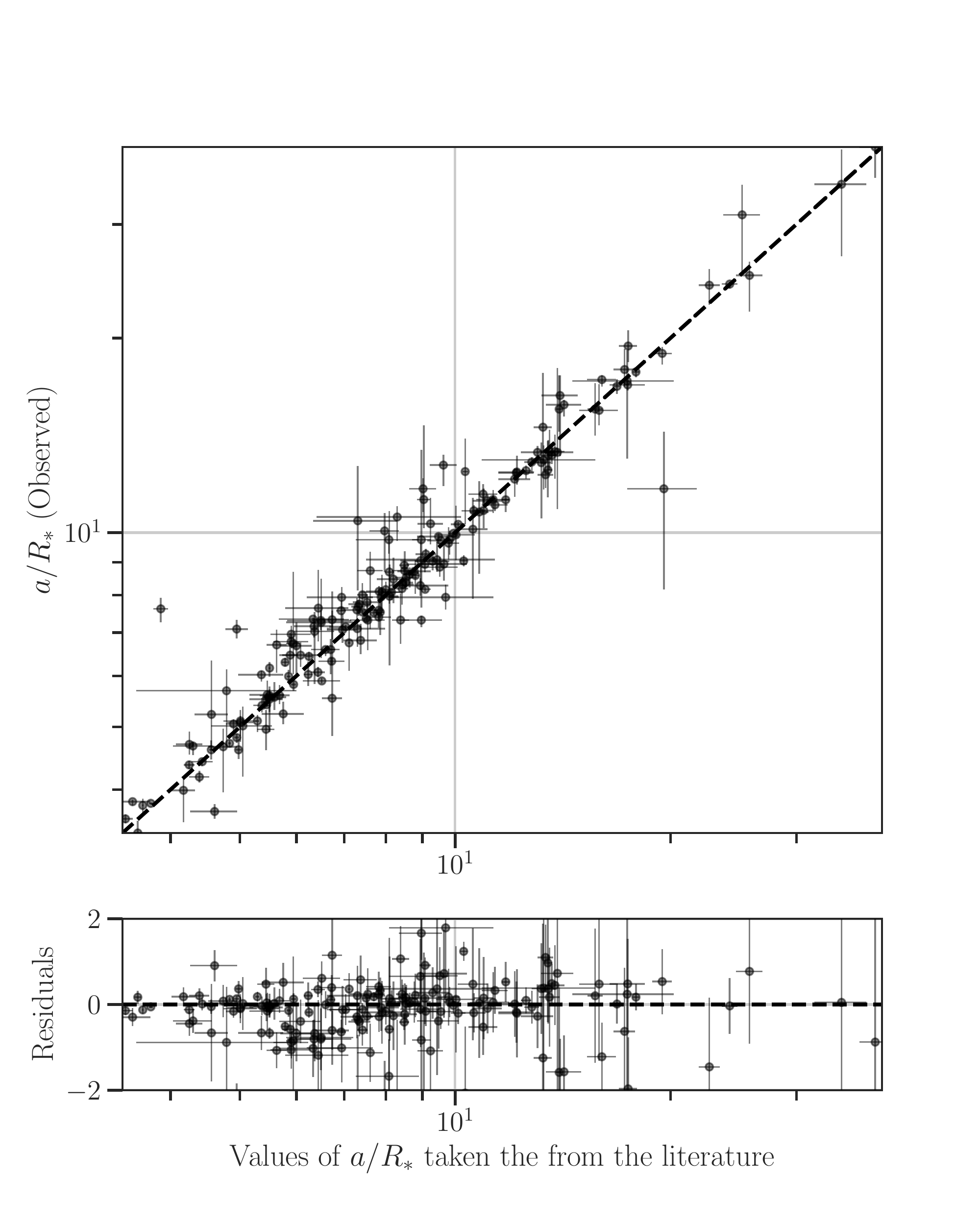}
    \caption{Comparison of the scaled semi-major axis ($a/R_*$) calculated using \texttt{juliet} with its corresponding literature values. The observed values and the values from the literature are on the y-axis and x-axis respectively in the upper panel, while the lower panel shows the residuals between those two values.}
    \label{fig:fig6}
\end{figure}

\begin{figure}
	\includegraphics[width=\columnwidth]{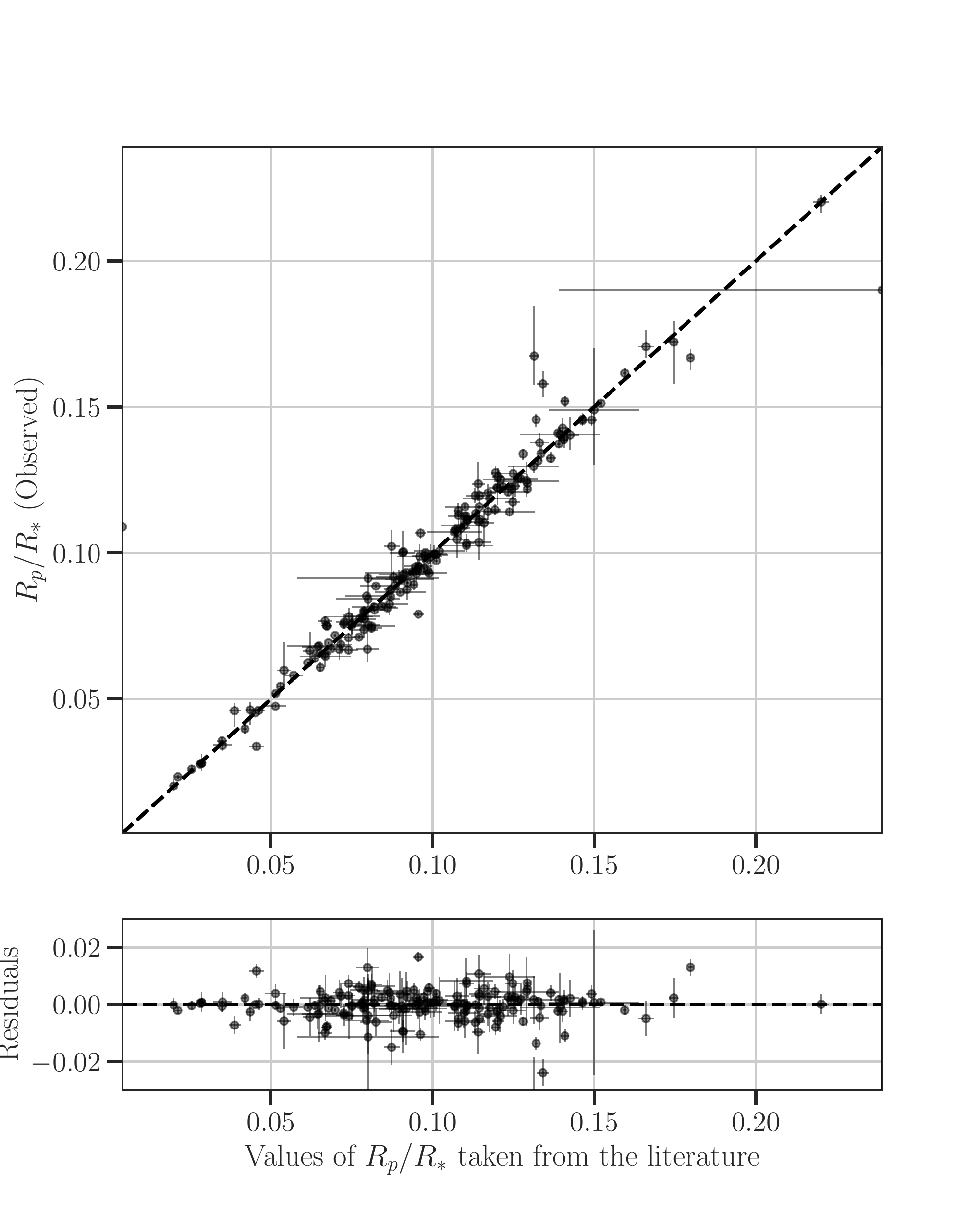}
    \caption{Comparison of the planet-to-star radius ratio ($R_p/R_*$) calculated using \texttt{juliet} with its corresponding literature values. The observed values and the values from the literature are on the y-axis and x-axis respectively in the upper panel, while the lower panel shows the residuals between those two values.}
    \label{fig:fig7}
\end{figure}

\begin{figure}
	\includegraphics[width=\columnwidth]{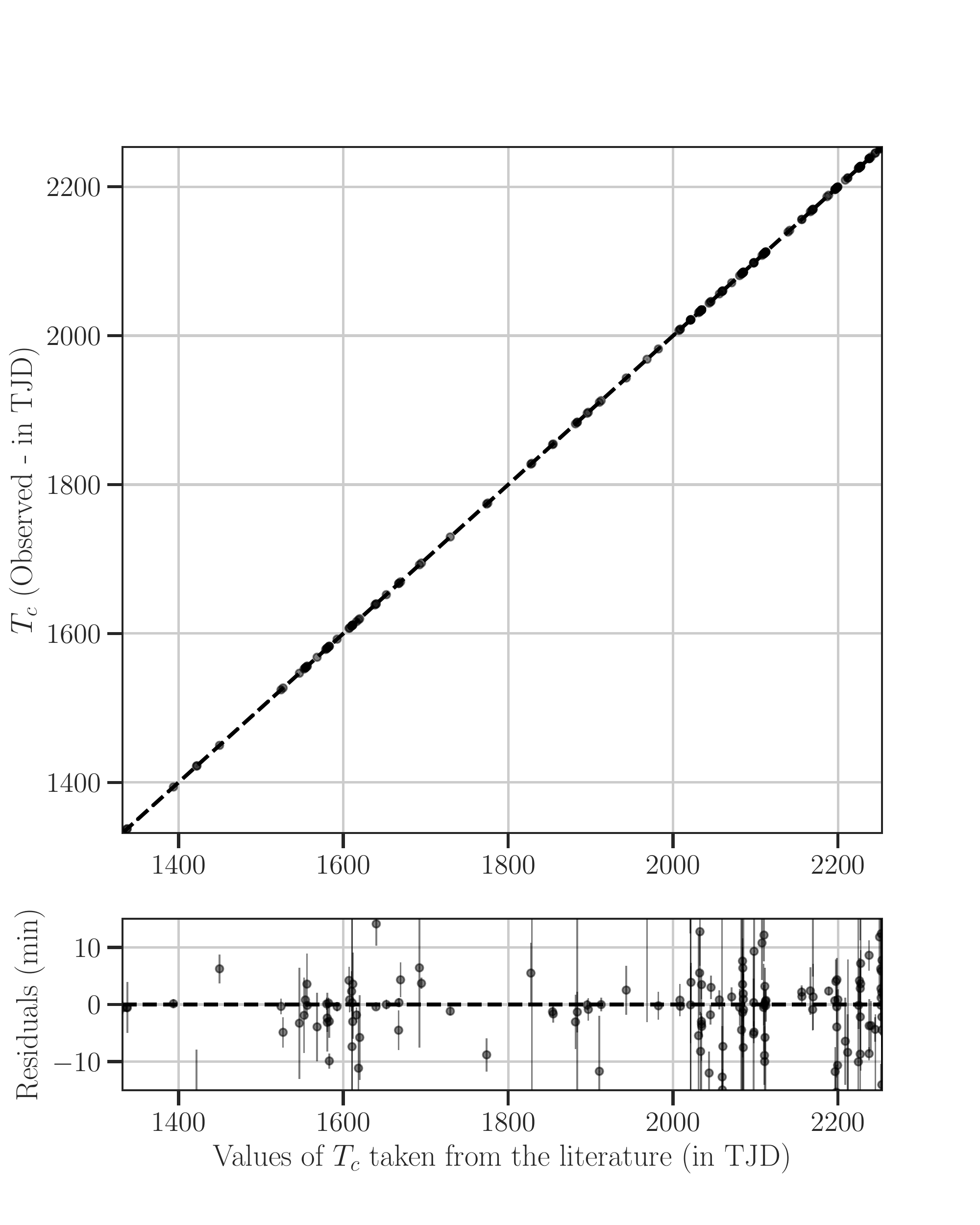}
    \caption{Comparison of the time of transit center ($t_c$) calculated using \texttt{juliet} with its corresponding literature values. The observed values and the values from the literature are on the y-axis and x-axis respectively in the upper panel, while the lower panel shows the residuals between those two values.}
    \label{fig:fig8}
\end{figure}

The same comparison {{on the retrieved transit parameters}} done above for WASP-62b is presented for \textit{all} the targets 
in our sample in Figures \ref{fig:fig6}, \ref{fig:fig7} and \ref{fig:fig8}. {{Literature values for the 
planetary parameters were obtained from the NASA Exoplanet Archive, as queried on February 23, 2021. The comparison 
is also }}presented in Table \ref{tab:Plan_prop}. Our results show in general {{very good} agreement} with literature values. There are, however, a set of {{25 exoplanetary systems in our 
sample for which we find that one or more than one of the planetary parameters compared in Figures 
\ref{fig:fig6}, \ref{fig:fig7} and \ref{fig:fig8} (i.e., the time-of-transit center, $a/R_*$ and/or $R_p/R_*$) are at 
least $3-\sigma$ away from what is published in the literature. A detailed, case-by-case analysis is presented 
in Appendix \ref{sec:appendix-discrepant} for those systems, but we found that for 19 of them we are confident our retrieved 
values are updates over previous published values for these parameters, and that for 5 of them the most likely explanation 
is either stellar activity variations producing slight transit-depth variations, real planetary variability in the transit 
depths and/or mismatches on the dilution factors applied by the \textit{TESS PDC} pipeline. The only target in which we consider 
the methods here presented failed to retrieve the correct properties is LTT9779~b, for which we sample a discrepant solution to the one described in \citet{Jenkins:2020} with our methodology. This solution was actually briefly discussed in \citet{Jenkins:2020}, but 
was discarded as the implied stellar density does not agree with the spectroscopic one observed in that work. Overall, we consider 
having only 1 outlier out of a sample of 176 targets is in fact a very good result, which in turns gives us confidence in our lightcurve 
analysis procedures.}}

{{Before moving on with our results, we would like to highlight the}} curious case of the 
WASP-140~b exoplanetary system, which has a nearby but not physically associated bright star{{. There are many systems in our sample that have such companions, and}} our hypothesis on such targets {{was that, while 
they would}} dilute the transit signal, the \texttt{PDC} lightcurves should have 
accounted for these dilutions in the final {{reported}} fluxes. {{However}}, it seems 
the correction in {{the case of WASP-140~b}} was either not appropriate or not sufficient to account 
for it {{based on a comparison of our original lightcurve fits with those published in the literature 
by \cite{2017MNRAS.465.3693H}}}. We thus decided to take care of this dilution via the so-called dilution factor in the modelling process \citep[see][]{juliet}. Instead of fixing this dilution factor in the modelling procedure to 1, as done for the rest of the targets, we fit it together with the rest of the transit parameters. {{Our results, 
however, still were inconsistent with those of \cite{2017MNRAS.465.3693H} even if accounting directly for this 
dilution. Our fixed dilution fit}} gives rise to a planet-to-star radius ratio which is over 
60\% larger than the one reported by \cite{2017MNRAS.465.3693H} --- and to a much larger radius 
if the dilution is left as a free parameter. In both cases, the impact parameter we retrieve is 
$b>1$ at 5-sigma, which although a couple of sigmas away from the value reported in that study 
($b = 0.93^{+0.07}_{-0.03}$) is still consistent with it in our fixed dilution case. We believe 
that a full joint analysis of the entire photometric and radial-velocity datasets for this 
system is needed in order to solve this discrepancy, but we leave such an analysis for 
future work. Here, we simply discard this {{system}} from our analysis.

\subsection{Analysis of targets in multiple sectors}
\begin{figure*}
    \centering
    \includegraphics[width=2\columnwidth]{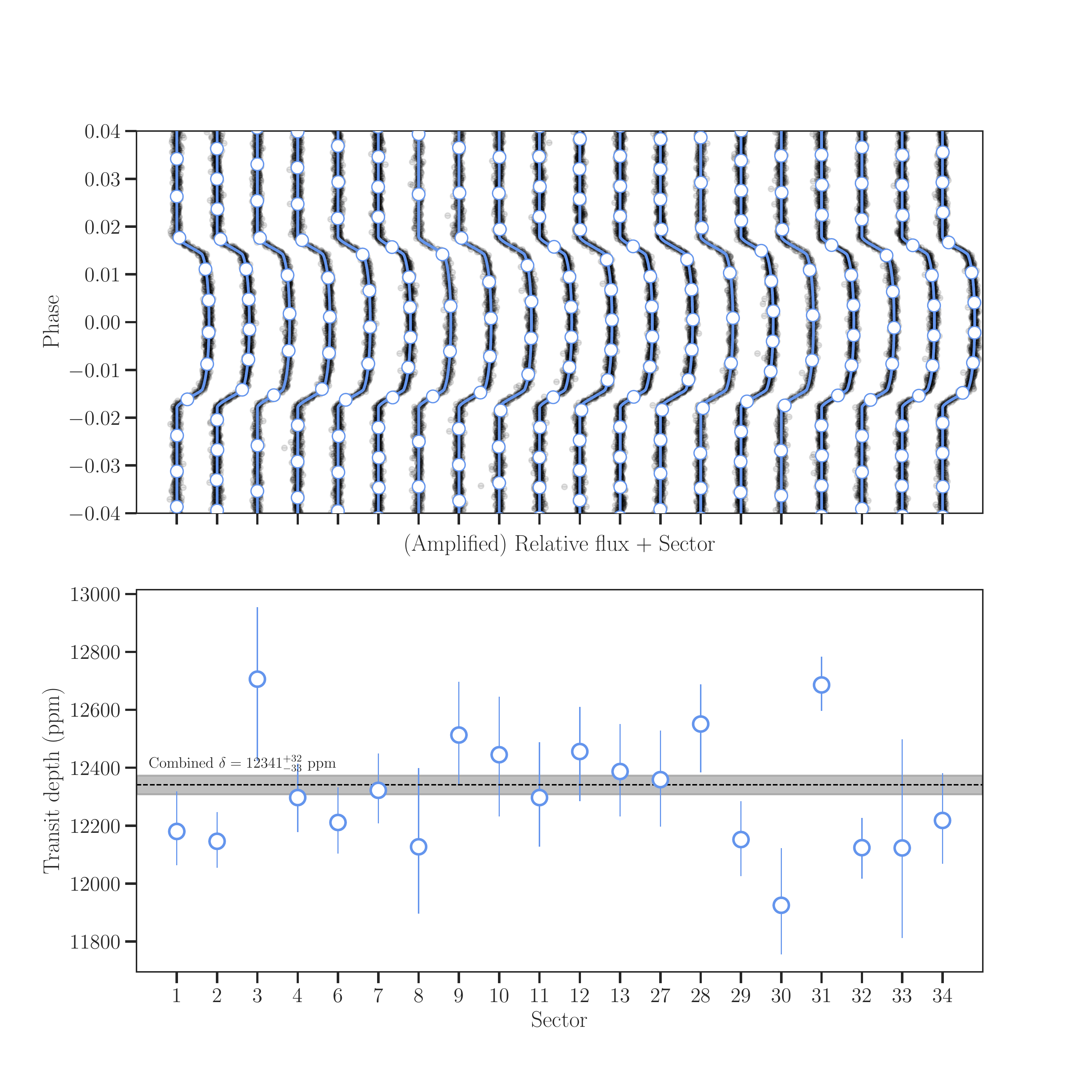}
    \caption{Transit lightcurves (top; black datapoints original 
    data, white points binned data, blue curves best-fit lightcurves) and transit depths (bottom; blue points with errorbars) on 
    each sector for one of our best constrained targets in this work: WASP-62b. As can be seen in the bottom plot, the retrieved 
    transit depths are consistent between sectors, and they converge 
    to a value ($12341^{+32}_{-33}$ ppm; grey band representing this 
    68\% credibility band) which is in agreement with 
    the literature value ($12210 \pm 66$ ppm).}
    \label{fig:wasp-126}
\end{figure*}

\begin{figure*}
    \centering
    \includegraphics[width=2\columnwidth]{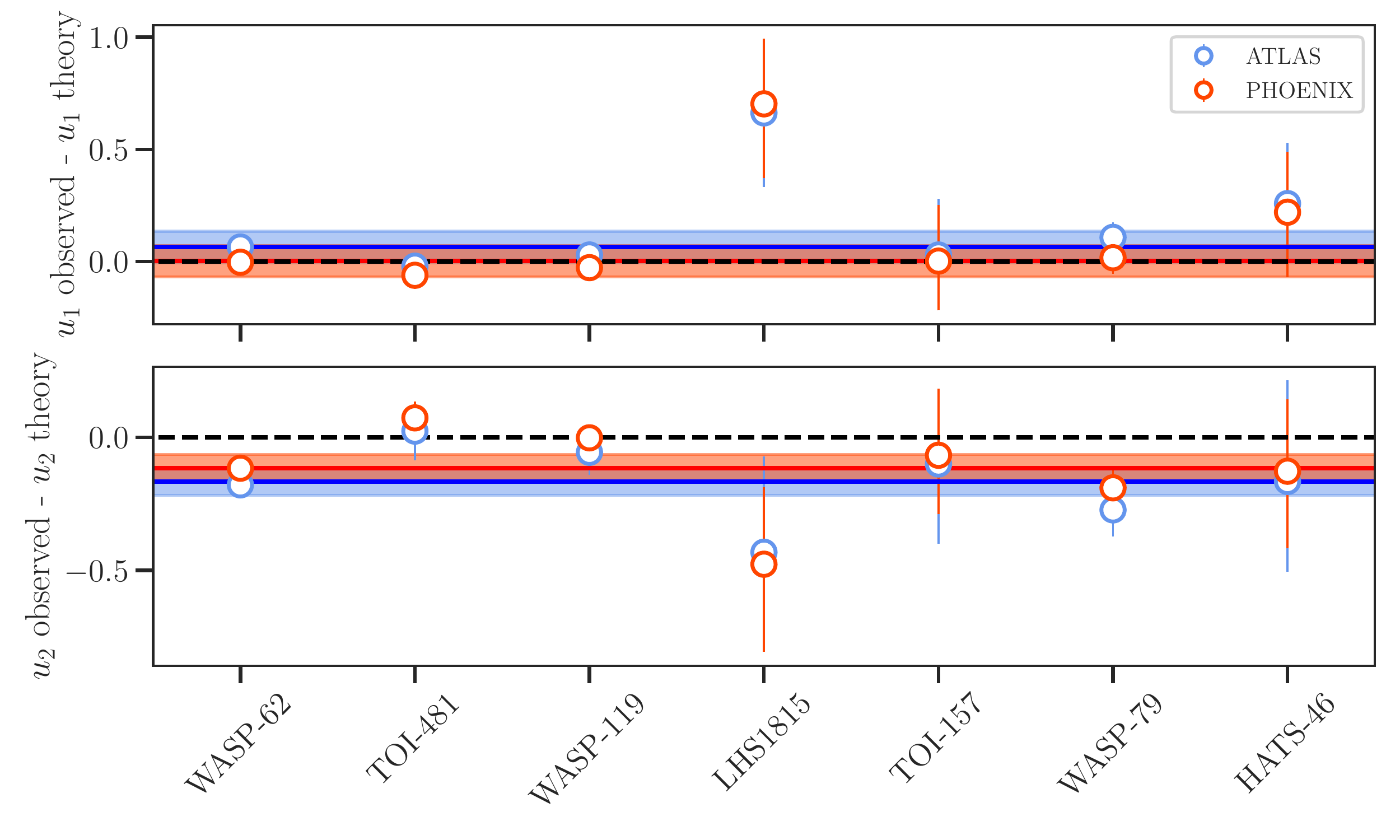}
    \caption{Limb-darkening coefficients extracted for 
    our best-precision targets (the ones spanning data in 
    multiple sectors), compared against their theoretical 
    predictions (blue for \textsc{atlas} models, red for \textsc{phoenix}). 
    As can be seen, a clear offset is observed 
    for the $u_2$ limb-darkening coefficient. The bands show the uncertainties in the mean offset with the red and blue colors representing the \textsc{phoenix} and \textsc{atlas} models. The method to compute theoretical LDCs is the one from \citet{2015MNRAS.450.1879E}.}
    \label{fig:multiplot}
\end{figure*}

An additional good consistency check for the results presented in this 
work is to perform a thorough analysis of exoplanetary 
systems that had data spanning multiple sectors. This 
allows us to test our methods in virtually independent 
datasets of the same systems which not only allows us to 
compare the retrieved transit parameters among different 
sectors, but also extract the most precise LDCs in our 
sample by combining the datasets at hand. For our analysis 
we decided to use all the data available for those systems up to Sector 34, 
which implies some of them had observations in sectors more recent than the ones 
we selected to define our target sample. In our case, 
these datasets were the ones for WASP-62 (20 sectors), TOI-481 (9 sectors), WASP-119 (12 sectors), 
LHS1815 (20 sectors), TOI-157 (12 sectors), WASP-79 (4 sectors) and HATS-46 (4 sectors).

In Figure \ref{fig:wasp-126}, we present our lightcurve 
fit results for WASP-62b in different sectors, alongside 
with the corresponding transit depths. {{The sector-by-sector 
fits were performed with the Exponential-Mat\`ern GP kernel}}. As can be seen, 
the transit depths are {{mostly}} consistent between 
sectors and they actually converge to the same value that 
is found in the literature. {{The only significantly 
discrepant value is that of Sector 31, which has a transit depth of 
$12687 \pm 97$ ppm --- $\sim 3-$sigma away from the combined transit 
depth. Interestingly, our analyses show that the discrepancy in this 
case is mostly driven by the selection of the GP kernel: a quasi-periodic kernel 
fit on this particular sector (which gives a much better bayesian evidence) gives 
back a transit depth of $12376 \pm 129$ ppm; consistent with the combined depth. 
While this would hint that we should perhaps allow different GP kernels to be 
fit on different sectors, we found that for the combined multi-sector analysis we 
perform in this work this extra complexity is not important, as we obtain the same 
results either way. Not considering Sector 31, we ran a chi-square test comparing the 
observed depths and errors from all sectors to that of the mean depth across all of them. 
This gave back a p-value of 0.34 --- with which we fail to reject the null hypothesis that 
the data is consistent with gaussian noise. This is good evidence that, indeed, the transit 
depths are constant accross sectors, and consistent with the value 
found in the literature}}.  As can be seen in Table 
\ref{tab:Plan_prop}, the same applies to all the other 
planets observed in multiple sectors. This gives us 
confidence that our analysis is also well behaved between 
sectors.

The corresponding limb-darkening coefficients extracted 
from the transit lightcurves for those targets are 
presented in Figure \ref{fig:multiplot}, where we 
plot the observed minus the predicted theoretical 
limb-darkening coefficients, which we obtained from \citet{2015MNRAS.450.1879E}, for our targets both using 
\textsc{phoenix} and \textsc{atlas} models. As can be seen, the limb-darkening 
coefficients are largely consistent for the $u_1$ coefficient, for both \textsc{atlas} and \textsc{phoenix} models though a small offset is still present. However, for the $u_2$ coefficients, both models are evidently a poor 
match to the observations, producing a comparatively large offset. As we will see below, such discrepancy is 
not exclusive of the multiple-sector data: this happens 
with the vast majority of the systems we analyzed.

\subsection{Comparison between theoretical and empirical LDCs from TESS}\label{sec33}

\begin{figure*}
    \centering
    \includegraphics[width=2\columnwidth]{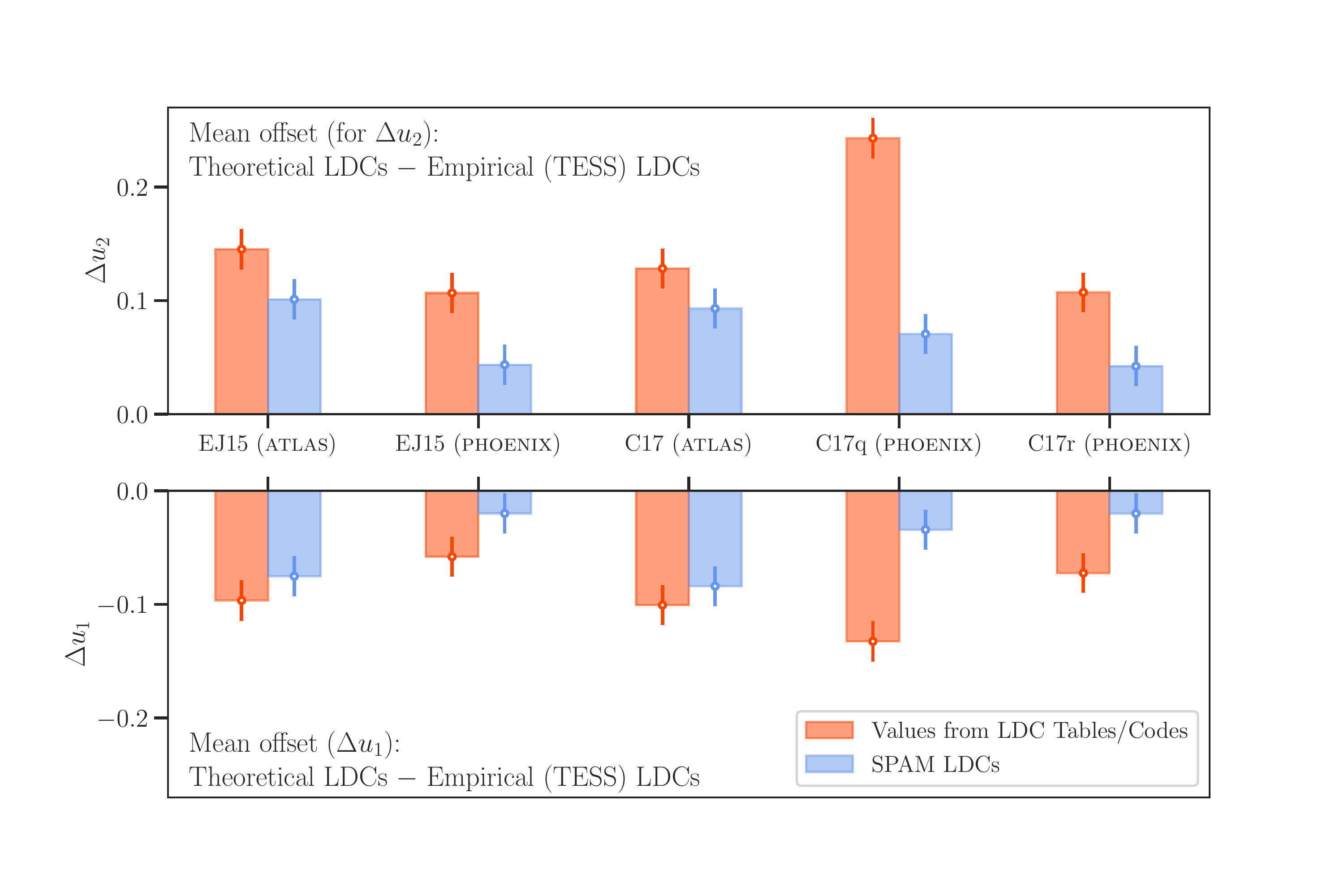}
    \caption{{Mean offset between empirically determined LDCs (through \textit{TESS} transiting exoplanet lightcurves) and the theoretically derived LDCs when using a quadratic law. Different colors of the bars represent the different methodologies to compute theoretical LDCs, from LDCs Tables/Codes (Section \ref{sec:LDC_tables}, orange) and SPAM LDCs (described in Section \ref{sec:LDC_SPAM}, blue). The x-axis shows the various origins of these theoretical LDCs: the work of \citet[][EJ15]{2015MNRAS.450.1879E} 
    and the work of \citet[][C17]{2017AA...600A..30C}. The additional sufix for this latter work  (\textit{q} and \textit{r}) 
    represent two different sets of assumptions used by \citet{2017AA...600A..30C} to obtain these LDCs (see text for 
    details). The top panel shows the offset for the $u_2$ coefficient while the lower panel shows the offset for the 
    $u_1$ coefficient of the quadratic law.}}
    \label{fig:mean-off}
\end{figure*}

\begin{deluxetable*}{lcc}
 \tablecaption{Mean offset present in the {{Tabular/Code and SPAM}} LDCs using \textsc{phoenix} and \textsc{atlas} model stellar atmospheres with different calculation methods.}
 \label{tab:mean_off}
 \tablehead{ \colhead{Method} & \colhead{$\Delta u_1$} & \colhead{$ \Delta u_2$}}
 \startdata
    \multicolumn{3}{c}{\text{Tabular/Code LDCs}}\\ \hline
    \textsc{atlas} {\citep{2015MNRAS.450.1879E}} & $-0.096 \pm 0.018$ & $0.145 \pm 0.022$ \\
    \textsc{phoenix} {\citep{2015MNRAS.450.1879E}} & $-0.058 \pm 0.018$ & $0.107 \pm 0.022$ \\
    \textsc{atlas} {\citep{2017AA...600A..30C}} & $-0.101 \pm 0.018$ & $0.128 \pm 0.021$ \\
    \textsc{phoenix} - \textit{q-method} {\citep{2017AA...600A..30C}} & $-0.132 \pm 0.018$ & $0.243 \pm 0.022$ \\ 
    \textsc{phoenix} - \textit{r-method} {\citep{2017AA...600A..30C}} & $-0.072 \pm 0.018$ & $0.107 \pm 0.022$ \\ \hline
    \multicolumn{3}{c}{\text{SPAM LDCs}}\\ \hline
    \textsc{atlas} {\citep{2015MNRAS.450.1879E}} & $-0.075 \pm 0.018$ & $0.101 \pm 0.022$ \\
    \textsc{phoenix} {\citep{2015MNRAS.450.1879E}} & $-0.019 \pm 0.018$ & $0.044 \pm 0.021$ \\
    \textsc{atlas} {\citep{2017AA...600A..30C}} & $-0.084 \pm 0.017$ & $0.093 \pm 0.022$ \\
    \textsc{phoenix} - \textit{q-method} {\citep{2017AA...600A..30C}} & $-0.034 \pm 0.018$ & $0.071 \pm 0.022$ \\ 
    \textsc{phoenix} - \textit{r-method} {\citep{2017AA...600A..30C}} & $-0.020 \pm 0.018$ & $0.042 \pm 0.022$ \\ \hline
\enddata
\end{deluxetable*}

In the previous sections we have performed a detailed comparison between the 
retrieved planetary parameters with their corresponding literature 
values, finding {{very good}} agreement between the two. We consider these results as a 
validation that our fits to \textit{TESS} data are indeed adequately constraining 
the transit lightcurve shapes.

Having validated our results with literature data on the physical planetary 
parameters (relative to those of the star), we now switch our focus in order 
to compare the retrieved LDCs with their corresponding theoretical values, obtained 
as described in Section \ref{sec:theo_LDC}. The empirically determined LDCs for each 
of the targets analyzed in this work as well as their theoretical predictions are 
presented in Table \ref{tab:ldcs} and Table \ref{tab:ldcs_SPAM}.

We present a {{system-by-system}} comparison between the empirical and the 
theoretical LDCs using {{the various here outlined}} procedures in Figures \ref{fig:u1_code} through \ref{fig:u1_cla_SPAM_r}{: from Figure \ref{fig:u1_code} through \ref{fig:u1_cla_r} we provide the comparison against theoretical LDCs obtained directly 
from previously published tables/codes (i.e., following the method 
described in Section \ref{sec:LDC_tables}), and from Figures \ref{fig:u1_code_SPAM} through \ref{fig:u1_cla_SPAM_r} we present the results following the SPAM methodology for 
obtaining the LDCs (i.e., following the method 
described in Section \ref{sec:LDC_SPAM}). To further summarize those results, in Figure \ref{fig:mean-off} we present the mean offset between the \textit{TESS} retrieved LDCs and the theoretically predicted ones for both sets of LDCs; these mean 
offsets are in turn also presented in Table \ref{tab:mean_off}.}

{{As can be observed from Figure \ref{fig:mean-off}, the general behaviour of the offsets in the LDCs remain the same independent of which method one decides to use to compute the theoretical LDCs. The $u_1$ coefficients are consistently 
under-predicted by those theoretical calculations, while the $u_2$ coefficients are over-predicted. This latter coefficient, 
however, is the one that shows the largest (absolute) offset accross methods. Perhaps one of the most interesting features of 
these results, however, is the significantly lower offset the SPAM LDCs show when compared against the empirically 
obtained LDCs --- similarly to what was observed by \textit{Kepler} in the analyses of \citet{2011MNRAS.418.1165H} 
and \citet{2015MNRAS.450.1879E}. As in those works, this suggests that this is, indeed, on average the correct way of extracting 
theoretical LDCs for usage in transit lightcurve modelling. LDCs extracted without this SPAM algorithm all show 
significant offsets ($>3\sigma$) with 
respect to empirically determined LDCs in at least one coefficient, with the worst performing method being the 
\textit{q-method} of \citet{2017AA...600A..30C} which shows a mean offset on the $u_2$ coefficient of 
$\Delta u_2 = 0.243 \pm 0.022$ --- a very large offset when one considers the space of all possible coefficients for $u_2$ 
spans the range from -1 to 1. When applying the SPAM methodology, however, most of these offsets become consistent with zero 
at about 2-3 $\sigma$ levels for both $u_1$ and $u_2$. The exceptions are the LDCs calculated using the \textit{q-method} 
of \citet[][$\Delta u_2 = 0.071 \pm 0.022$]{2017AA...600A..30C} and the \texttt{ATLAS} LDCs, which show significant 
offsets on both coefficients for both, the calculations made using the \citet{2017AA...600A..30C} tables or the \texttt{limb-darkening} library of \citet{2015MNRAS.450.1879E}.}}

\section{Discussion}\label{sec:discussion}

{{The results presented in the previous section and summarized in Figure \ref{fig:mean-off} seem to suggest 
two main take-home messages. First, when using quadratic LDCs extracted directly from limb-darkening tables, large 
offsets as large as $\Delta u_1 \approx -0.1$ and $\Delta u_2 \approx 0.2$ can be expected, on average, with empirically 
determined LDCs in the \textit{TESS} bandpass, with the offsets being worse for the \texttt{PHOENIX} models than for 
the \texttt{ATLAS} models (top set of values in Table \ref{tab:mean_off}). If one uses the SPAM methodology to obtain 
these LDCs, however, these offsets reduce, on average, to virtually zero for LDCs obtained using the \texttt{PHOENIX} 
stellar models and the \textit{r-method} with both the tables published by \citet{2017AA...600A..30C} and the 
\texttt{limb-darkening} library of \citet{2015MNRAS.450.1879E} (see bottom set of values in Table \ref{tab:mean_off}). 
This, in turn, suggests the SPAM methodology is the one that should be adopted by default when using theoretical LDCs 
to perform transit lightcurve modelling in the \textit{TESS} bandpass, in particular using that technique together 
with \texttt{PHOENIX} stellar model atmospheres. 

These take-home messages, however, do not tell the entire story of the observed offsets between theoretically and 
empirically determined LDCs. Given different stars have different limb-darkening profiles, we must be careful with 
interpreting these results especially considering the overabundance of F and G-type stars in our sample. With 
this motivation in mind, we discuss the variation of these offsets with host star temperature in the next sub-section.}}

\subsection{Variation in offset with effective temperature of the host star}

\begin{figure*}
    \centering
    \textbf{For Tabular/Code LDCs}\par\medskip
    \includegraphics[width=\columnwidth]{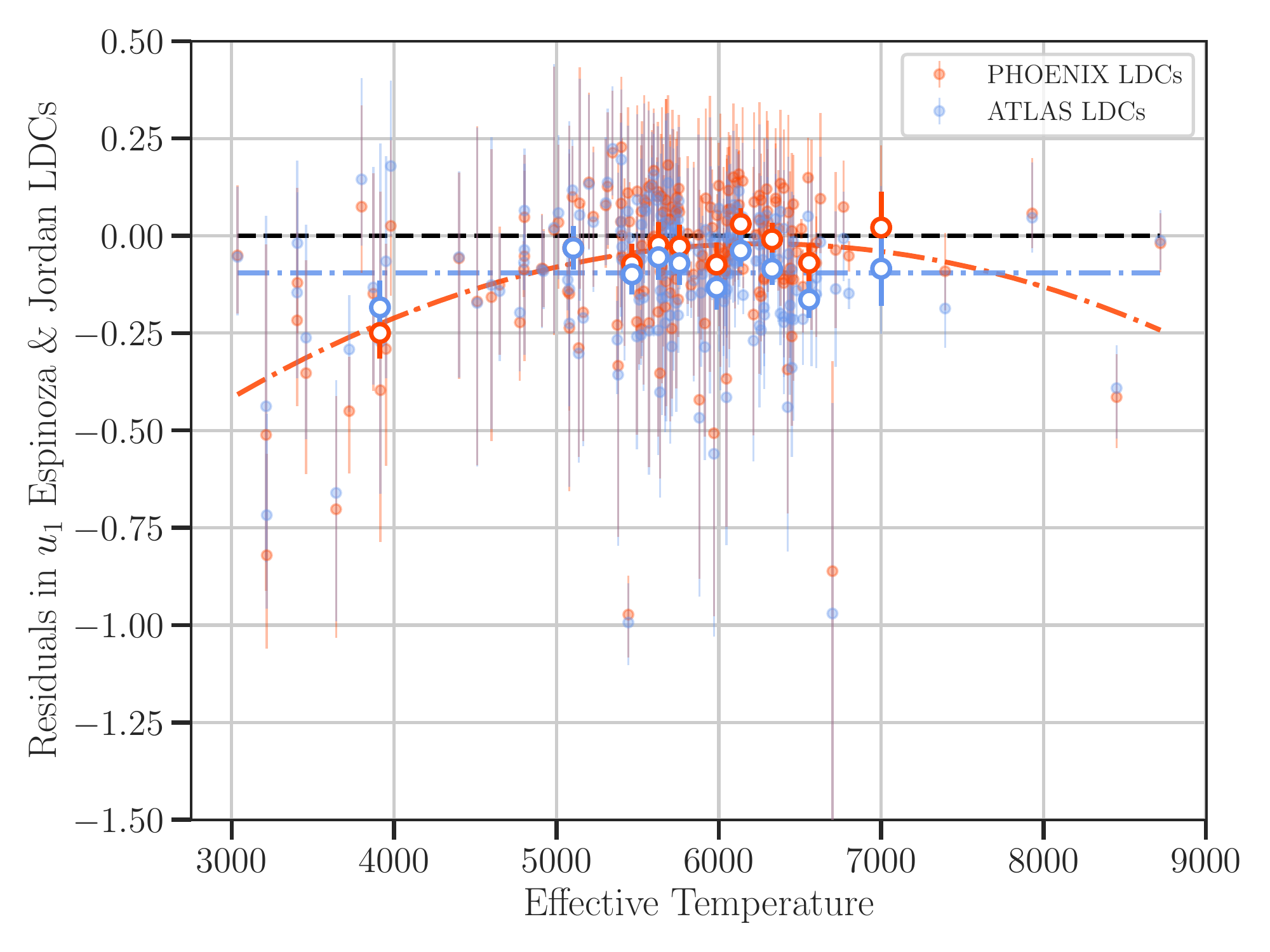}
    \includegraphics[width=\columnwidth]{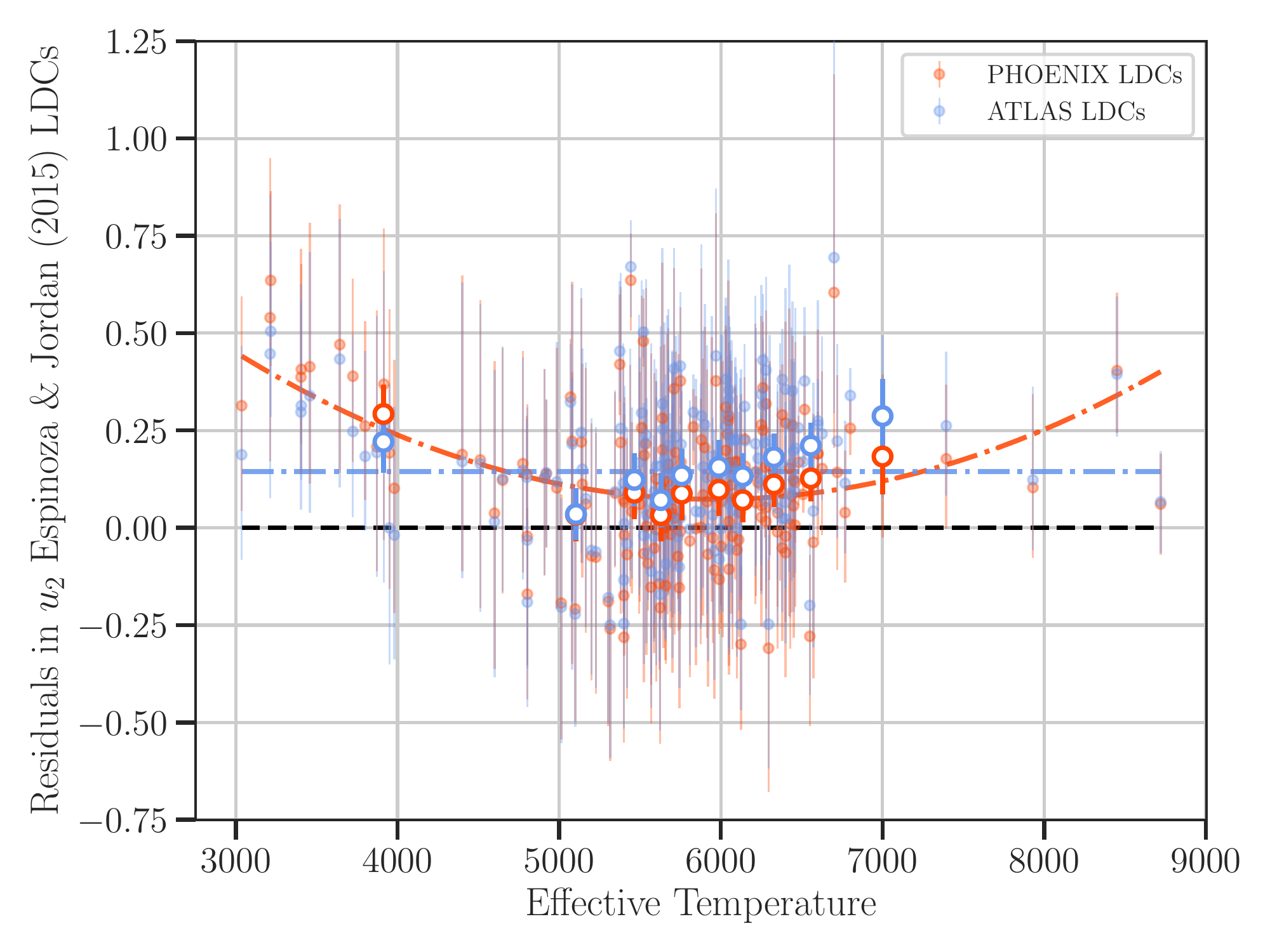}
    \caption{Temperature variation of offsets in $u_1$ and $u_2$, when, theoretically, LDCs are calculated using code provided by \citet{2015MNRAS.450.1879E}. The dashed-dotted lines show the best fitted model to the residuals.}
    \label{fig:u1_code_te}
\end{figure*}


\begin{figure*}
    \centering
    \textbf{For SPAM LDCs}\par\medskip
    \includegraphics[width=\columnwidth]{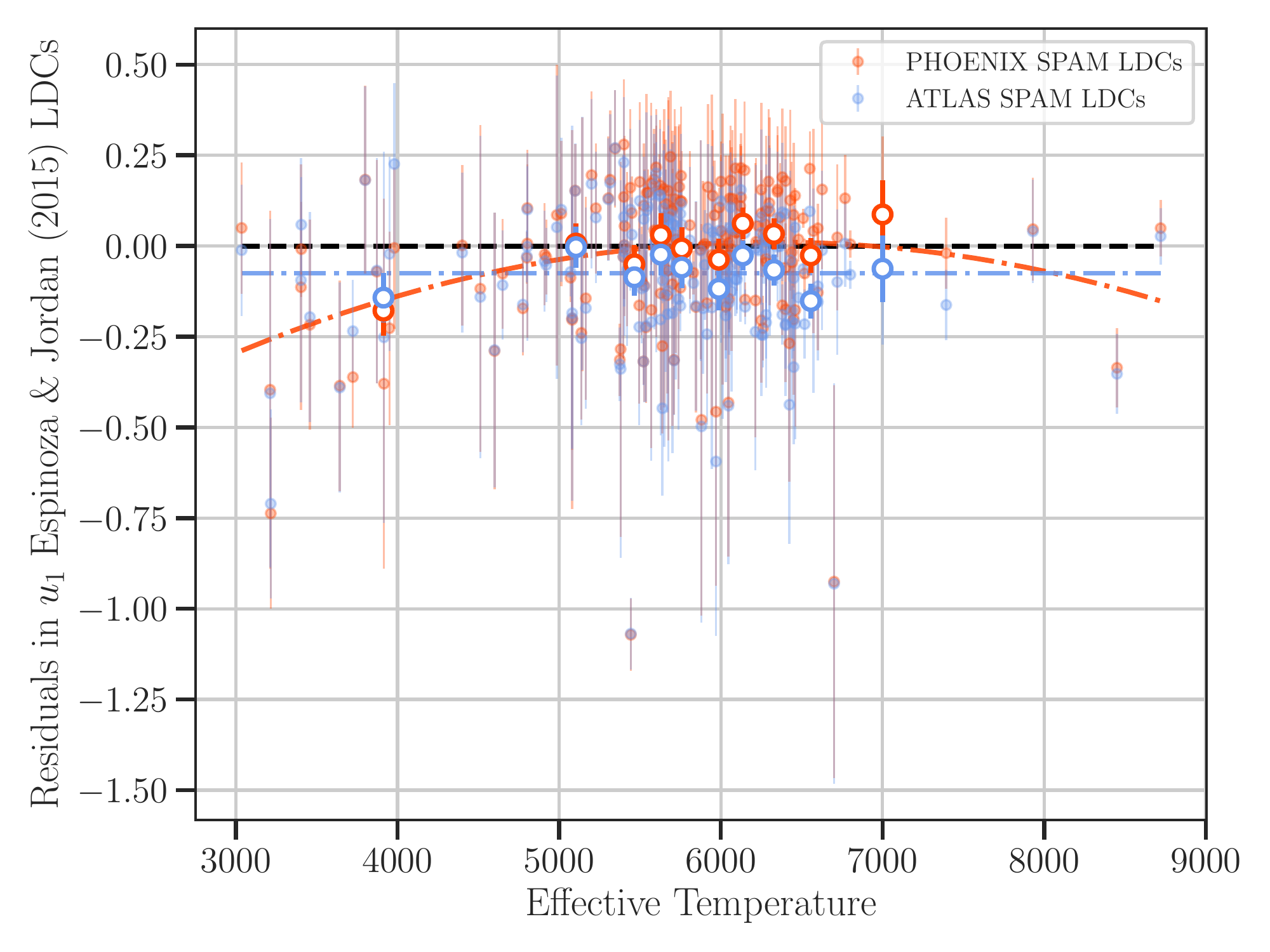}
    \includegraphics[width=\columnwidth]{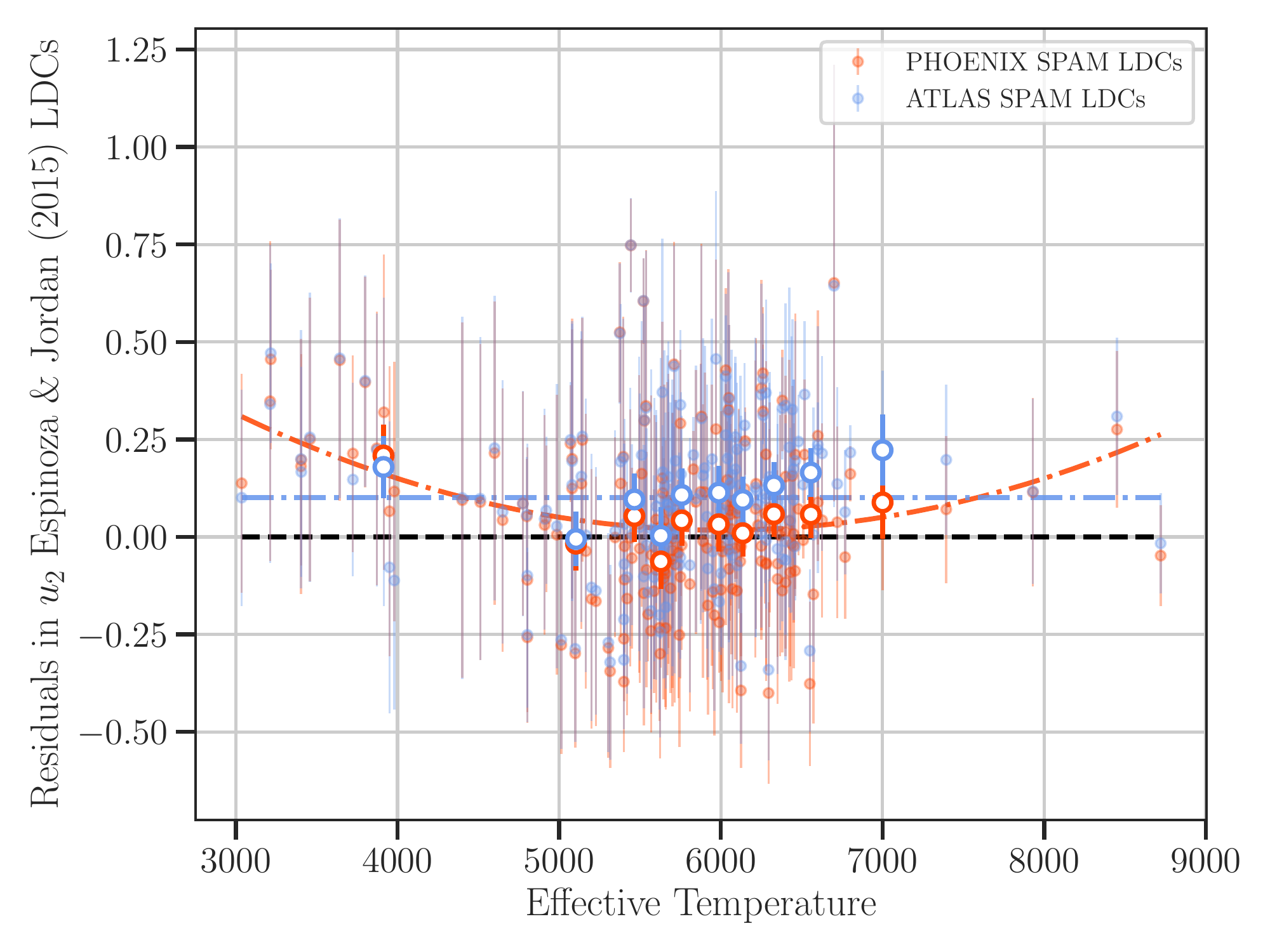}
    \caption{Same as Figure \ref{fig:u1_code_te}, but now using SPAM LDCs.}
    \label{fig:u1_code_te_SPAM}
\end{figure*}


As described in the previous section, {{there are suggestions of mean offsets when comparing 
empirical to theoretically determined LDCS. The works of \citet{muller:2013} and \citet{2015MNRAS.450.1879E}, in turn, 
when performing a similar analysis on \textit{Kepler} lightcurves, observed}} a possible 
dependence of those offsets with stellar temperature. Motivated by this, we here explore that 
dependence using our \textit{TESS} LDCs in what follows. 

To check for any such possible dependencies, we plot the residuals between the theoretical and the TESS retrieved LDCs as 
a function of the effective temperature for both {{sets of theoretical LDCs (those derived from limb-darkening Tables/Codes and the SPAM LDCs) in Figures \ref{fig:u1_code_te} and \ref{fig:u1_code_te_SPAM} for the representative case of the \citet{2015MNRAS.450.1879E} LDCs (which are fairly similar to the ones tabulated by \citet{2017AA...600A..30C} using the \textit{r-method}\footnote{For completeness, we show the same plots for the LDCs obtained using the tables in 
\citet{2017AA...600A..30C} in Figures \ref{fig:u1_cla_te} through Figure \ref{fig:u1_cla_te_r_SPAM}) in the Appendix}.}}) To find any possible correlation 
between these offsets and the effective temperature of the host stars, we fit those residuals with polynomials ranging 
from zeroth (i.e., a constant model) to second order. We used the Bayesian Information Criterion 
\citep[BIC;][]{Schwarz:1978} to determine the best fitting model among those three models, and present the best of them 
(i.e., the one with the minimum BIC) in Figures \ref{fig:u1_code_te} {{and \ref{fig:u1_code_te_SPAM}}} with dashed-dotted lines.

{{In almost all of those residual plots we find a quadratic function seems to fit the residuals best, with the 
vertex of the parabola touching (or being close to) zero at about 6000 K. The residuals, in turn, seem to increase their 
absolute values as they go to cooler and hotter stars. For instance, for the \texttt{PHOENIX} models, these quadratic 
functions predict offsets of order $\approx 0.2$ for both SPAM and non-SPAM coefficients. Indeed, the SPAM LDCs seem 
to significantly flatten those parabolas, but indeed, the LDCs especially for cooler stars don't seem to be consistent 
with zero. This suggest that independent of the methods, LDCs for stars cooler than about 5000 K are not predicted very 
precisely by our modelling efforts, and thus care must be taken when using theoretical LDCs for stars at those temperatures.}}

While interpreting these correlation between the effective temperature and the offsets, one needs to keep in mind that the distribution of our sample is not uniform in the effective temperature, as shown in Figure \ref{fig:fig2} -- these distribution shows the deficiency of targets at low and very high temperatures. Still, it is interesting to see how the residuals between theoretical and observed LDCs seem to be better at temperatures close to that of the Sun, which makes intuitive sense given most stellar model atmospheres have been extensively tested against the Sun.

{{
\subsection{Limitations of our study}

An important limitation of our study to have in mind is the fact that our work requires, by construction, a 
layered approach to obtaining limb-darkening coefficients from stellar model atmospheres. In other words, 
the comparison is not direct \textit{to} the models, but involves a series of assumptions and methods that 
lead to the final coefficients we compare against the data. As such, it is somewhat complicated to make 
direct statements about the \textit{actual validity of the stellar model atmospheres} tested in our work. 
For instance, while we observe that the PHOENIX coefficients obtained through the r-method behave better 
when compared against our retrieved empirical LDCs, we cannot confirm if this is because some bias cancels 
out between the methodology and the actual stellar model atmospheres, or because the stellar models are inherently
reasonably well behaved in the \textit{TESS} bandpass, and the r-method is intrinsically better. We can, however, 
make the \textit{practical} suggestion that, when confronted with the data, and if willing to use tabulated LDCs, 
one should prefer this method if using the PHOENIX models over the q-method proposed by \citet{2017AA...600A..30C}.

}}

\section{Conclusion and Future Work}\label{sec:conclusion}

In this paper, our main goal was to study the limb darkening of exoplanet host stars {{using \textit{TESS} 
transiting exoplanet lightcurves. To achieve this goal, we used precise exoplanet transit lightcurves obtained by \textit{TESS} for 176 known exoplanetary systems, for which we computed limb-darkening coefficients --- the 
primary products needed to meet our work's goal. A secondary but not less important product of our work are updated transit parameters for these exoplanetary systems. A subset of those ($R_p/R_*$, $a/R_*$ and the time-of-transit centers) were }}compared with their corresponding values from the literature {{with which we found very good agreement. This not only provides a valuable validation 
of our transit lightcurve fits, but also a rich dataset for the community to use to, e.g., use as prior in precise 
follow-up observations of these systems, or to even plan observations of these systems given the improved precision 
of our ephemerides for most of our studied systems.}}

We used these results to test how well the limb darkening effect is modelled by {{current methodologies and }} stellar model atmospheres. We compared the retrieved LDCs with both \textsc{ATLAS} and \textsc{PHOENIX} model predictions calculated by two authors, \citet{2017AA...600A..30C} and \citet{2015MNRAS.450.1879E} {{for one of the 
most widely use limb-darkening laws: the quadratic law}}. We found {{significant offsets on the limb-darkening 
coefficients $u_1$ and $u_2$ of the quadratic law between empirically and theoretically determined LDCs, which can be as 
large as $\Delta u_1\approx 0.1$ and $\Delta u_2\approx 0.2$, depending on the methodology and the stellar model atmospheres 
used for the calculations. Our main take-home message is that the most accurate LDCs for the \textit{TESS} bandpass are 
the ones that use the SPAM algorithm proposed in \citet{2011MNRAS.418.1165H}, and which use the \texttt{PHOENIX} model 
atmospheres using the technique introduced in \cite{2004AA...413..711W} and used by \citet{2015MNRAS.450.1879E} to 
calculate limb-darkening coefficients for these stellar model atmospheres --- the \textit{r-method} in the terminology 
used in the published tables of \citet{2017AA...600A..30C}}}.

{{Our analysis, however, suggests that even with the ``right" methodology, large offsets between theoretical 
and empirically determined LDCs could be observed for stars cooler than about 5000 K. Below these temperatures, there 
seem to be offsets as large as $\Delta u_i \sim 0.2$ for the LDCs of the quadratic law. We believe an important future 
avenue would be to extend our work to a larger sample of exoplanetary systems orbiting cooler stars in order to confirm 
this trend. The newest of \textit{TESS} discoveries are indeed paving the way to form such a sample, which we believe 
would be fundamental not only to understand the physical properties and architectures of these exoplanetary systems, but 
also to understand the limitations of current models and methodologies at predicting the imprint of the limb-darkening effect 
on real transit lightcurves.}}

\section*{Acknowledgements}
We would like to thank an anonymous referee for their
thorough feedback which significantly improved the presentation of our results. JAP wants to thank Dr. N\'{e}stor Espinoza and Mr. Ashok Patel for providing the financial support to visit the Max-Planck-Institut f\"ur Astronomie (MPIA), Heidelberg in Germany to perform the present work. A significant part of the present work constitute the final year project work of JAP's Master's degree. JAP wants to thank his parent institute -- Sardar Vallabhbhai National Institute of Technology, Surat-7, Gujarat, India -- and his academic supervisor Prof. K. N. Pathak to give permission to work at MPIA during this period. NE would like to thank M. G\"unther and T. Daylan for useful discussions regarding uniform analyses of TESS known exoplanets.

This research has made use of the NASA Exoplanet Archive, which is operated by the California Institute of Technology, under contract with the National Aeronautics and Space Administration under the Exoplanet Exploration Program.

This research made use of the open source Python package exoctk, the Exoplanet Characterization Toolkit \citep[][]{matthew_bourque_2021_4556063}.

\appendix

{{
\section{Discrepant planetary systems with literature values}\label{sec:appendix-discrepant}
As noted in Section \ref{sec:comp_lit}, our lightcurve analysis showed that for 25 
exoplanetary systems, one or more than one of the planetary parameters are at least $3-\sigma$ away from what is published in the literature. These systems are LTT9779~b, XO-3~b, WASP-161~b, WASP-121~b, HAT-P-2~b, HAT-P-69~b, KELT-11~b, CoRoT-18~b, WASP-4~b, K2-237~b, WASP-131~b, WASP-17~b, KELT-20~b, WASP-31~b, WASP-46~b, WASP-19~b, WASP-92~b, XO-6~b, TOI-157~b, WASP-22~b, K2-260~b, WASP-7~b, WASP-95~b, CoRoT-32~b and HATS-18~b. Here, we perform a more in-depth discussion on what might be producing these offsets on the planetary parameters.

For 11 of those ``discrepant" systems (HAT-P-69~b, KELT-11~b, WASP-4~b, WASP-131~b, KELT-20~b, WASP-31~b, WASP-46~b, WASP-22~b, K2-260~b, WASP-95~b and CoRoT-32~b) the offset is in the predicted 
time-of-transit center versus our computed time-of-transit center, which could either point to 
possible transit-timing variations (TTVs) in these systems \citep[e.g., WASP-4b is known to have TTVs, which we recover here; see ][]{bouma19} or simply act as updates with respect to previous ephemerides which might be 
outdated. {{Similar offsets have been found for HAT-P-69~b and WASP-95~b in \cite{shan2021}; however, 
the timing 
offsets for KELT-11~b, WASP-131~b, KELT-20~b, WASP-31~b, WASP-46~b, WASP-22~b, K2-260~b and 
CoRoT-32~b reported here have not been reported elsewhere to our knowledge.}}

The remaining 14 exoplanetary systems (LTT9779~b, XO-3~b, WASP-161~b, WASP-121~b, HAT-P-2~b, CoRoT-18~b, 
K2-237~b, WASP-17~b, WASP-19~b, WASP-92~b, XO-6~b, TOI-157~b, WASP-7~b and HATS-18~b) all show planet-to-star radius ratio ($R_p/R_*$) and/or semi-major axis-to-stellar 
radius ratio ($a/R_*$) discrepancies between 3 to 5 sigma from the literature values; some also show 
time-of-transit center offsets together with those. 

From the planetary systems that show discrepancies on $a/R_*$ 
(5 exoplanetary systems: XO-3~b, HAT-P-2~b, K2-237~b, XO-6~b and TOI-157~b), there are different explanations for the offsets:
\begin{itemize}
    \item For XO-3~b, while the reported value in this work of $a/R_* = 7.09^{+0.24}_{-0.23}$ is inconsistent with the value in the 
    discovery paper of \cite[][$a/R_* = 4.95^{+0.18}_{-0.18}$]{JK:2008}, our value is consistent with that of \cite{Wong:2014} of $a/R_* = 7.052^{+0.076}_{-0.097}$. The value in \cite{Wong:2014} has much better precision as the orbit is constrained through the entire 
    phase-curve of the exoplanet.
    
    \item For HAT-P-2~b, we obtain $a/R_* = 9.04^{+0.19}_{-0.18}$. This value is, in fact, consistent with various values in the literature \citep{Pal:2010, S:2010, S:2017}, including the discovery paper \citep[$a/R_* = 9.770^{+1.100}_{-0.020}$;][]{Bakos:2007}. However, it is inconsistent with the value found by \citet{L:2008} of $a/R_* = 10.28^{+0.12}_{-0.19}$. That value, however, was 
    obtained by fitting a radial-velocity only dataset for HAT-P-2~b which includes a measurement of the Rossiter-McLaughlin (RM) effect. \cite{Pal:2010} used the same dataset but not considering the radial-velocities obtained in-transit, and considering a 
    wide array of ground-based photometric follow-up transits to obtain their value of $a/R_* = 8.99^{+0.39}_{-0.41}$ which is fully consistent with our 
    solution --- this suggests the \citet{L:2008} might be biased given its constrain mainly comes from the RM effect. We consider our parameter here an update to the parameters of this system given these previous attempts.
    
    \item K2-237~b was observed by \textit{Kepler/K2} back in 2016, and independant analyses were performed by the teams 
of \citet[][ $a/R_* = 5.50^{+0.15}_{-0.11}$]{Soto:2018} and 
\citet[][ $a/R_* = 5.503^{+0.015}_{-0.207}$]{Smith:2019}, both of which give consistent $a/R_*$ values with each 
other, but inconsistent values at 3-sigma with the ones reported in our present work 
($a/R_* = 6.17^{+0.13}_{-0.19}$; see Table \ref{tab:Plan_prop}), especially when compared to the work 
of \cite{Smith:2019}. 
Interestingly, our value of $a/R_*$ is perfectly matched with the recent full re-analysis of the system made by 
\citet[][ $a/R_* = 6.07^{+0.14}_{-0.18}$]{Ikwut-Ukwa:2020}. We believe that our offset with the work of 
\citet{Smith:2019} is the product of a simple typo in their upper errorbars ($0.015$), which are one order of 
magnitude better than what can be reasonable achieved with the \textit{K2} data-quality; it is interesting, however, 
that the two analyses made on the (long-cadence) \textit{K2} photometry are systematically lower than the \textit{TESS} 
short-cadence analyses presented in this work and that of \citet{Ikwut-Ukwa:2020}. 

\item For XO-6~b, we also believe 
our value corresponds to an update with respect to previous constraints on this parameter. First, the value we obtain 
in our work ($a/R_* = 8.17^{+0.07}_{-0.07}$) is consistent with an independent analysis made on the same 
\textit{TESS} data by \citet[][$a/R_* = 8.383 \pm 0.074$]{RH:2020}. These values are however inconsistent with the 
one reported in \cite{Crouzet:2017} who from the photometry alone conclude on $a/R_* = 9.20 \pm 0.19$, but 
when combining that data with Doppler tomographic results settle on $a/R_* = 9.08 \pm 0.17$. Interestingly, the 
constrain on this parameter using only the Doppler tomography in that work is $a/R_* = 8.30^{+1.2}_{-0.8}$, which 
is fully consistent with the value that both our work and that of \cite{RH:2020} retrieve using the \textit{TESS} 
photometry. It is likely, thus, that the photometric analysis in \cite{Crouzet:2017} is somewhat biasing their 
result towards a larger $a/R_*$, but it is unclear which part of their analysis could give rise to such a 
large bias. 

\item For TOI-157~b, our value for $a/R_*$ of $6.30\pm0.09$ is inconsistent at 3-sigma with that reported in 
\cite{Nielsen:2020} of $5.79^{+0.07}_{-0.07}$. One of the differences between our analyses is that 
\cite{Nielsen:2020} use, along some ground-based photometric follow-up, 8-sectors-worth of 30-min cadence data 
and 4-sectors-worth of 2-min cadence data, which arguably mostly define the planetary properties. In our analysis, however, we use 12-sectors-worth of 2-minute cadence 
data. It is interesting to note, in addition, that the differences 
on this parameter between the \cite{Nielsen:2020} estimate and our work go in the same 
direction as those observed and discussed for K2-237~b before: our analysis obtains a larger value for $a/R_*$. The only 
observational similarity between the \textit{TESS} observations of TOI-157~b analyzed by \cite{Nielsen:2020} and the \textit{K2} 
observations of K2-237~b analyzed by the teams of \cite{Soto:2018} and \cite{Smith:2019} is that both datasets rely heavily on long-cadence data. This 
is suspiciously consistent with what is expected by morphological lightcurve distortions due to finite integration time 
\citep{binningsinning}: not properly accounting/resampling the lightcurves in long-cadence observations would give rise to 
smaller values of $a/R_*$ for a fixed period. We suggest, thus, that perhaps the value obtained in \cite{Nielsen:2020} is due 
to the fact that lightcurve resampling following procedures similar to those outlined in \cite{binningsinning} were either 
not applied or not performed with sufficient precision to properly account for the effect. Similarly, perhaps this latter 
option is the case as well for the \textit{K2} observations of K2-237~b analyzed by both \cite{Soto:2018} and 
\cite{Smith:2019}.
\end{itemize}

Finally, for the 9 systems that show $R_p/R_*$ discrepancies with literature values (LTT9779~b, WASP-161~b, WASP-121~b, 
CoRoT-18~b, WASP-17~b, WASP-19~b, WASP-92~b, WASP-7~b and HATS-18~b), different 
explanations exist:

\begin{itemize}
    \item For LTT9779~b, we retrieve $R_p/R_* = 0.0337^{+0.0011}_{-0.0009}$, which 
    is inconsistent with the value found by \cite{Jenkins:2020} of $R_p/R_* = 0.0455^{+0.0022}_{-0.0017}$. In fact, for this system, our retrieved value for $a/R_* = 7.62^{+0.30}_{-0.36}$ is also significantly discrepant with that of \citet[][$a/R_*=3.88 \pm 0.09$]{Jenkins:2020}. Our retrieved parameter values for this system, however, are most likely wrong and should not be used --- we 
    only present them here for completeness and transparency of our process. While 
    the transit fits from both our solutions and the ones reported in \cite{Jenkins:2020} both fit the data equally well, the combination of our retrieved $a/R_*$ with the planetary period yield a stellar density of $13342 \pm 1744$ kg/m$^3$ which is completely inconsistent with that obtained through spectroscopy by \cite{Jenkins:2020} of $1810 \pm 130$ kg/m$^3$. This two-solution problem was 
    in fact briefly studied by \cite{Jenkins:2020} as well in their Methods section.
    
    \item For WASP-161~b, we retrieve $R_p/R_* = 0.0751^{+0.0009}_{-0.0008}$, which 
    is much more precise but discrepant at 3-sigma with the one reported by \citet[][$R_p/R_* = 0.0671^{+0.0017}_{-0.0017}$]{Barkaoui:2019}. While 
    this target does have a nearby companion about 16'' to the SE, the PDC 
    algorithm takes this dilution into account on the photometry and as such it is 
    unlikely this is the cause of the discrepancy. The work of \cite{Barkaoui:2019} 
    has only one full ground-based transit (with the rest of the follow-up photometry 
    being partial transits) and, as such, we believe our value for $R_p/R_*$ is 
    effectively an update on this parameter. In addition, we also find a discrepant 
    time-of-transit center, which has also been reported by \cite{shan2021}.
    
    \item For WASP-121~b, we retrieve $R_p/R_* = 0.1217^{+0.0003}_{-0.0003}$, which is discrepant at about 3-sigma with the value of 
    $R_p/R_* = 0.1245^{+0.0005}_{-0.0005}$ reported in \cite{Delerez:2016}. Interestingly, the 
    \cite{Delerez:2016} value is consistent with analysis of Sector 7 \textit{TESS} 
    data from other teams \citep[see, e.g.,][]{Yang:2020, Bourrier:2020, Daylan:2021}. 
    In our work, however, we use additional data from Sectors 33 and 34. If we run 
    our analyses on Sector 7 \textit{TESS} data only, our resulting value of $R_p/R_* = 0.12394^{+0.00047}_{-0.00043}$ is consistent with both, the value of \cite{Delerez:2016} and the rest of the \textit{TESS} analyses in the literature. However, individual analyses on Sectors 33 and 34 reveal that the transit depth 
    on Sector 7 is significantly larger than the ones observed in those: 500 ppm 
    larger. This might, indeed, be true variability in the transit depth caused 
    either by the star or the planetary atmosphere itself --- this has already 
    been suggested by ground-based observations by \cite{Wilson:2021}. We note 
    that the transit depth reported in this work is consistent with the average 
    transit depth in the HST/WFC3 transit spectrum presented by \cite{Evans:2018}. We also observe a 3-sigma time-of-transit offset from that of \cite{Delerez:2016}, 
    which might also hint to possible long-term TTVs.
    
    \item For CoRoT-18~b, we obtain $R_p/R_* = 0.1579^{+0.0042}_{-0.0046}$, which is 
    significantly discrepant with the $0.1341^{+0.0019}_{-0.0019}$ value reported in \citet{Hebrard:2011}. This latter reported value, in turn, agrees well with 
    follow-up ground-based photometry of \cite{Southworth}. While the discrepant 
    transit depth in the \textit{TESS} photometry could be due to variability either 
    in the star or the planet itself, given the field containing this target is 
    so crowded, it is also possible that the discrepancy in the planet-to-star radius 
    ratio is due to an over-correction of the dilution from nearby sources, similar 
    to the case if WASP-140~b in Section \ref{sec:comp_lit}.
    
    \item For WASP-17~b, we retrieve $R_p/R_* = 0.1218^{+0.0016}_{-0.0014}$, 
    which is inconsistent at 3-sigma with the value reported in \citet[][$R_p/R_* = 0.1293^{+0.0008}_{-0.0008}$]{Anderson:2010}. We believe, however, 
    that our value is an update on this parameter given the agreement of our 
    value with recent precise HST and Spitzer transit spectroscopy reported by \citet{Saba:2021}.
    
    \item For WASP-19~b, we obtain $R_p/R_* = 0.1519^{+0.0018}_{-0.0020}$, 
    which agrees with some results in the literature, but not with others. In 
    particular, this value agrees with the originally reported value by 
    \cite{Hebb:2010} and the recent \textit{TESS} analysis performed by \cite{Wong:2020}, but does not with the values reported by \cite{TR:2013}, \cite{Mancini:2013} and \cite{WASP19} by more than 3-sigma. This is, in turn, not surprising, given 
    the high levels of stellar activity the star is known to have, which has been 
    observed to directly contaminate the observed transit depths both due to 
    occulted and unnoculted spots \citep[see, e.g.,][]{WASP19}.
    
    \item For WASP-92~b, we retrieve $R_p/R_* = 0.1068^{+0.0015}_{-0.0022}$, 
    which is inconsistent at 3-sigma with the value reported in \cite{Gajdo:2019} of $R_p/R_* = 0.0963^{+0.0017}_{-0.0017}$. While the there seems to be an 
    extended object (the galaxy LEDA 2387340; TIC 10000706526) about 14.8'' to 
    the NW of WASP-92, this object is indeed in the TIC catalog, and correctly 
    identified as an extended object in it, which means it is correctly being 
    introduced in the PDC dilution correction. It is, 
    thus, unlikely this is the source of the discrepancy. We believe our value 
    is an update to the one reported in \cite{Gajdo:2019}, however, because that 
    work does not have full coverage of the transit event with all their 
    ground-based photometric follow-up. The TESS dataset we use here, however, 
    is composed of 3 sectors full of data which comprise a couple of tens of 
    transits in total.
    
    \item For WASP-7~b, we obtain $R_p/R_* = 0.0790^{+0.0007}_{-0.0006}$, which 
    is inconsistent at more than 5-sigma from the $R_p/R_* = 0.0956^{+0.0016}_{-0.0016}$ value reported by \citet{Southworth:2011}. Interestingly, our value is 
    consistent with the discovery paper $R_p/R_*$ reported by \citet{Hellier:2009}. 
    The larger $R_p/R_*$ reported by \citet{Southworth:2011} comes from only 
    one (high-precision) transit, whereas the value reported in this work comes 
    from four high-precision transit events from Sector 27. We would, thus, be 
    tempted to think of our value as the ``correct" one. However, it is important to note that the third transit observed by \textit{TESS} shows a possible 
    spot-crossing event feature (see Figure \ref{fig:w7}), suggesting that perhaps the varying transit 
    depths accross different studies is indeed real, and produced by occulted 
    and unnoculted stellar heterogeneities such as the ones observed for WASP-19~b. 
    
    \item Finally, for HATS-18~b, we obtain $R_p/R_* = 0.1456^{+0.0021}_{-0.0024}$, 
    which is inconsistent with the $R_p/R_* = 0.1320^{+0.0004}_{-0.0004}$ value 
    reported in \citet{Penev:2016} by more than 5-sigma. This target is quite 
    special because while only observed by \textit{TESS} in Sector 10, its 
    short, 20-hour period implies 26 full transits were used for the analysis 
    presented in this work. All the rest of the parameters are consistent with those of \citet{Penev:2016}. This target 
    does not have any significant, nearby contaminant --- so it is isn't likely the 
    offset in $R_p/R_*$ is due to miscalculated dilution. The target is, however, 
    active --- \citet{Penev:2016} measure rotational variability with a period of 
    about 10 days. Our best explanation for this discrepancy, thus, is that 
    the different transit depths could arise due to different levels of stellar 
    activity in the \textit{TESS} observations as compared to those observed 
    by \citet{Penev:2016} --- i.e., varying transit depths due to both occulted 
    and unnoculted spots, similar to the WASP-19~b case. There is evidence, in fact, 
    of varying levels of stellar activity between epochs for HATS-18. While \citet{Penev:2016} 
    measure a peak-to-peak amplitude of this variability of order 20 mmag (20,000 ppm) 
    in the HAT-South photometry, we see peak-to-peak amplitudes of this variability 
    in our \textit{TESS} photometry at least 1/3 of that --- with a peak-to-peak amplitude of about 6,000 ppm. HATS-18~b transiting 
    brighter regions of the star during the \textit{TESS} observations as compared 
    to the previous transits observed by \citet{Penev:2016} could explain, for 
    instance, the larger transit depths observed in our work.
    
\end{itemize}

\begin{figure}
	\includegraphics[width=\columnwidth]{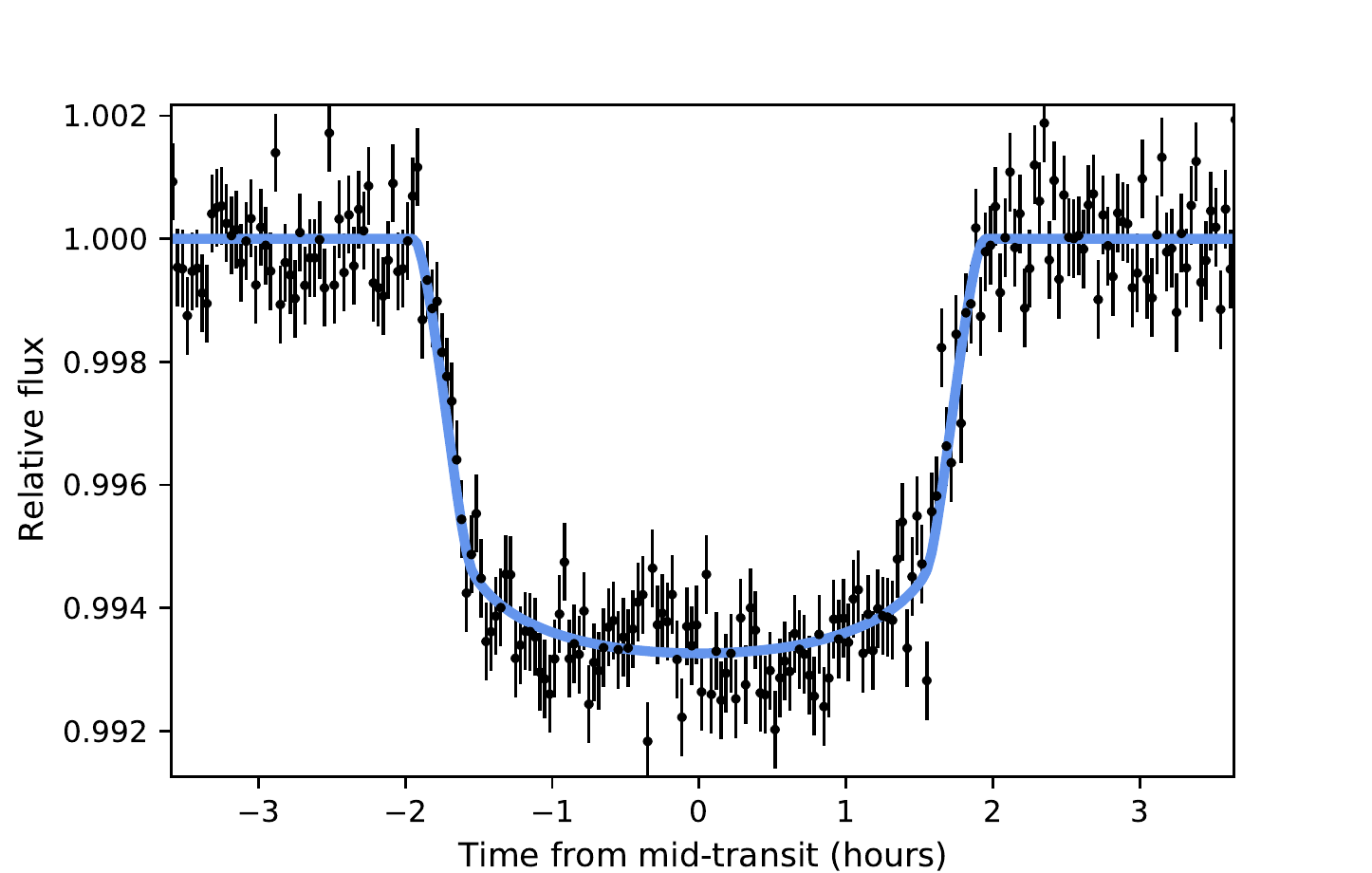}
    \caption{Third transit of WASP-7b observed by \textit{TESS} in Sector 27. An event, suggestive of a spot crossing event, is seen at about 24 minutes before mid-transit.}
    \label{fig:w7}
\end{figure}

From the above analyses, thus, it seems that 24 out of 25 discrepant systems are really mostly updates to existing reported planetary 
parameters in the literature for the exoplanets under consideration in this work. Our only real outlier is LTT9799~b. We consider having 
one confirmed outlier out of a sample of 176 targets is perfectly consistent with random chance and, thus, we consider this a very good check that our procedures are giving consistent (and/or updated) results to previous studies.
}}

\section{Tables}

\startlongtable
\begin{deluxetable}{lcccl}
\tablecaption{Various stellar properties of our targets. The data was retrieved from the NASA Exoplanet Achieve on February 23, 2021. {{Only first 20 rows are shown here; the full table is available on the website in a machine readable ASCII form.}}}
\label{tab:stellar_properties}
\tablehead{\colhead{Star Name} & \colhead{$T_{eff}$} & \colhead{[M/H]} & \colhead{$\log{g}$} & \colhead{$V_{turb}$}\\
& \colhead{(K)} &  & \colhead{(cgs)} & \colhead{(km/s)} }
\startdata
    WASP-61 & 6250.0 & -0.10 & 4.26 & ... \\ 
	WASP-130 & 5625.0 & 0.26 & 4.49 & ... \\ 
	HATS-13 & 5523.0 & 0.05 & 4.52 & ... \\ 
	WASP-156 & 4910.0 & 0.24 & 4.60 & ... \\ 
	NGTS-4 & 5143.0 & -0.28 & 4.50 & ... \\ 
	WASP-190 & 6400.0 & -0.02 & 4.17 & ... \\ 
	WASP-44 & 5420.0 & -0.003 & 4.49 & ... \\ 
	TOI-892 & 6261.0 & 0.24 & 4.26 & ... \\ 
	WASP-82 & 6480.0 & 0.12 & 3.96 & ... \\ 
	TOI-540 & 3216.0 & 0.00 & 4.44 & ... \\ 
	WASP-22 & 6000.0 & -0.05 & 4.50 & ... \\ 
	TOI-905 & 5570.0 & 0.14 & 4.50 & ... \\ 
	HAT-P-42 & 5743.0 & 0.27 & 4.14 & ... \\ 
	WASP-62 & 6230.0 & 0.04 & 4.45 & ... \\ 
	HATS-3 & 6351.0 & -0.16 & 4.22 & ... \\ 
	LTT9779 & 5443.0 & 0.27 & 4.35 & ... \\ 
	K2-260 & 6367.0 & -0.14 & 4.15 & ... \\ 
	TrES-3 & 5650.0 & -0.20 & 4.57 & ... \\ 
	Qatar-10 & 6124.0 & 0.016 & 4.30 & ... \\ 
	WASP-144 & 5200.0 & 0.18 & 4.53 & ... \\ 
\enddata
\end{deluxetable}


\startlongtable
\clearpage
\begin{longrotatetable}
\begin{deluxetable*}{lllllllllllll}
\tablecaption{Limb darkening coefficients calculated using different methods for quadratic law; here EJ15 means the LDCs calculated using the method by \citet{2015MNRAS.450.1879E} and C17 implies the tabulated values from \citet{2017AA...600A..30C}, with their \textit{r-method} (C17r) and \textit{q-method} (C17q) when using \textsc{phoenix} models \\ $^*$ {{dash indicates that the LDCs for the given target could not be calculated because the effective temperature of the host star is out of the range of \citet{2017AA...600A..30C} \textsc{ATLAS} tables.}} \\ {{Only first 20 rows are shown here; the full table is available on the website in a machine readable ASCII form.}}}
\label{tab:ldcs}
\tablehead{\colhead{Star name} & \colhead{$u_1$} & \colhead{$u_2$} & \colhead{$u_1$} & \colhead{$u_2$} & \colhead{$u_1$} & \colhead{$u_2$} & \colhead{$u_1^*$} & \colhead{$u_2^*$} & \colhead{$u_1$} & \colhead{$u_2$} & \colhead{$u_1$} & \colhead{$u_2$}  \\
    & \colhead{(Empirical)} & \colhead{(Empirical)} & \colhead{(EJ15} & \colhead{(EJ15} & \colhead{(EJ15} & \colhead{(EJ15} & \colhead{(C17} & \colhead{(C17} & \colhead{(C17q} & \colhead{(C17q} & \colhead{(C17r} & \colhead{(C17r} \\
    &  &  & \colhead{\textsc{atlas})} & \colhead{\textsc{atlas})} & \colhead{\textsc{phoenix})} & \colhead{\textsc{phoenix})} & \colhead{\textsc{atlas})} & \colhead{\textsc{atlas})} & \colhead{\textsc{phoenix})} & \colhead{\textsc{phoenix})} & \colhead{\textsc{phoenix})} & \colhead{\textsc{phoenix})}
}
\startdata
    WASP-61 & $0.26^{+0.12}_{-0.13}$ & $0.19^{+0.24}_{-0.22}$ & 0.23 & 0.31 & 0.30 & 0.25 & 0.23 & 0.31 & 0.24 & 0.38 & 0.31 & 0.25 \\ 
	WASP-130 & $0.25^{+0.18}_{-0.16}$ & $0.43^{+0.27}_{-0.35}$ & 0.36 & 0.26 & 0.36 & 0.23 & 0.33 & 0.27 & 0.30 & 0.34 & 0.36 & 0.23 \\ 
	HATS-13 & $0.31^{+0.17}_{-0.17}$ & $0.29^{+0.34}_{-0.33}$ & 0.34 & 0.26 & 0.37 & 0.23 & 0.34 & 0.26 & 0.31 & 0.33 & 0.37 & 0.22 \\ 
	WASP-156 & $0.52^{+0.14}_{-0.15}$ & $0.08^{+0.28}_{-0.25}$ & 0.43 & 0.21 & 0.44 & 0.21 & 0.45 & 0.20 & 0.39 & 0.30 & 0.43 & 0.21 \\ 
	NGTS-4 & $0.31^{+0.35}_{-0.22}$ & $0.10^{+0.31}_{-0.24}$ & 0.36 & 0.25 & 0.40 & 0.21 & 0.38 & 0.23 & 0.34 & 0.33 & 0.40 & 0.22 \\ 
	WASP-190 & $0.42^{+0.20}_{-0.20}$ & $0.30^{+0.30}_{-0.32}$ & 0.21 & 0.32 & 0.31 & 0.24 & 0.22 & 0.32 & 0.22 & 0.39 & 0.30 & 0.25 \\ 
	WASP-44 & $0.44^{+0.21}_{-0.22}$ & $0.30^{+0.30}_{-0.37}$ & 0.34 & 0.26 & 0.38 & 0.23 & 0.36 & 0.25 & 0.32 & 0.33 & 0.38 & 0.22 \\ 
	TOI-892 & $0.21^{+0.12}_{-0.12}$ & $-0.00^{+0.17}_{-0.11}$ & 0.24 & 0.32 & 0.30 & 0.25 & 0.24 & 0.32 & 0.24 & 0.38 & 0.31 & 0.25 \\ 
	WASP-82 & $0.35^{+0.06}_{-0.07}$ & $0.07^{+0.12}_{-0.11}$ & 0.21 & 0.33 & 0.30 & 0.24 & 0.21 & 0.33 & 0.20 & 0.44 & 0.30 & 0.24 \\ 
	TOI-540 & $0.96^{+0.26}_{-0.24}$ & $-0.11^{+0.23}_{-0.22}$ & 0.24 & 0.39 & 0.14 & 0.52 & --- & --- & 0.02 & 0.72 & 0.12 & 0.53 \\ 
	WASP-22 & $0.19^{+0.11}_{-0.11}$ & $0.30^{+0.21}_{-0.21}$ & 0.26 & 0.30 & 0.32 & 0.25 & 0.26 & 0.30 & 0.26 & 0.36 & 0.32 & 0.25 \\ 
	TOI-905 & $0.59^{+0.38}_{-0.37}$ & $0.19^{+0.41}_{-0.44}$ & 0.34 & 0.26 & 0.36 & 0.23 & 0.34 & 0.27 & 0.31 & 0.33 & 0.37 & 0.22 \\ 
	HAT-P-42 & $0.25^{+0.17}_{-0.15}$ & $0.39^{+0.29}_{-0.31}$ & 0.31 & 0.29 & 0.35 & 0.24 & 0.31 & 0.28 & 0.27 & 0.39 & 0.34 & 0.24 \\ 
	WASP-62 & $0.30^{+0.01}_{-0.01}$ & $0.13^{+0.03}_{-0.03}$ & 0.24 & 0.31 & 0.30 & 0.25 & 0.24 & 0.31 & 0.24 & 0.37 & 0.31 & 0.25 \\ 
	HATS-3 & $0.22^{+0.15}_{-0.13}$ & $0.20^{+0.24}_{-0.24}$ & 0.23 & 0.31 & 0.31 & 0.24 & 0.22 & 0.31 & 0.23 & 0.39 & 0.31 & 0.24 \\ 
	LTT9779 & $1.35^{+0.10}_{-0.11}$ & $-0.41^{+0.12}_{-0.13}$ & 0.36 & 0.26 & 0.38 & 0.23 & 0.36 & 0.25 & 0.31 & 0.35 & 0.37 & 0.23 \\ 
	K2-260 & $0.29^{+0.15}_{-0.16}$ & $0.14^{+0.31}_{-0.23}$ & 0.23 & 0.31 & 0.31 & 0.24 & 0.22 & 0.31 & 0.23 & 0.39 & 0.31 & 0.24 \\ 
	TrES-3 & $0.45^{+0.41}_{-0.30}$ & $0.11^{+0.35}_{-0.33}$ & 0.29 & 0.29 & 0.36 & 0.23 & 0.31 & 0.28 & 0.30 & 0.33 & 0.36 & 0.23 \\ 
	Qatar-10 & $0.15^{+0.10}_{-0.08}$ & $0.55^{+0.20}_{-0.22}$ & 0.27 & 0.30 & 0.31 & 0.25 & 0.25 & 0.31 & 0.25 & 0.38 & 0.32 & 0.25 \\ 
	WASP-144 & $0.26^{+0.23}_{-0.17}$ & $0.29^{+0.34}_{-0.32}$ & 0.39 & 0.23 & 0.40 & 0.22 & 0.40 & 0.23 & 0.34 & 0.32 & 0.40 & 0.22 \\ 
\enddata
\end{deluxetable*}
\end{longrotatetable}


\startlongtable
\begin{deluxetable*}{ccccccccccc}
\tablecaption{SPAM limb darkening coefficients calculated using different methods for quadratic law; here EJ15 means the LDCs calculated using the method by \citet{2015MNRAS.450.1879E} and C17 implies the tabulated values from \citet{2017AA...600A..30C}, with their \textit{r-method} (C17r) and \textit{q-method} (C17q) when using \textsc{phoenix} models. \\ {{Only first 20 rows are shown here; the full table is available on the website in a machine readable ASCII form.}}}
\label{tab:ldcs_SPAM}
\tablehead{\colhead{Star name} & \colhead{$u_1$} & \colhead{$u_2$} & \colhead{$u_1$} & \colhead{$u_2$} & \colhead{$u_1$} & \colhead{$u_2$} & \colhead{$u_1$} & \colhead{$u_2$} & \colhead{$u_1$} & \colhead{$u_2$}  \\
    & \colhead{(EJ15} & \colhead{(EJ15} & \colhead{(EJ15} & \colhead{(EJ15} & \colhead{(C17} & \colhead{(C17} & \colhead{(C17q} & \colhead{(C17q} & \colhead{(C17r} & \colhead{(C17r} \\
    & \colhead{\textsc{atlas})} & \colhead{\textsc{atlas})} & \colhead{\textsc{phoenix})} & \colhead{\textsc{phoenix})} & \colhead{\textsc{atlas})} & \colhead{\textsc{atlas})} & \colhead{\textsc{phoenix})} & \colhead{\textsc{phoenix})} & \colhead{\textsc{phoenix})} & \colhead{\textsc{phoenix})}
}
\startdata
    WASP-61 & 0.30 & 0.20 & 0.35 & 0.17 & 0.30 & 0.20 & 0.37 & 0.16 & 0.37 & 0.15 \\ 
	WASP-130 & 0.40 & 0.19 & 0.42 & 0.14 & 0.37 & 0.20 & 0.40 & 0.16 & 0.41 & 0.14 \\ 
	HATS-13 & 0.38 & 0.18 & 0.42 & 0.14 & 0.38 & 0.19 & 0.42 & 0.15 & 0.42 & 0.14 \\ 
	WASP-156 & 0.48 & 0.13 & 0.50 & 0.11 & 0.50 & 0.11 & 0.49 & 0.12 & 0.49 & 0.11 \\ 
	NGTS-4 & 0.31 & 0.35 & 0.32 & 0.35 & 0.32 & 0.35 & 0.32 & 0.36 & 0.32 & 0.35 \\ 
	WASP-190 & 0.21 & 0.29 & 0.24 & 0.31 & 0.21 & 0.29 & 0.25 & 0.32 & 0.24 & 0.31 \\ 
	WASP-44 & 0.38 & 0.20 & 0.43 & 0.14 & 0.38 & 0.20 & 0.42 & 0.16 & 0.42 & 0.14 \\ 
	TOI-892 & 0.23 & 0.31 & 0.24 & 0.32 & 0.23 & 0.31 & 0.25 & 0.32 & 0.37 & 0.15 \\ 
	WASP-82 & 0.21 & 0.32 & 0.37 & 0.14 & 0.21 & 0.32 & 0.36 & 0.17 & 0.37 & 0.14 \\ 
	TOI-540 & 0.25 & 0.36 & 0.22 & 0.35 & -0.04 & 0.20 & 0.21 & 0.34 & 0.21 & 0.34 \\ 
	WASP-22 & 0.31 & 0.21 & 0.37 & 0.16 & 0.32 & 0.21 & 0.37 & 0.17 & 0.37 & 0.16 \\ 
	TOI-905 & 0.37 & 0.21 & 0.41 & 0.14 & 0.36 & 0.21 & 0.39 & 0.18 & 0.41 & 0.15 \\ 
	HAT-P-42 & 0.26 & 0.32 & 0.41 & 0.14 & 0.26 & 0.32 & 0.38 & 0.18 & 0.41 & 0.14 \\ 
	WASP-62 & 0.29 & 0.22 & 0.36 & 0.16 & 0.29 & 0.21 & 0.35 & 0.18 & 0.36 & 0.16 \\ 
	HATS-3 & 0.22 & 0.29 & 0.38 & 0.14 & 0.21 & 0.29 & 0.35 & 0.18 & 0.37 & 0.15 \\ 
	LTT9779 & 0.28 & 0.34 & 0.28 & 0.34 & 0.28 & 0.34 & 0.28 & 0.34 & 0.28 & 0.33 \\ 
	K2-260 & 0.29 & 0.21 & 0.37 & 0.14 & 0.28 & 0.21 & 0.36 & 0.16 & 0.36 & 0.15 \\ 
	TrES-3 & 0.31 & 0.24 & 0.41 & 0.15 & 0.33 & 0.23 & 0.38 & 0.19 & 0.40 & 0.16 \\ 
	Qatar-10 & 0.31 & 0.22 & 0.37 & 0.16 & 0.30 & 0.22 & 0.37 & 0.17 & 0.37 & 0.15 \\ 
	WASP-144 & 0.43 & 0.16 & 0.46 & 0.13 & 0.44 & 0.16 & 0.44 & 0.15 & 0.45 & 0.13 \\ 
\enddata
\end{deluxetable*}


\startlongtable
\begin{deluxetable*}{lcccccc}
\tablecaption{Retrieved planetary parameters along with their literature values. The data of literature planetary parameters was extracted from the NASA Exoplanet Archive on February 23, 2021.\\ {{Only first 20 rows are shown here; the full table is available on the website in a machine readable ASCII form.}}}
\label{tab:Plan_prop}
\tablehead{\colhead{Planet name} & \colhead{$R_p/R_*$} & \colhead{$a/R_*$} & \colhead{$t_c$ - 2458000} & \colhead{$R_p/R_*$} & \colhead{$a/R_*$} & \colhead{$t_c$ - 2458000} \\
    & \colhead{(This work)} & \colhead{(This work)} & \colhead{(This work)} & \colhead{(Literature)} & \colhead{(Literature)} & \colhead{(Literature)}
}
\startdata
    WASP-61b & $0.0934^{+0.0008}_{-0.0009}$ & $8.09^{+0.11}_{-0.24}$ & $1198.73545^{+0.00033}_{-0.00034}$ & $0.0937^{+0.0031}_{-0.0031}$ & $8.15^{+0.24}_{-0.24}$ & $1198.738 \pm 0.003$ \\ 
	WASP-130b & $0.0955^{+0.0025}_{-0.0021}$ & $24.15^{+1.42}_{-1.62}$ & $607.58558^{+0.00042}_{-0.00041}$ & $0.0955^{+0.0044}_{-0.0044}$ & $22.67^{+0.77}_{-0.77}$ & $607.586 \pm 0.001$ \\ 
	HATS-13b & $0.1427^{+0.0033}_{-0.0028}$ & $9.76^{+0.30}_{-0.51}$ & $1083.00912^{+0.00053}_{-0.00053}$ & $0.1402^{+0.0016}_{-0.0016}$ & $9.82^{+0.18}_{-0.18}$ & $1083.006 \pm 0.002$ \\ 
	WASP-156b & $0.0672^{+0.0009}_{-0.0008}$ & $12.86^{+0.23}_{-0.47}$ & $1169.86001^{+0.00030}_{-0.00028}$ & $0.0685^{+0.0012}_{-0.0012}$ & $12.80^{+0.30}_{-0.30}$ & $1169.861 \pm 0.004$ \\ 
	NGTS-4b & $0.0341^{+0.0017}_{-0.0018}$ & $5.69^{+0.45}_{-1.01}$ & $1227.52707^{+0.00111}_{-0.00098}$ & $0.0350^{+0.0030}_{-0.0030}$ & $4.79^{+1.21}_{-1.21}$ & $1227.529 \pm 0.010$ \\ 
	WASP-190b & $0.0781^{+0.0029}_{-0.0020}$ & $9.05^{+0.68}_{-0.88}$ & $1098.12518^{+0.00064}_{-0.00065}$ & $0.0741^{+0.0075}_{-0.0075}$ & $8.95^{+0.57}_{-0.57}$ & $1098.122 \pm 0.001$ \\ 
	WASP-44b & $0.1174^{+0.0040}_{-0.0036}$ & $8.47^{+0.67}_{-0.70}$ & $393.84978^{+0.00035}_{-0.00036}$ & $0.1248^{+0.0023}_{-0.0023}$ & $8.20^{+0.43}_{-0.43}$ & $393.8498 \pm 0.0004$ \\ 
	TOI-892b & $0.0800^{+0.0009}_{-0.0009}$ & $15.77^{+0.27}_{-0.64}$ & $1208.92591^{+0.00064}_{-0.00062}$ & $0.0790^{+0.0010}_{-0.0010}$ & $14.20^{+0.80}_{-0.80}$ & $1208.922 \pm 0.005$ \\ 
	WASP-82b & $0.0773^{+0.0004}_{-0.0004}$ & $4.42^{+0.04}_{-0.07}$ & $449.78954^{+0.00017}_{-0.00016}$ & $0.0788^{+0.0037}_{-0.0037}$ & $4.43^{+0.15}_{-0.15}$ & $449.794 \pm 0.002$ \\ 
	TOI-540b & $0.0462^{+0.0028}_{-0.0051}$ & $13.31^{+4.68}_{-2.43}$ & $1199.92424^{+0.00030}_{-0.00029}$ & $0.0436^{+0.0012}_{-0.0012}$ & $13.90^{+0.72}_{-0.72}$ & $1199.925 \pm 0.001$ \\ 
	WASP-22b & $0.1001^{+0.0011}_{-0.0011}$ & $8.28^{+0.22}_{-0.30}$ & $422.50014^{+0.00023}_{-0.00023}$ & $0.0978^{+0.0012}_{-0.0012}$ & $8.94^{+0.85}_{-0.85}$ & $422.468 \pm 0.008$ \\ 
	TOI-905b & $0.1296^{+0.0027}_{-0.0024}$ & $11.47^{+0.42}_{-0.40}$ & $639.56977^{+0.00021}_{-0.00021}$ & $0.1312^{+0.0079}_{-0.0079}$ & $10.94^{+0.50}_{-0.50}$ & $639.5695 \pm 0.0002$ \\ 
	HAT-P-42b & $0.0811^{+0.0013}_{-0.0012}$ & $9.75^{+0.23}_{-0.48}$ & $1252.87532^{+0.00052}_{-0.00052}$ & $0.0860^{+0.0033}_{-0.0033}$ & $8.08^{+0.82}_{-0.82}$ & $1252.901 \pm 0.023$ \\ 
	WASP-62b & $0.1111^{+0.0001}_{-0.0001}$ & $9.72^{+0.03}_{-0.03}$ & $1252.58570^{+0.00004}_{-0.00004}$ & $0.1105^{+0.0003}_{-0.0003}$ & $9.55^{+0.41}_{-0.41}$ & $1252.594 \pm 0.002$ \\ 
	HATS-3b & $0.0973^{+0.0011}_{-0.0012}$ & $8.01^{+0.34}_{-0.34}$ & $337.89608^{+0.00031}_{-0.00031}$ & $0.1011^{+0.0006}_{-0.0006}$ & $7.42^{+0.12}_{-0.12}$ & $337.896 \pm 0.003$ \\ 
	LTT9779b & $0.0337^{+0.0011}_{-0.0009}$ & $7.62^{+0.30}_{-0.36}$ & $1112.22021^{+0.00017}_{-0.00017}$ & $0.0455^{+0.0022}_{-0.0022}$ & $3.88^{+0.09}_{-0.09}$ & $1112.208 \pm 0.009$ \\ 
	K2-260b & $0.0947^{+0.0016}_{-0.0017}$ & $5.11^{+0.11}_{-0.20}$ & $1186.61999^{+0.00056}_{-0.00057}$ & $0.0973^{+0.0003}_{-0.0003}$ & $5.29^{+0.03}_{-0.03}$ & $1186.604 \pm 0.001$ \\ 
	TrES-3b & $0.1706^{+0.0058}_{-0.0039}$ & $5.82^{+0.12}_{-0.13}$ & $1034.47459^{+0.00009}_{-0.00009}$ & $0.1660^{+0.0024}_{-0.0024}$ & $5.95^{+0.05}_{-0.05}$ & $1034.472 \pm 0.001$ \\ 
	Qatar-10b & $0.1255^{+0.0015}_{-0.0015}$ & $5.05^{+0.08}_{-0.12}$ & $1034.37308^{+0.00021}_{-0.00020}$ & $0.1265^{+0.0010}_{-0.0010}$ & $4.90^{+0.12}_{-0.12}$ & $1034.371 \pm 0.005$ \\ 
	WASP-144b & $0.1128^{+0.0029}_{-0.0041}$ & $7.32^{+0.73}_{-0.60}$ & $1071.05883^{+0.00038}_{-0.00037}$ & $0.1079^{+0.0013}_{-0.0013}$ & $8.39^{+0.23}_{-0.23}$ & $1071.060 \pm 0.001$ \\ 
\enddata
\end{deluxetable*}

\section{Figures}


\begin{figure*}
    \centering
    \includegraphics[width=\columnwidth]{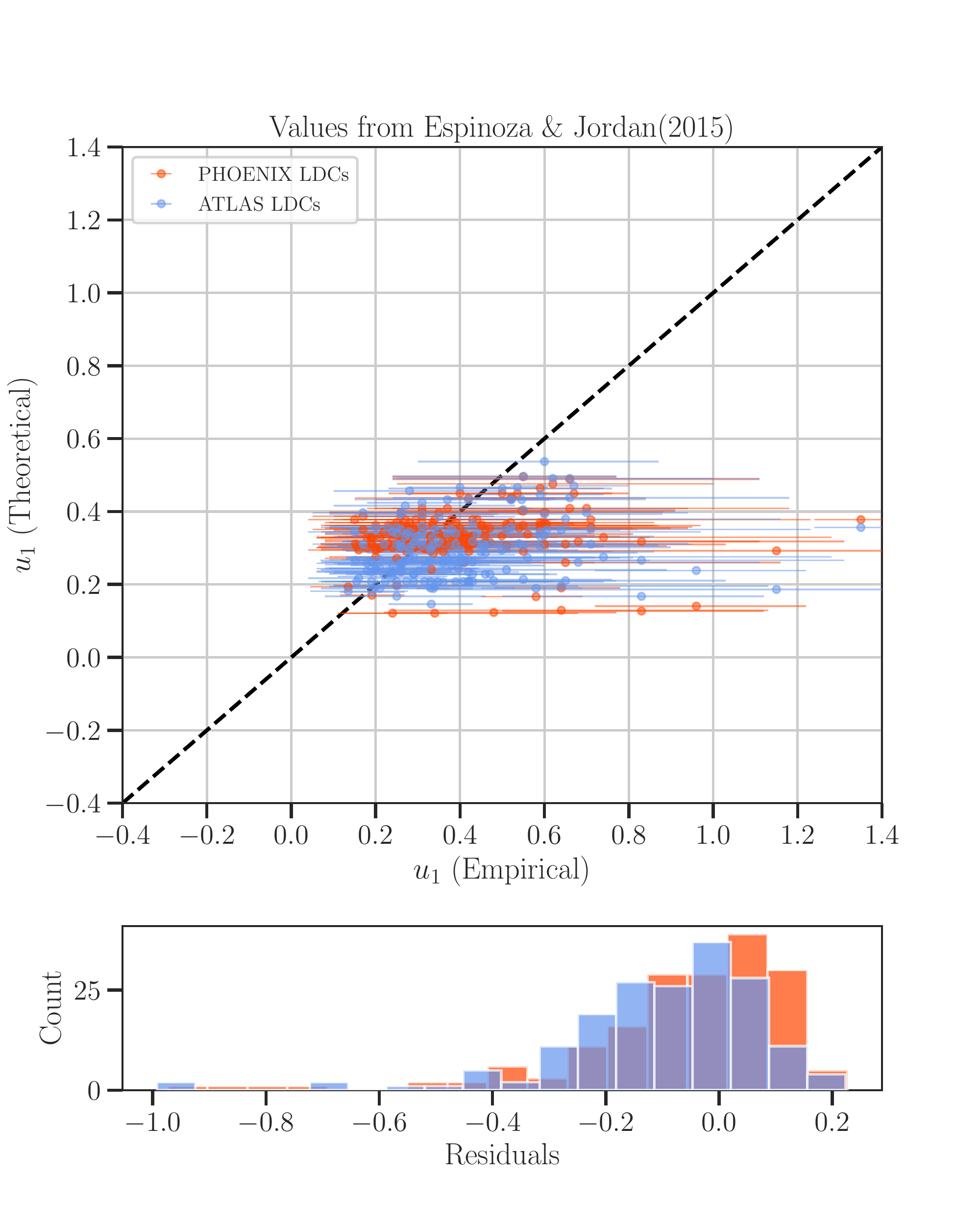}
    \includegraphics[width=\columnwidth]{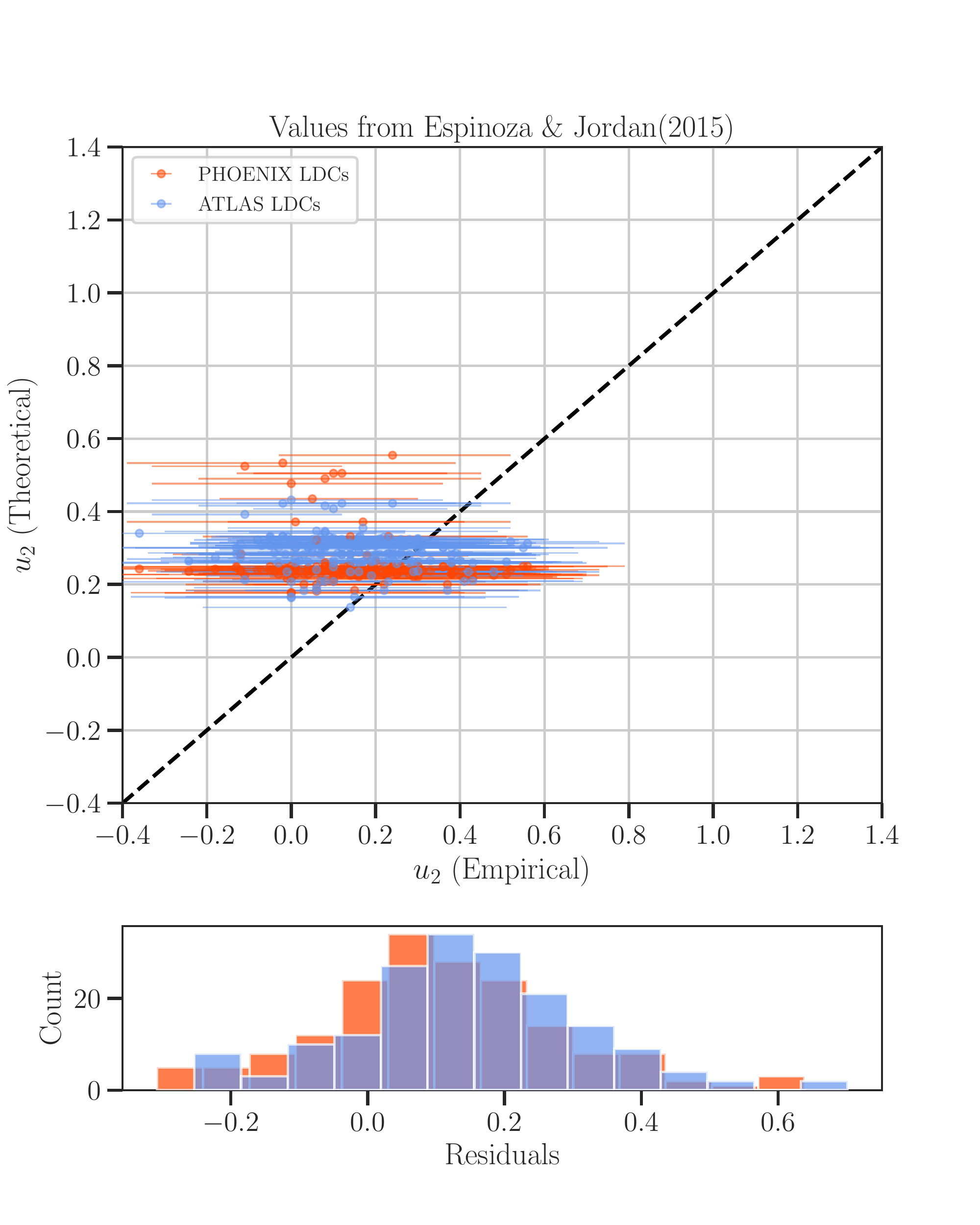}
    \caption{Comparison of the retrieved LDCs $u_1$ and $u_2$ from the TESS data with the one calculated from the model stellar atmospheres using the code provided by \citet{2015MNRAS.450.1879E}. In upper panel, y-axis represents the theoretical LDCs while x-axis shows the empirical values of LDCs. The black dashed line is the line of equality between the x- and y- axis. On the other hand, the lower panel shows the distribution of residuals between the theoretical and empirical LDCs.}
    \label{fig:u1_code}
\end{figure*}

\begin{figure*}
    \centering
    \includegraphics[width=\columnwidth]{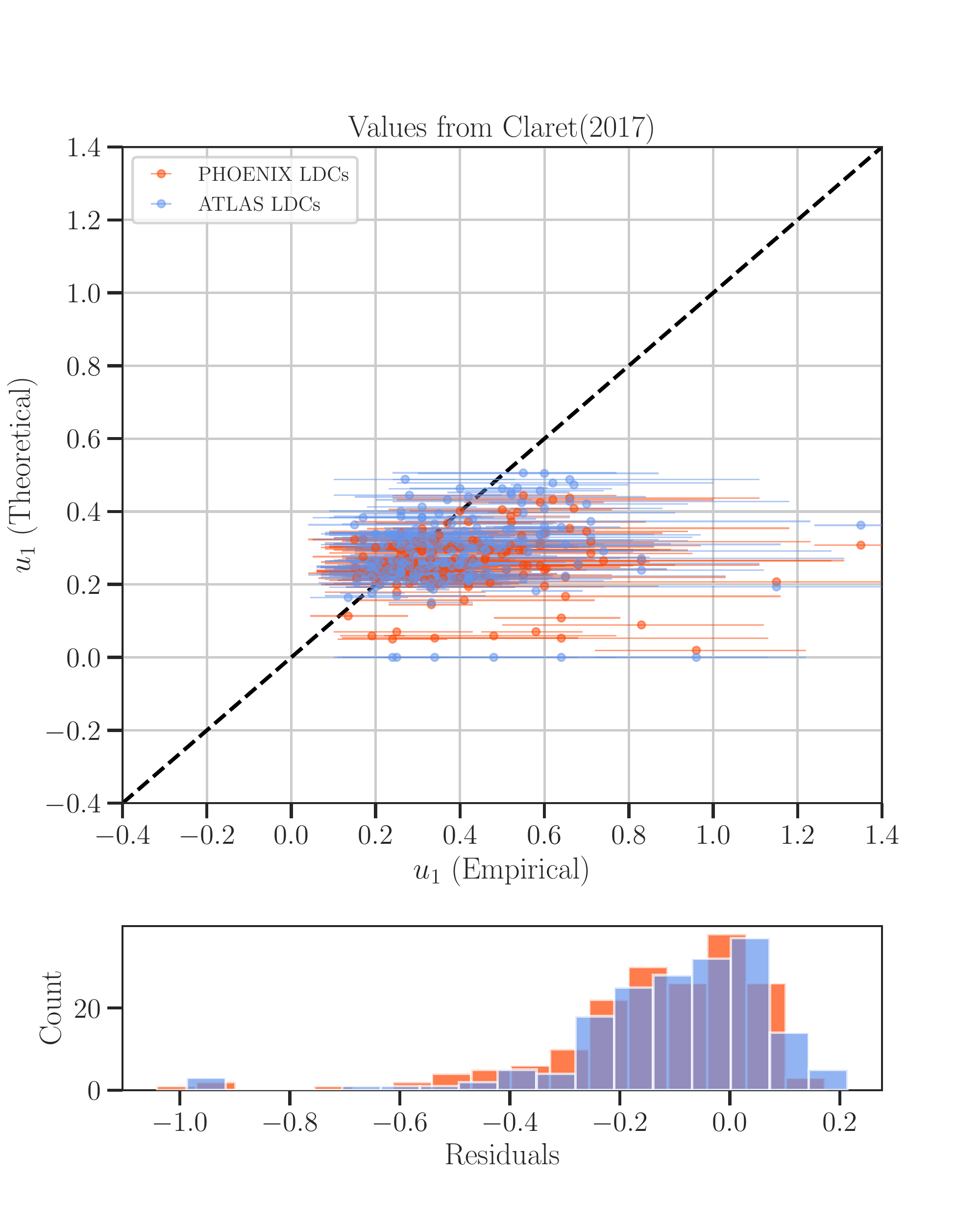}
    \includegraphics[width=\columnwidth]{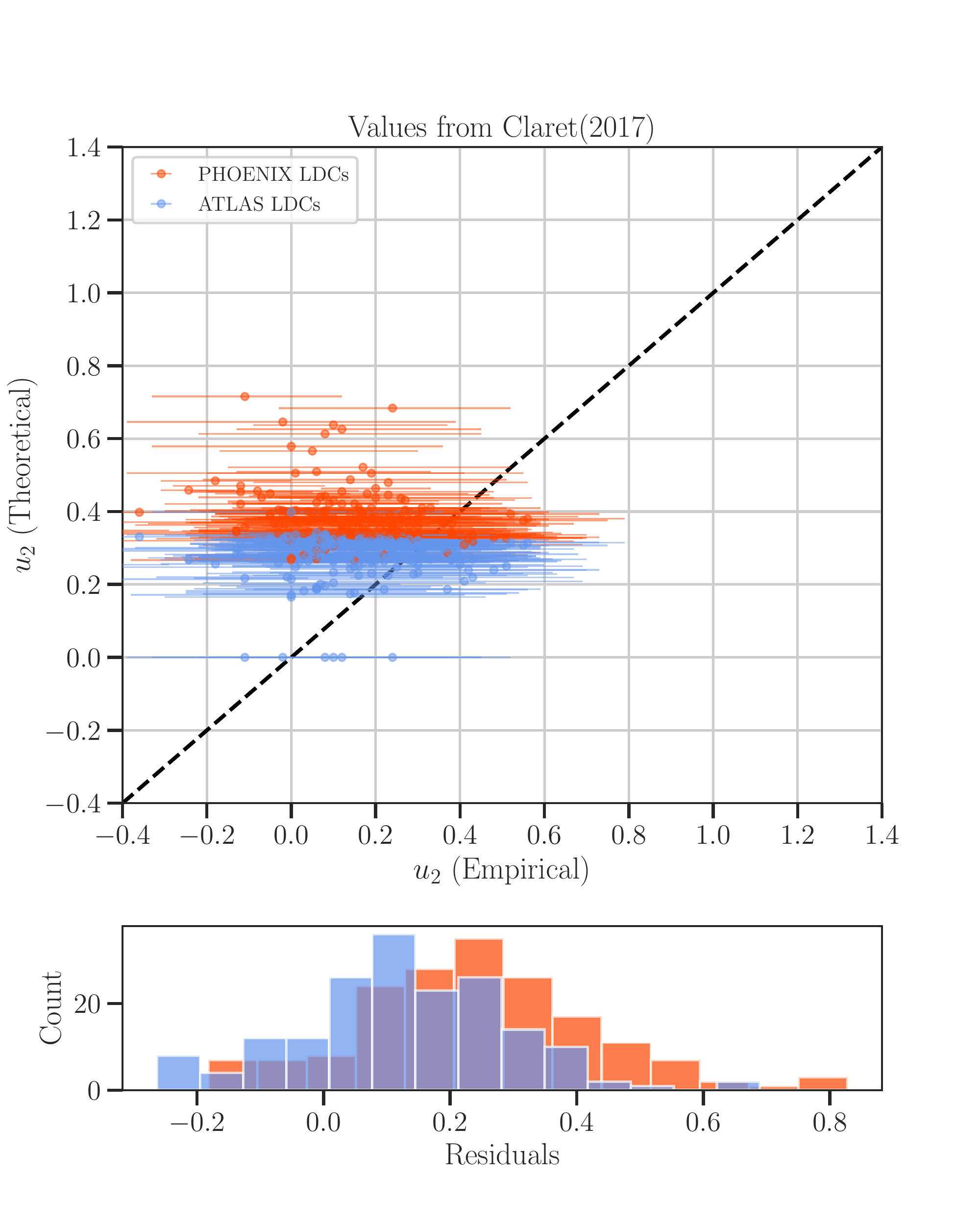}
    \caption{Comparison of the retrieved $u_1$ from the TESS data with the one calculated from the model stellar atmospheres using the tabulated values from \citet{2017AA...600A..30C}. In upper panel, y-axis represents the theoretical LDCs while x-axis shows the empirical values of LDCs. The black dashed line is the line of equality between the x- and y- axis. On the other hand, the lower panel shows the distribution of residuals between the theoretical and empirical LDCs.}
    \label{fig:u1_cla}
\end{figure*}

\begin{figure*}
    \centering
    \includegraphics[width=\columnwidth]{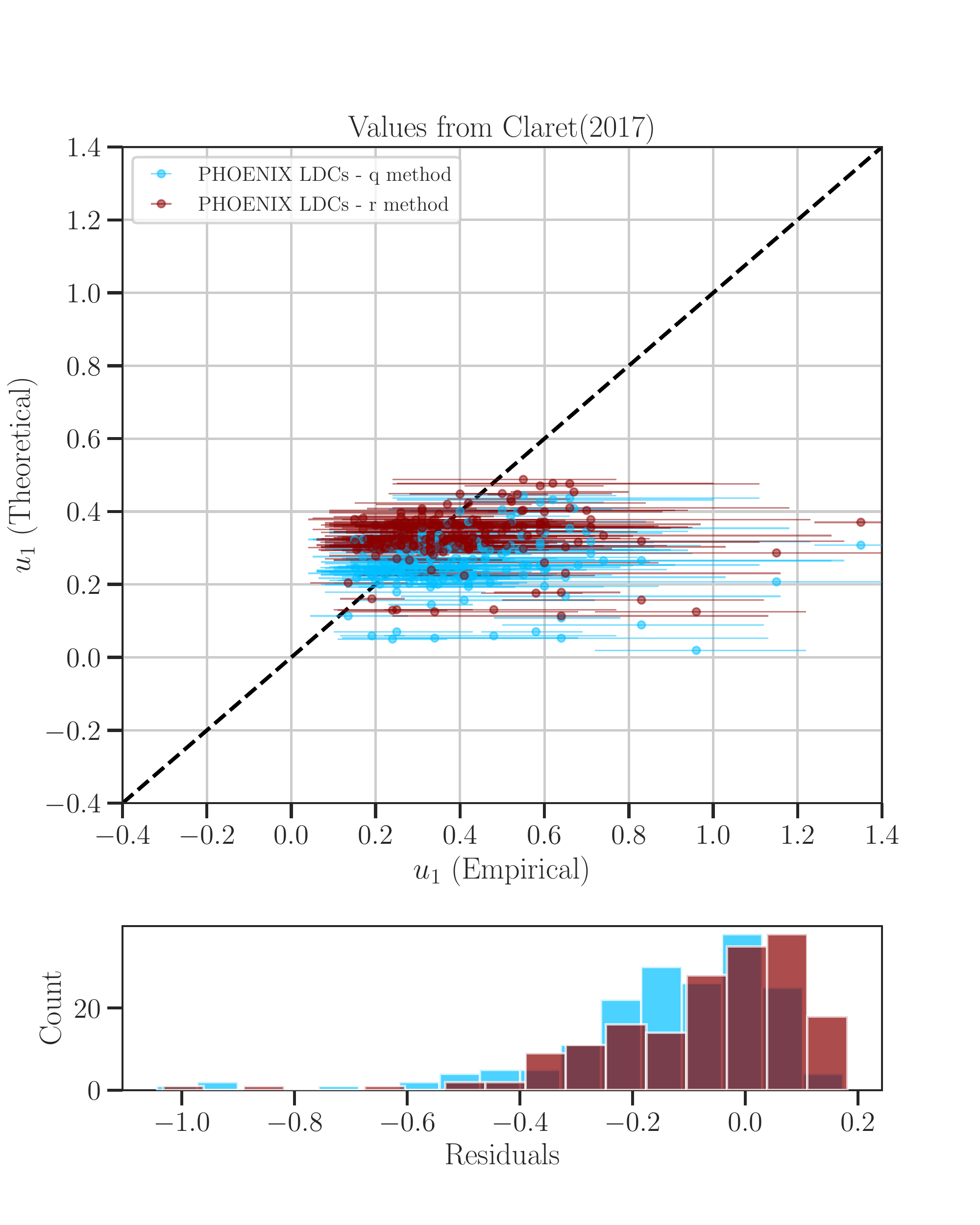}
    \includegraphics[width=\columnwidth]{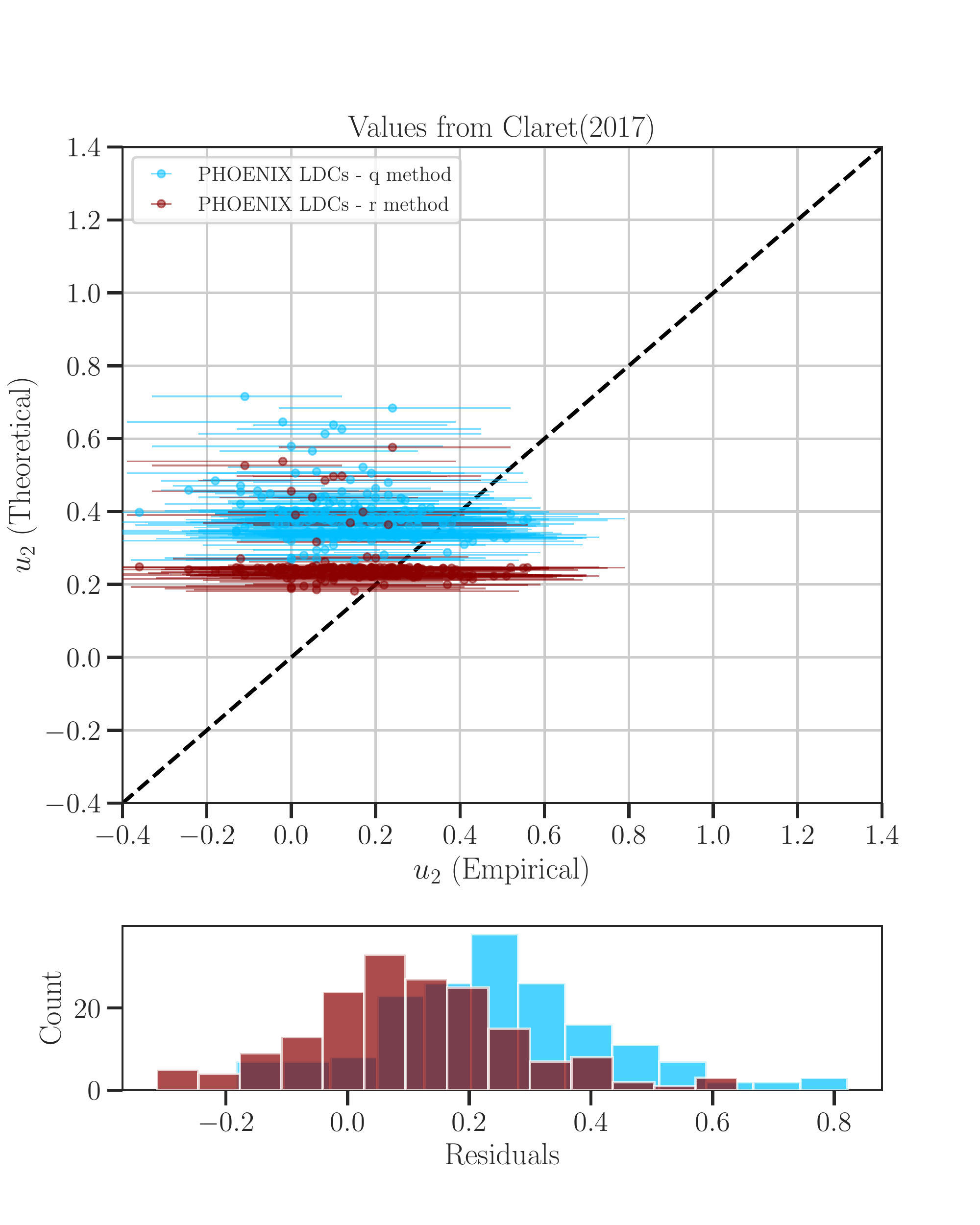}
    \caption{Comparison of the retrieved $u_1$ and $u_2$ coefficients from the TESS data with the one calculated from the model stellar atmospheres using tabulated values from \citet{2017AA...600A..30C} using \textit{r-method} and \textit{q-method}.}
    \label{fig:u1_cla_r}
\end{figure*}


\begin{figure*}
    \centering
    \includegraphics[width=\columnwidth]{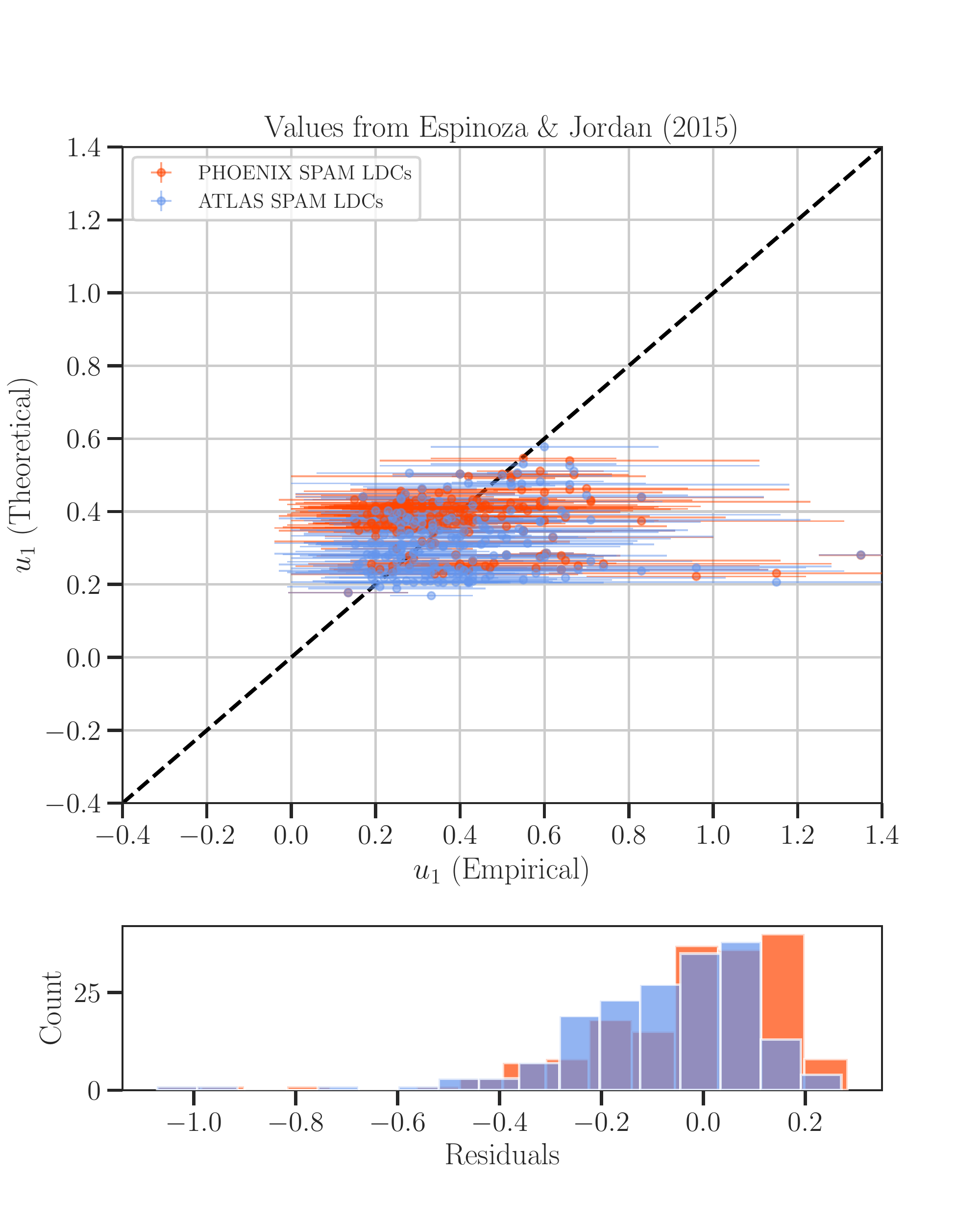}
    \includegraphics[width=\columnwidth]{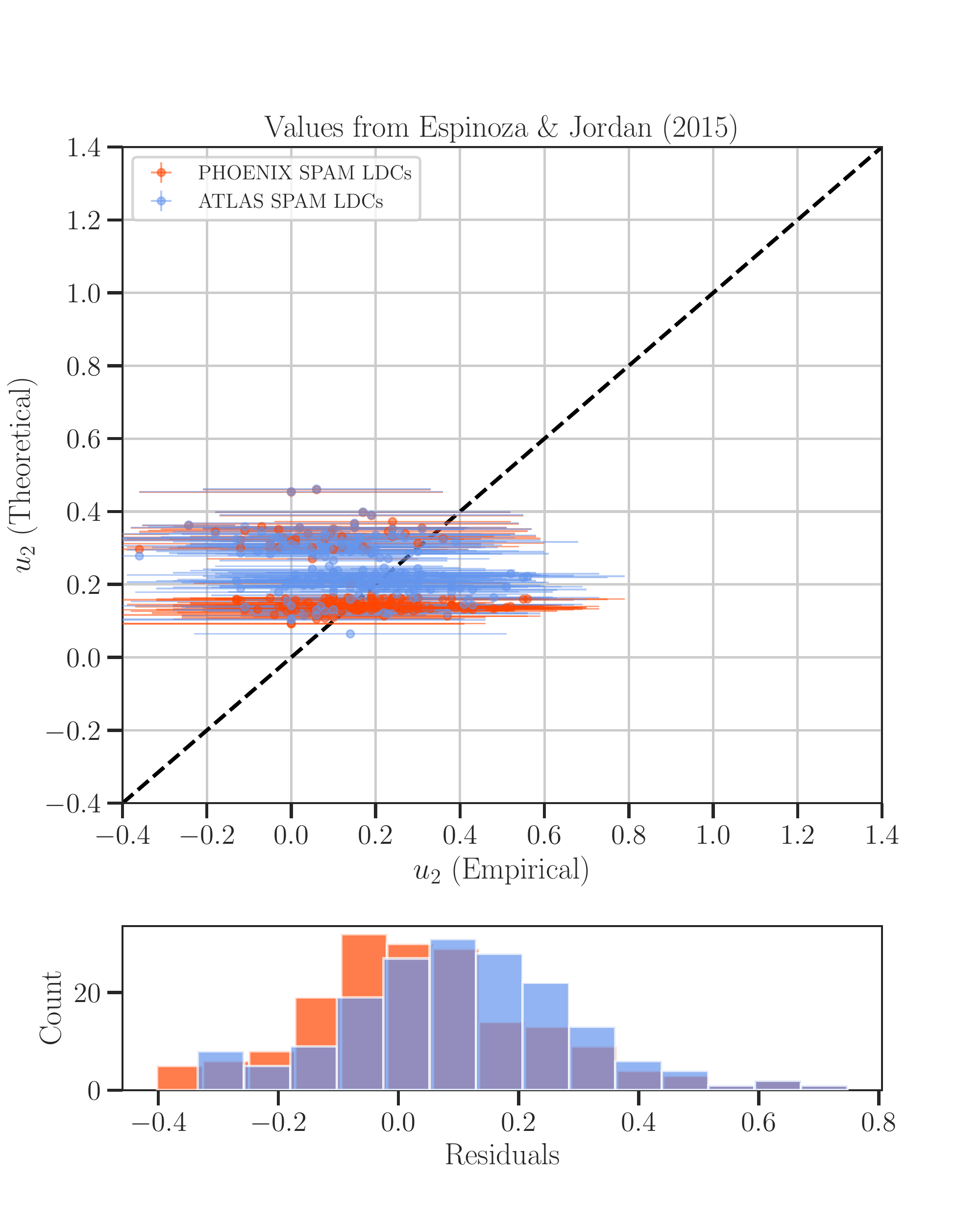}
    \caption{Comparison of the retrieved LDCs $u_1$ and $u_2$ from the TESS data with the SPAM LDCs. The later made use of non-linear LDCs from the code provided by \citet{2015MNRAS.450.1879E}. In upper panel, y-axis represents the SPAM LDCs while x-axis shows the empirical values of LDCs. The black dashed line is the line of equality between the x- and y- axis. On the other hand, the lower panel shows the distribution of residuals between the SPAM and empirical LDCs.}
    \label{fig:u1_code_SPAM}
\end{figure*}

\begin{figure*}
    \centering
    \includegraphics[width=\columnwidth]{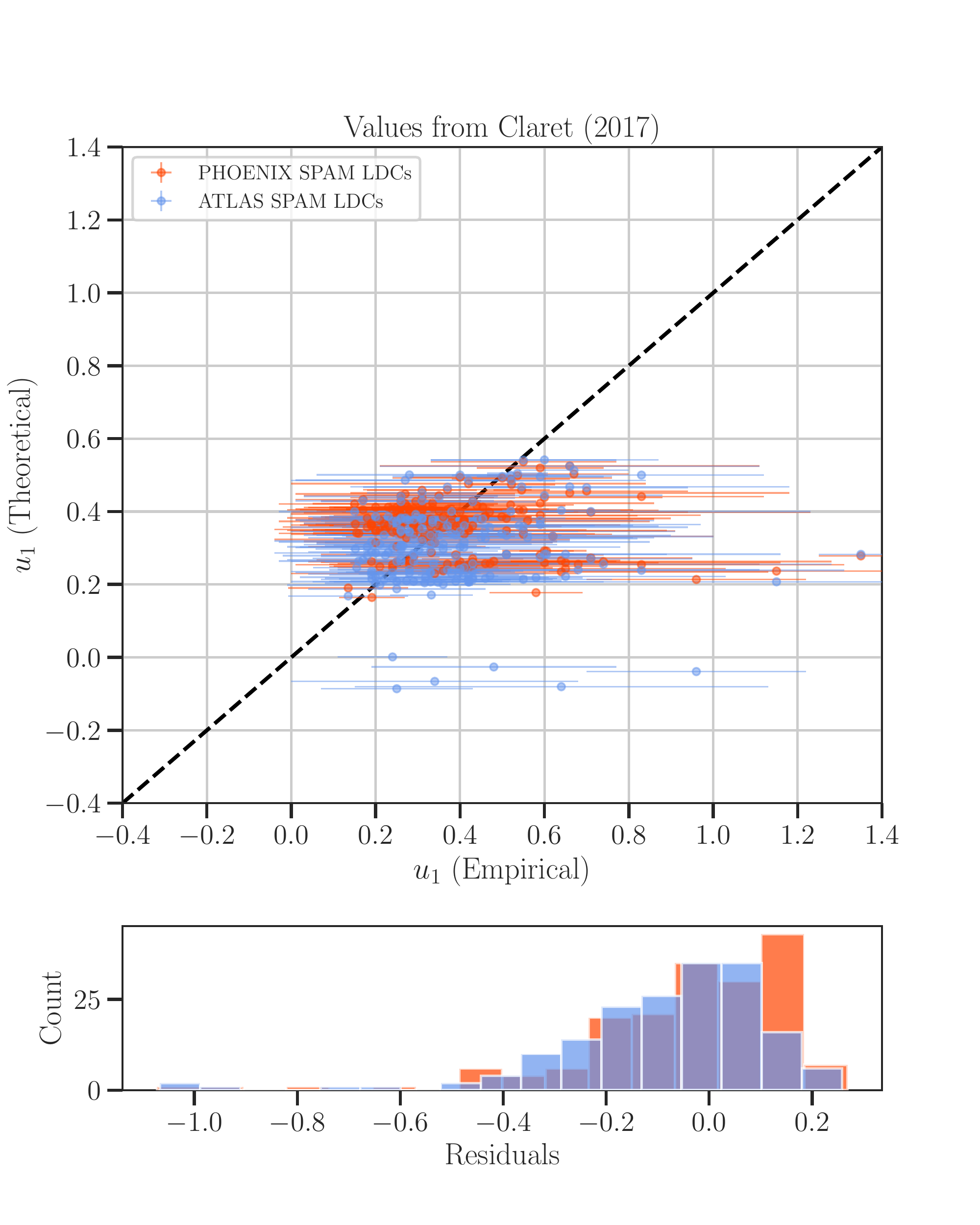}
    \includegraphics[width=\columnwidth]{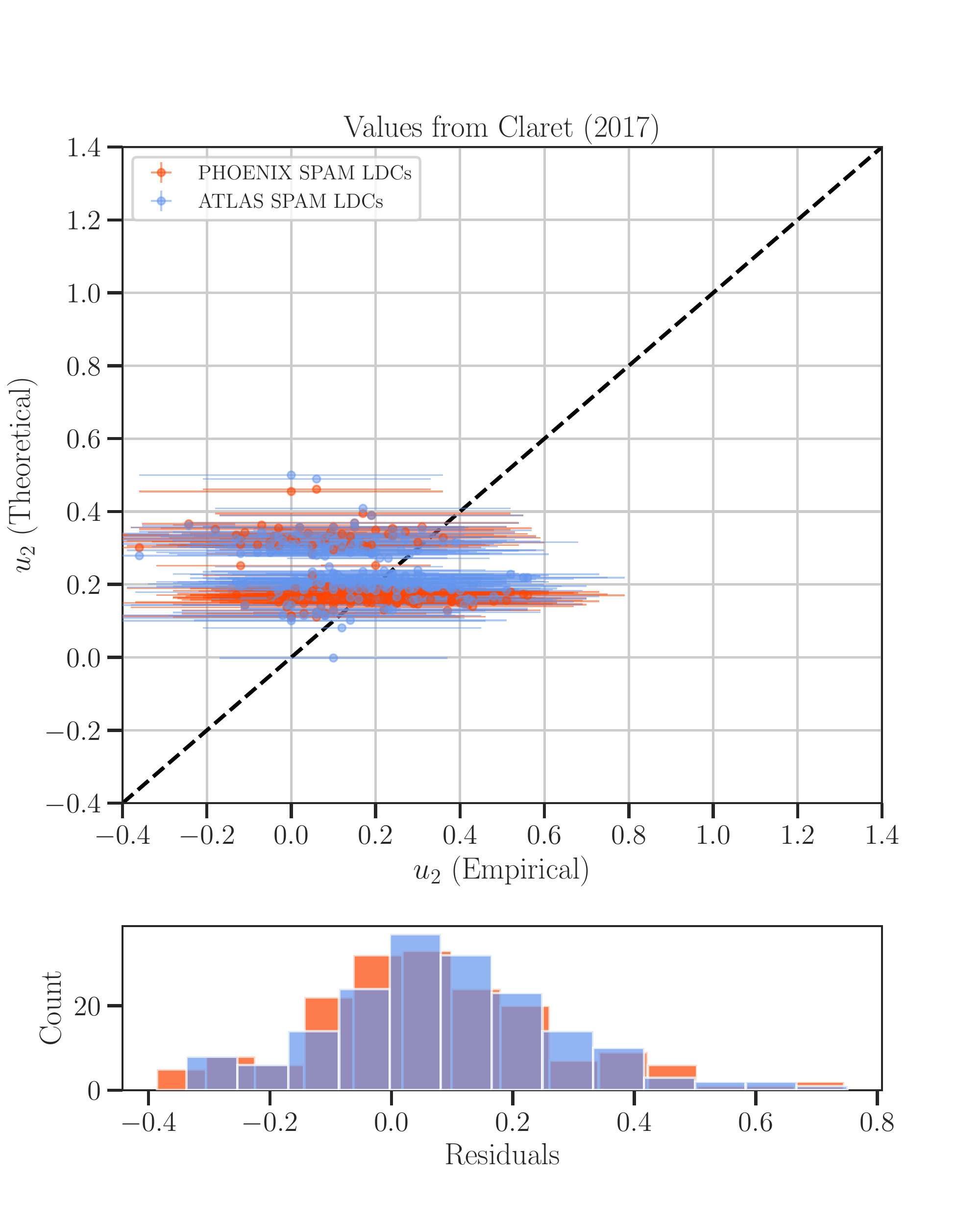}
    \caption{Comparison of the retrieved LDCs $u_1$ and $u_2$ from the TESS data with the SPAM LDCs. The later made use of non-linear LDCs from the code provided by \citet{2017AA...600A..30C}. In upper panel, y-axis represents the SPAM LDCs while x-axis shows the empirical values of LDCs. The black dashed line is the line of equality between the x- and y- axis. On the other hand, the lower panel shows the distribution of residuals between the SPAM and empirical LDCs.}
    \label{fig:u1_cla_SPAM}
\end{figure*}

\begin{figure*}
    \centering
    \includegraphics[width=\columnwidth]{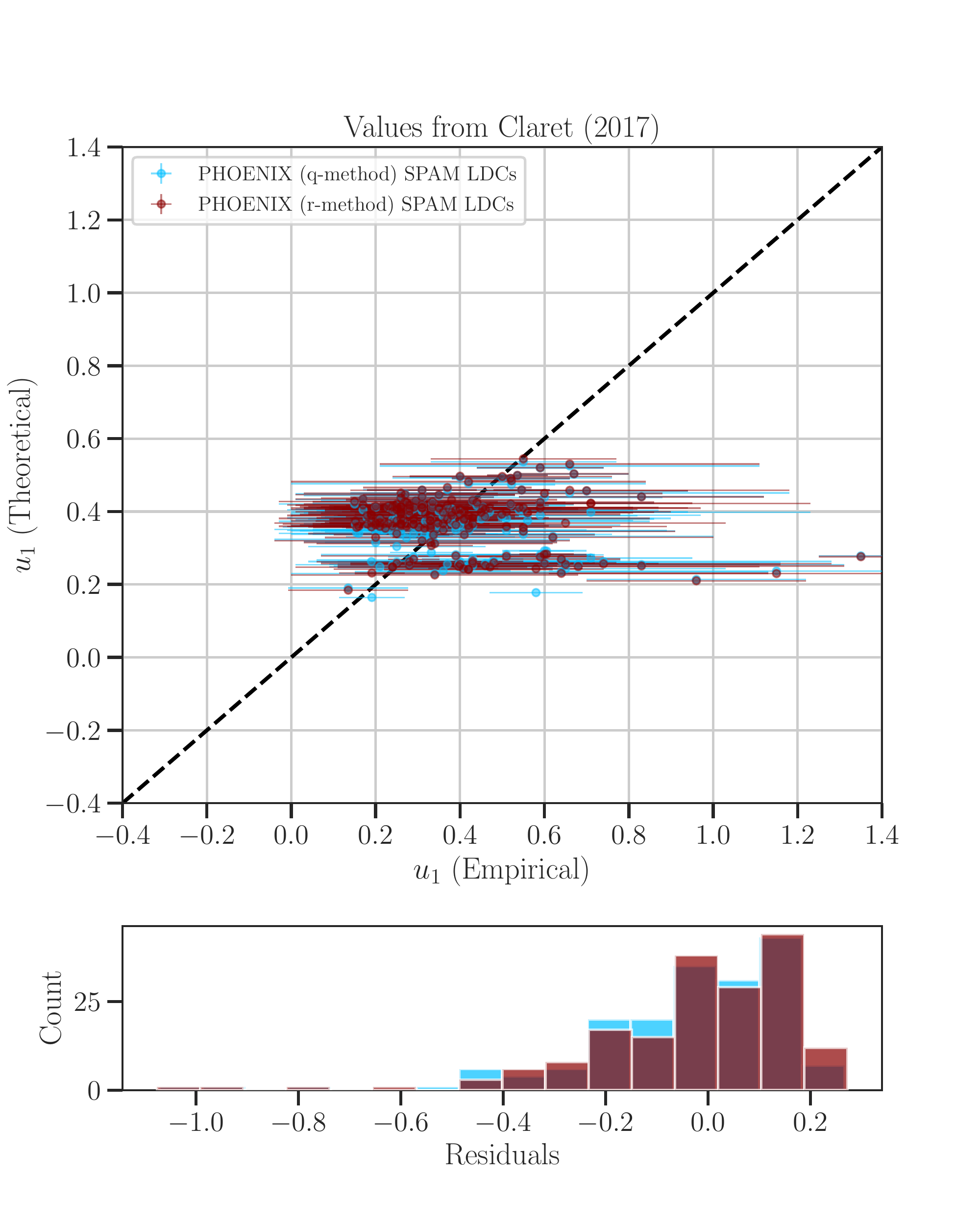}
    \includegraphics[width=\columnwidth]{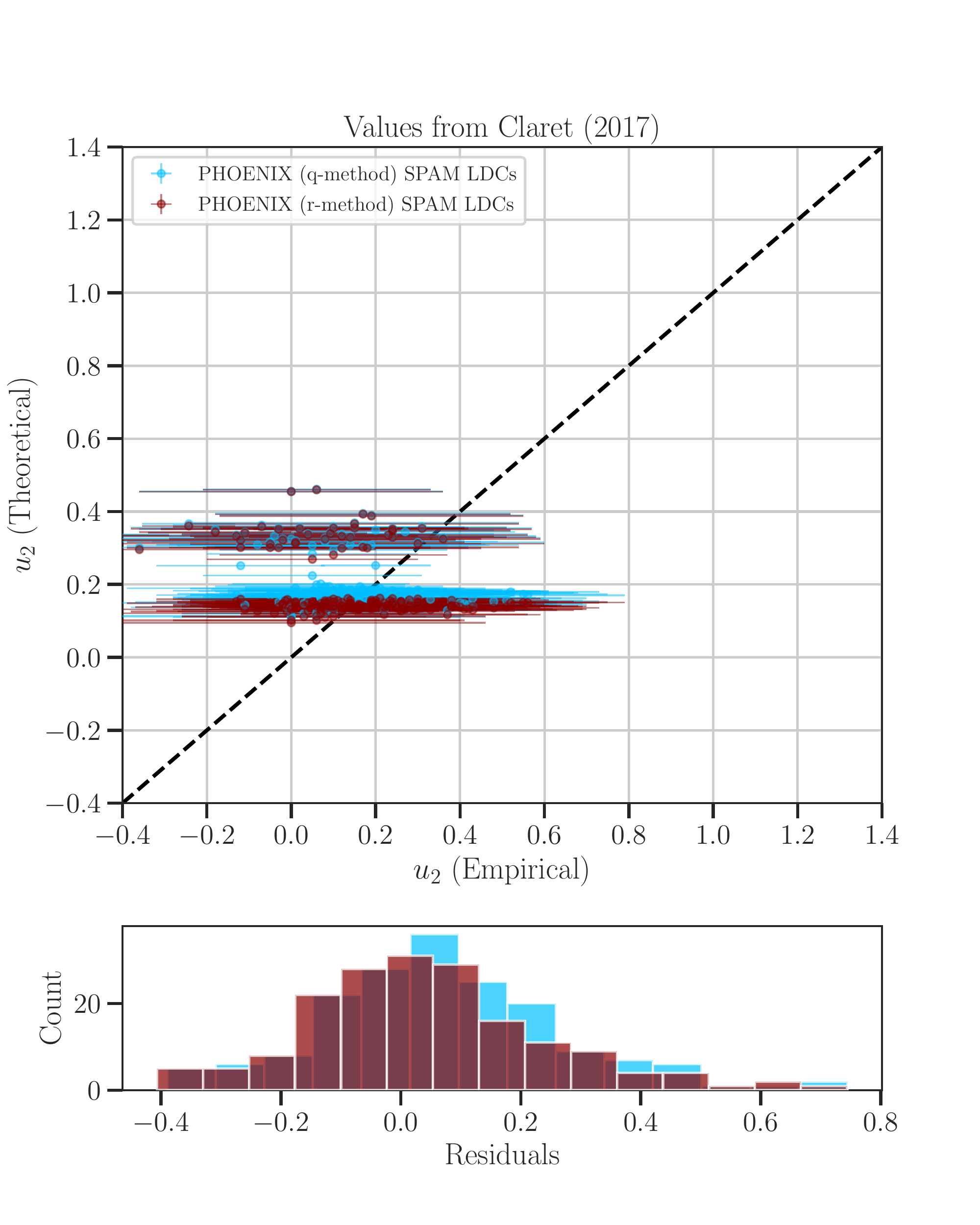}
    \caption{Comparison of the retrieved $u_1$ and $u_2$ coefficients from the TESS data with the SPAM LDCs which made use of tabulated values of non-linear LDCs from \citet{2017AA...600A..30C} using \textit{r-method} and \textit{q-method}.}
    \label{fig:u1_cla_SPAM_r}
\end{figure*}


\begin{figure*}
    \centering
    \textbf{For Tabular/Code LDCs}\par\medskip
    \includegraphics[width=\columnwidth]{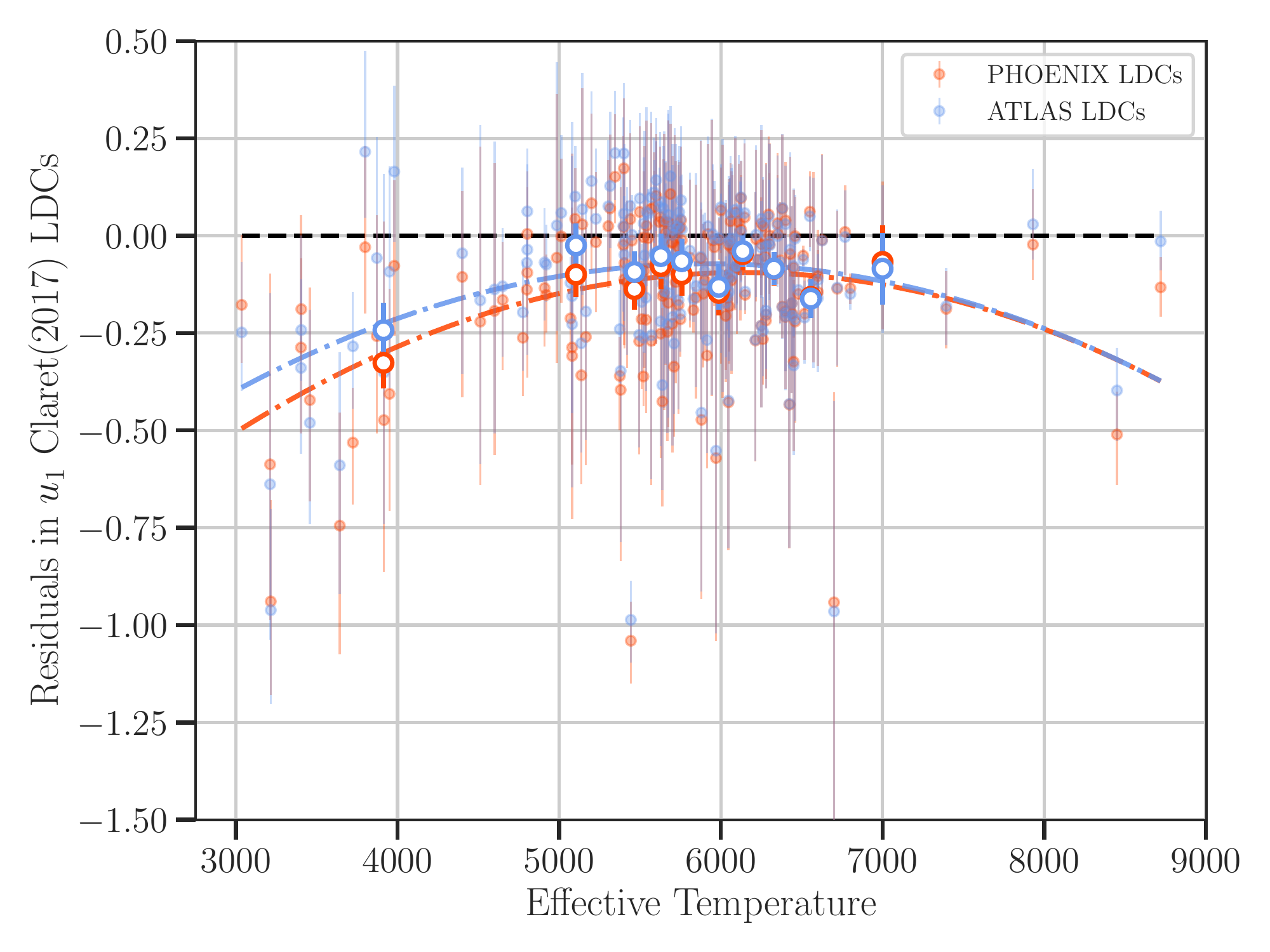}
    \includegraphics[width=\columnwidth]{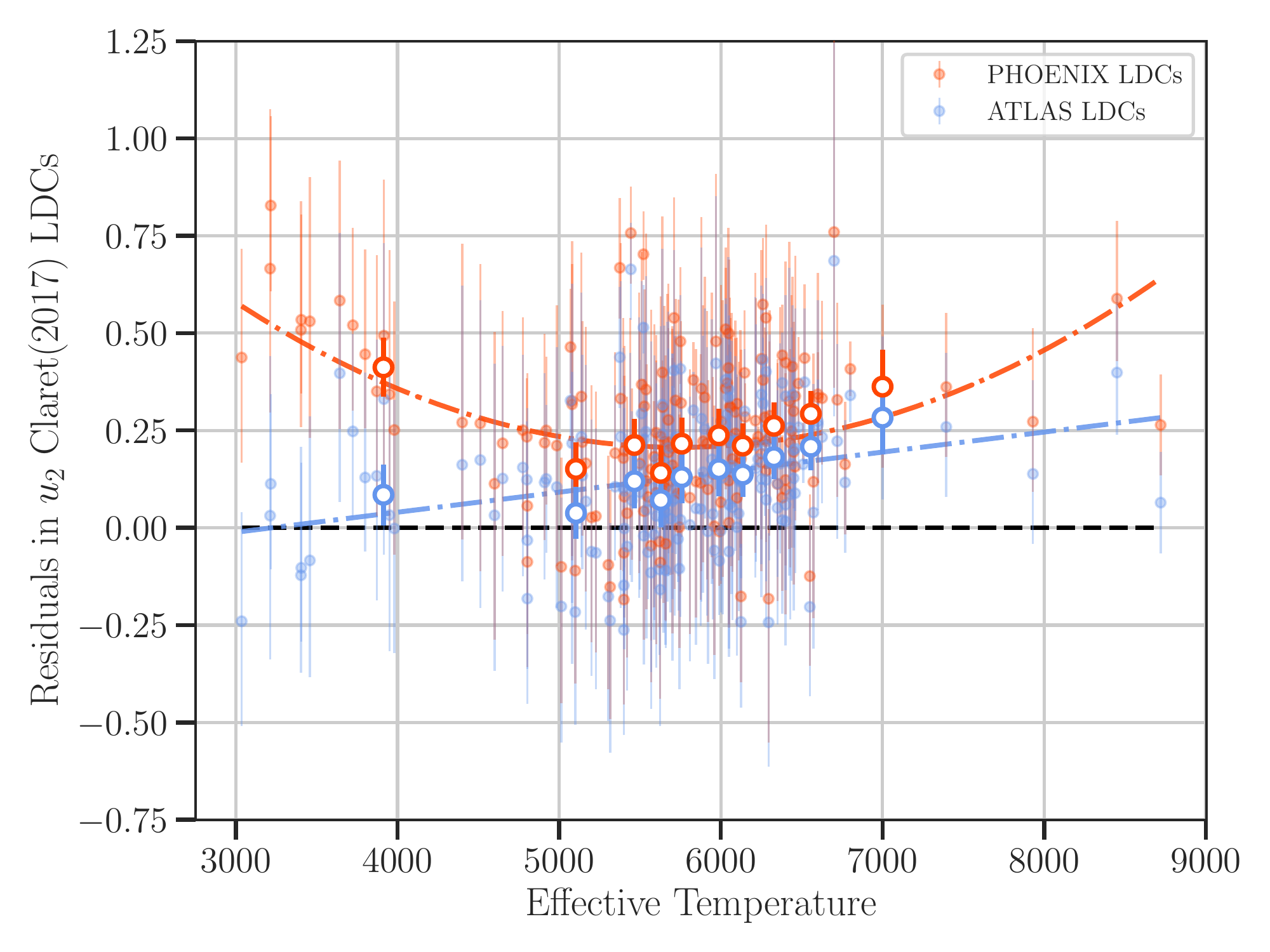}
    \caption{Temperature variation of offsets in $u_1$ and $u_2$, when, theoretically, LDCs are calculated using LDCs provided by \citet{2017AA...600A..30C}. The dashed-dotted lines show the best fitted model to the residuals.}
    \label{fig:u1_cla_te}
\end{figure*}

\begin{figure*}
    \centering
    \textbf{For Tabular/Code LDCs}\par\medskip
    \includegraphics[width=\columnwidth]{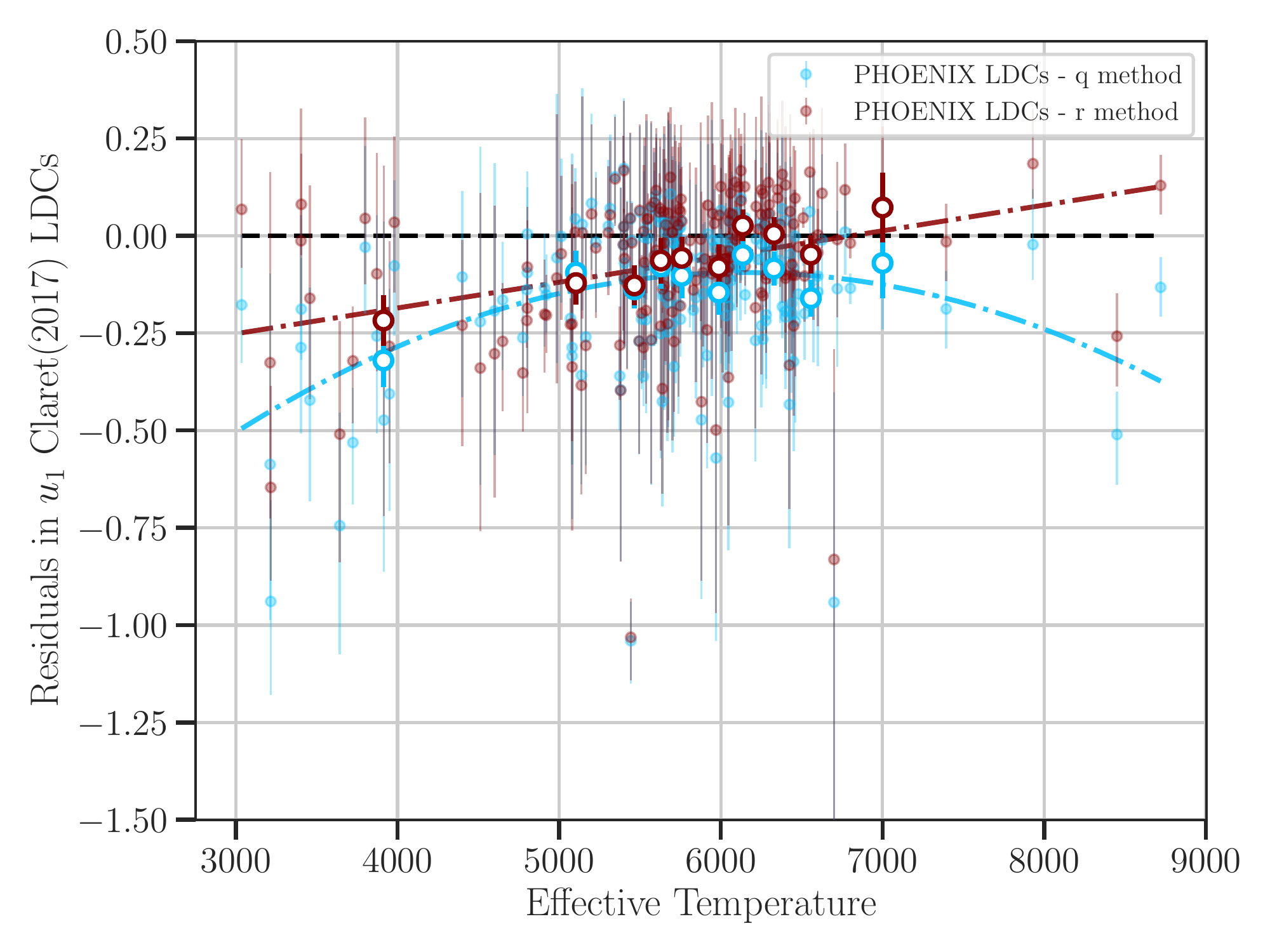}
    \includegraphics[width=\columnwidth]{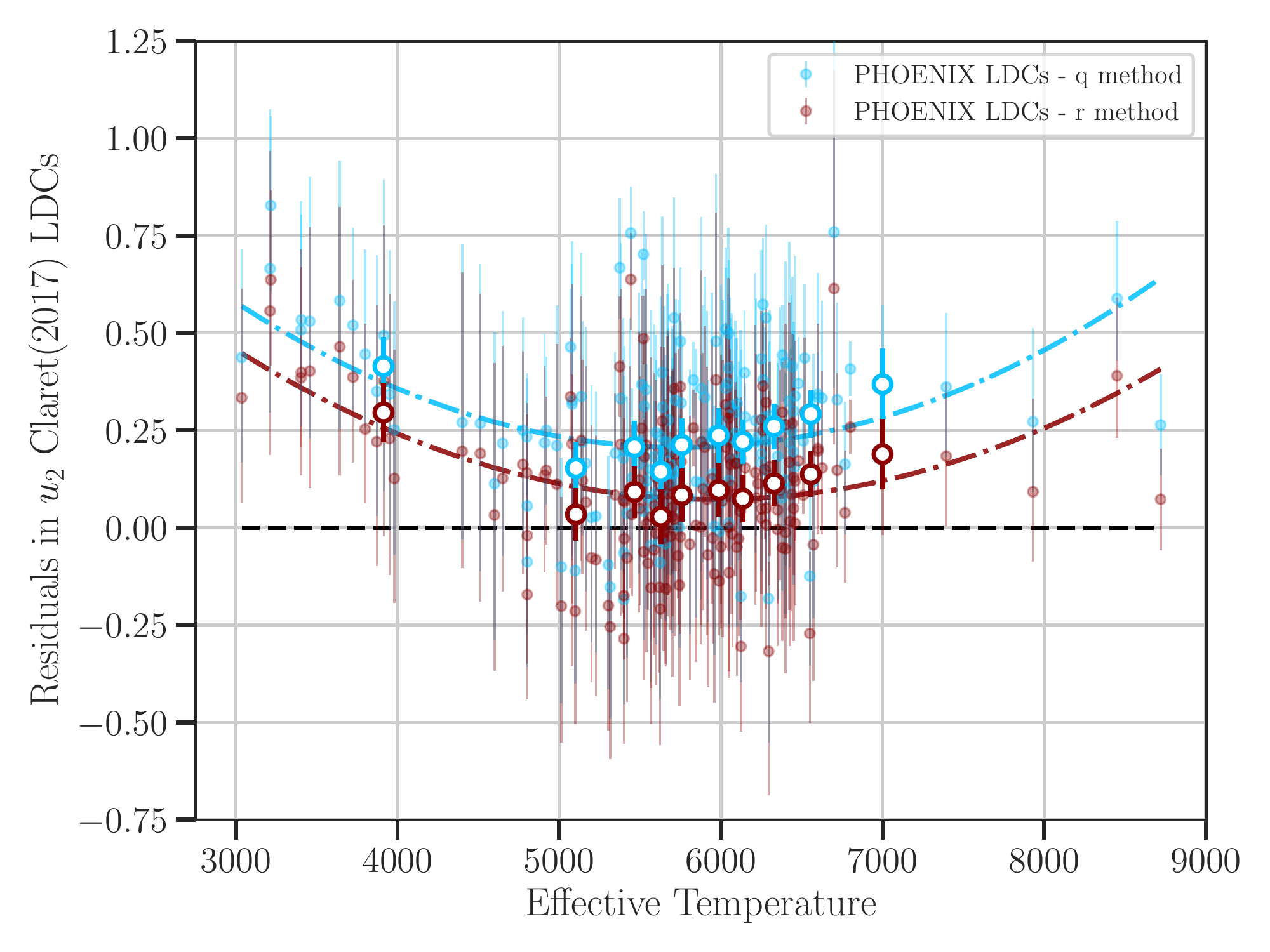}
    \caption{Temperature variation of offsets in $u_1$ and $u_2$, when, theoretically, LDCs are calculated using tables provided by \citet{2017AA...600A..30C}, using \textit{q-method} and \textit{r-method}. The dashed-dotted lines show the best fitted model to the residuals}
    \label{fig:u1_cla_te_r}
\end{figure*}


\begin{figure*}
    \centering
    \textbf{For SPAM LDCs}\par\medskip
    \includegraphics[width=\columnwidth]{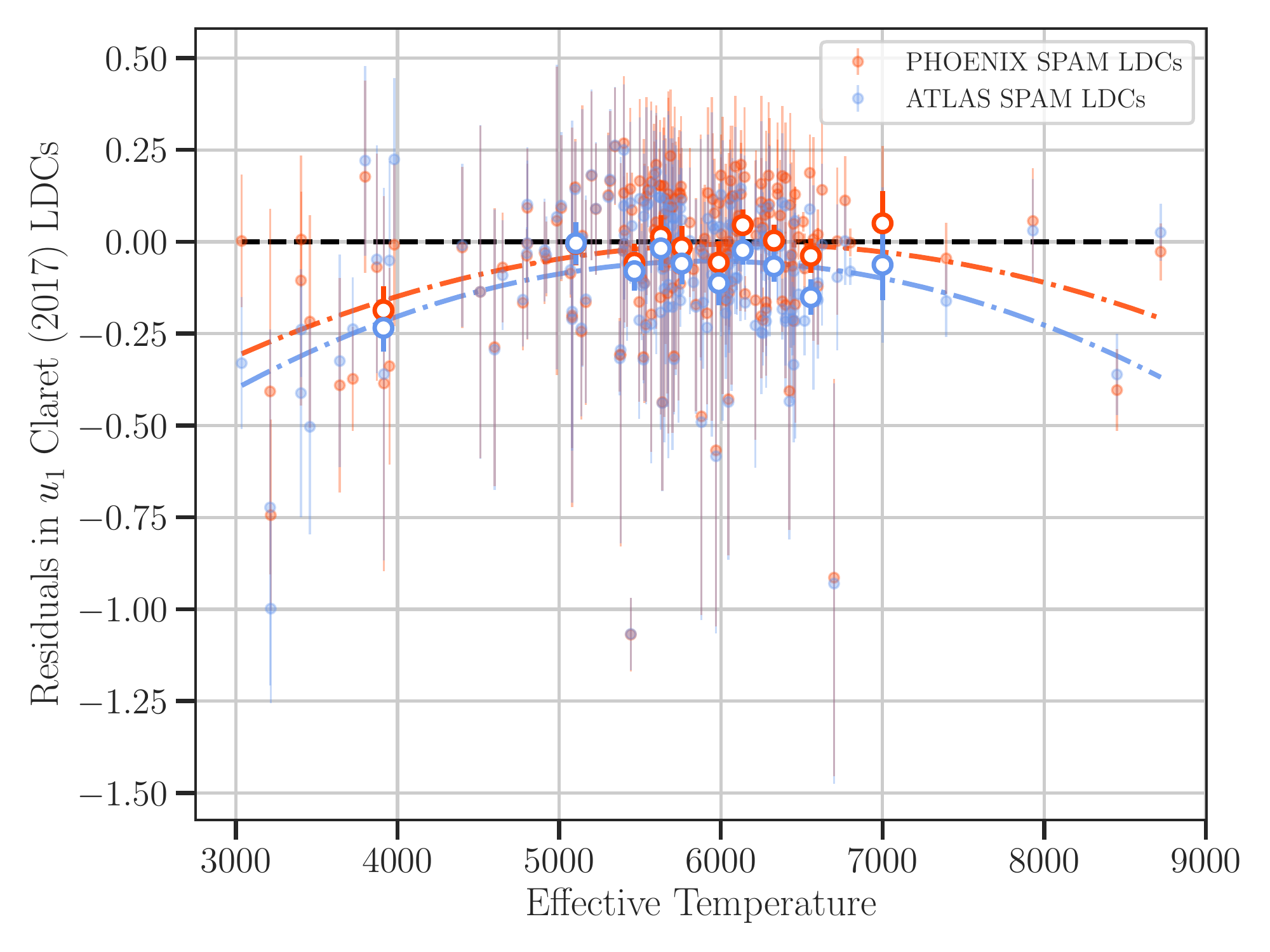}
    \includegraphics[width=\columnwidth]{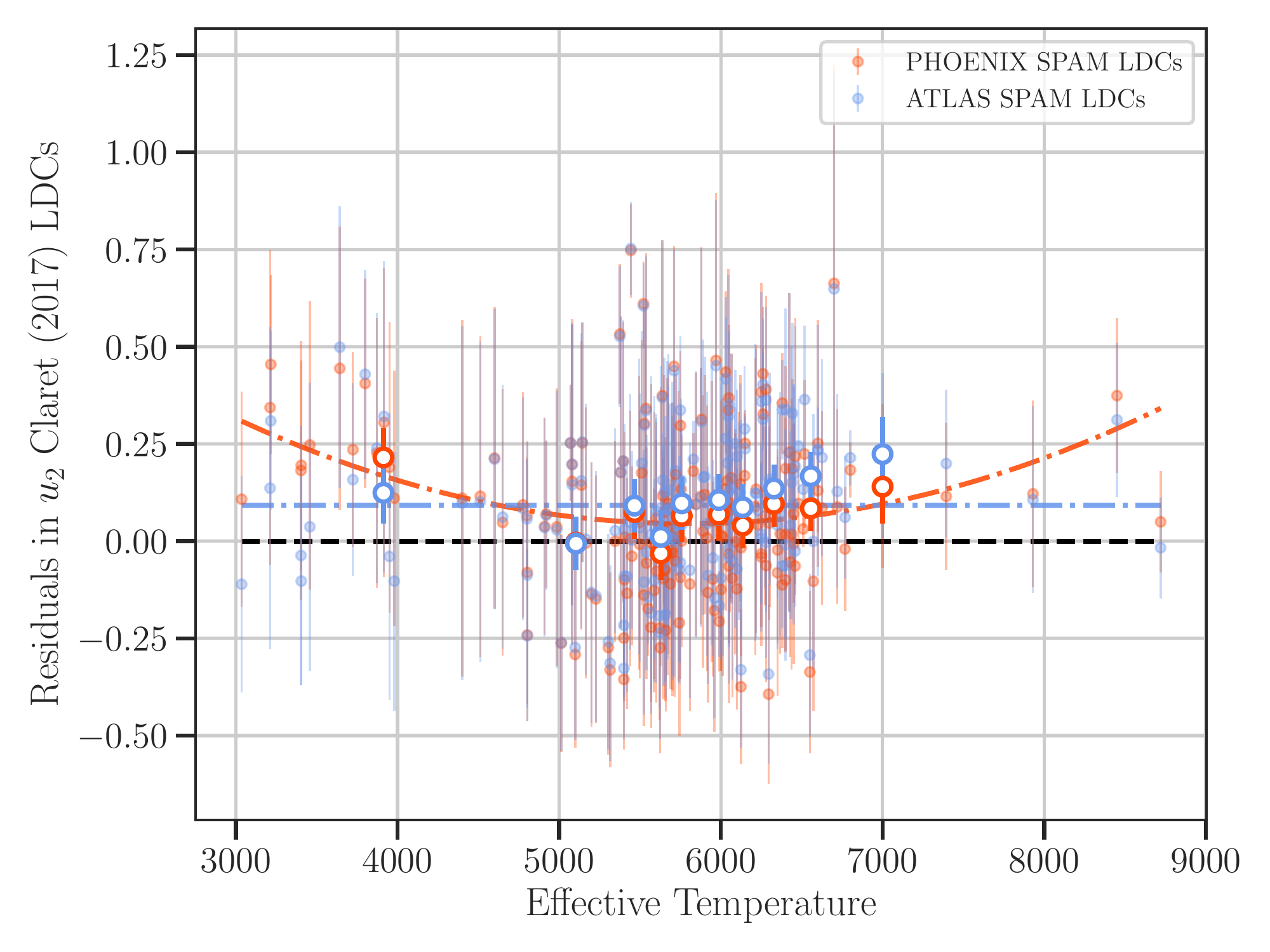}
    \caption{Same as Figure \ref{fig:u1_cla_te}, but now using SPAM LDcs.}
    \label{fig:u1_cla_te_SPAM}
\end{figure*}

\begin{figure*}
    \centering
    \textbf{For SPAM LDCs}\par\medskip
    \includegraphics[width=\columnwidth]{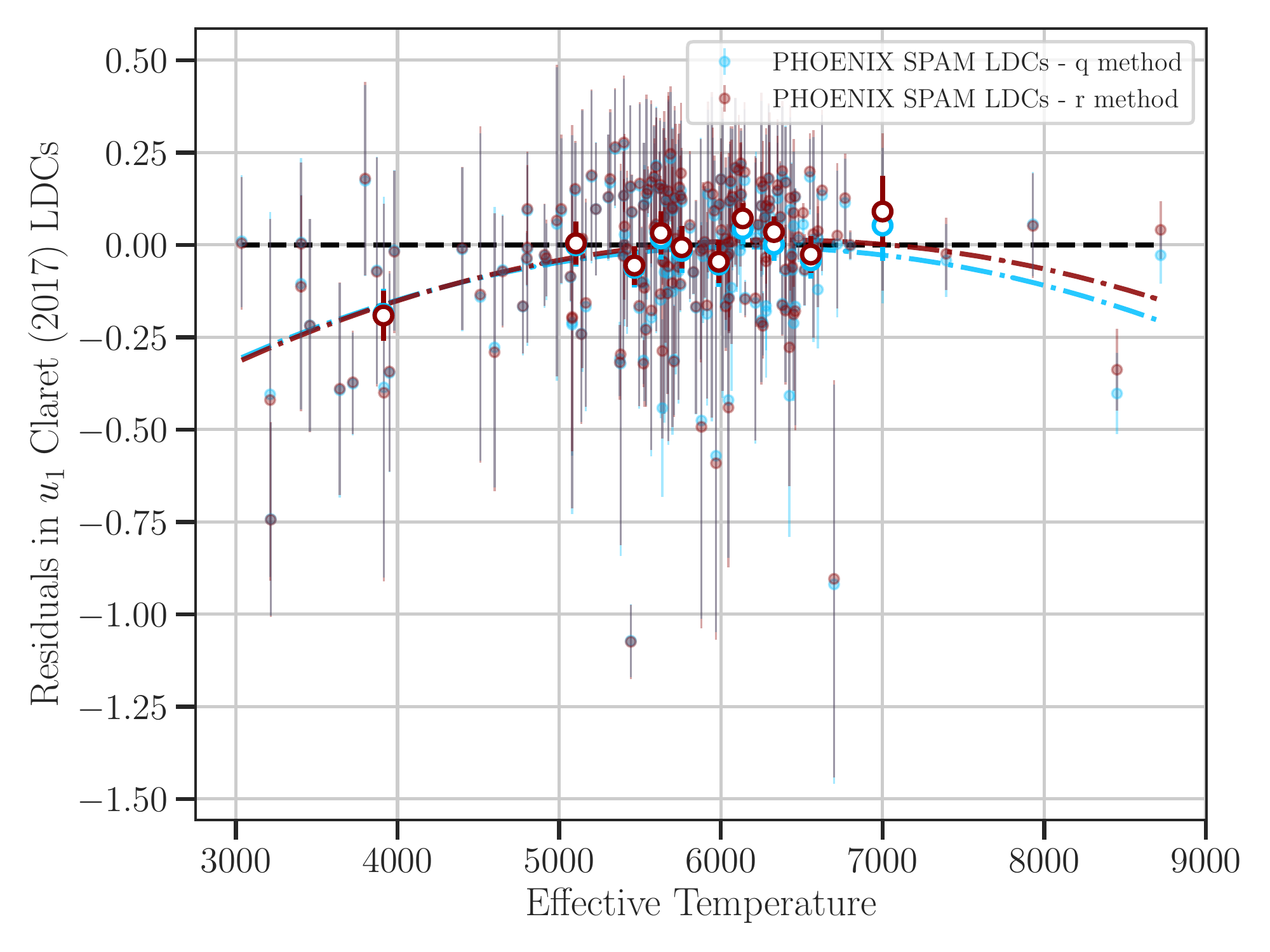}
    \includegraphics[width=\columnwidth]{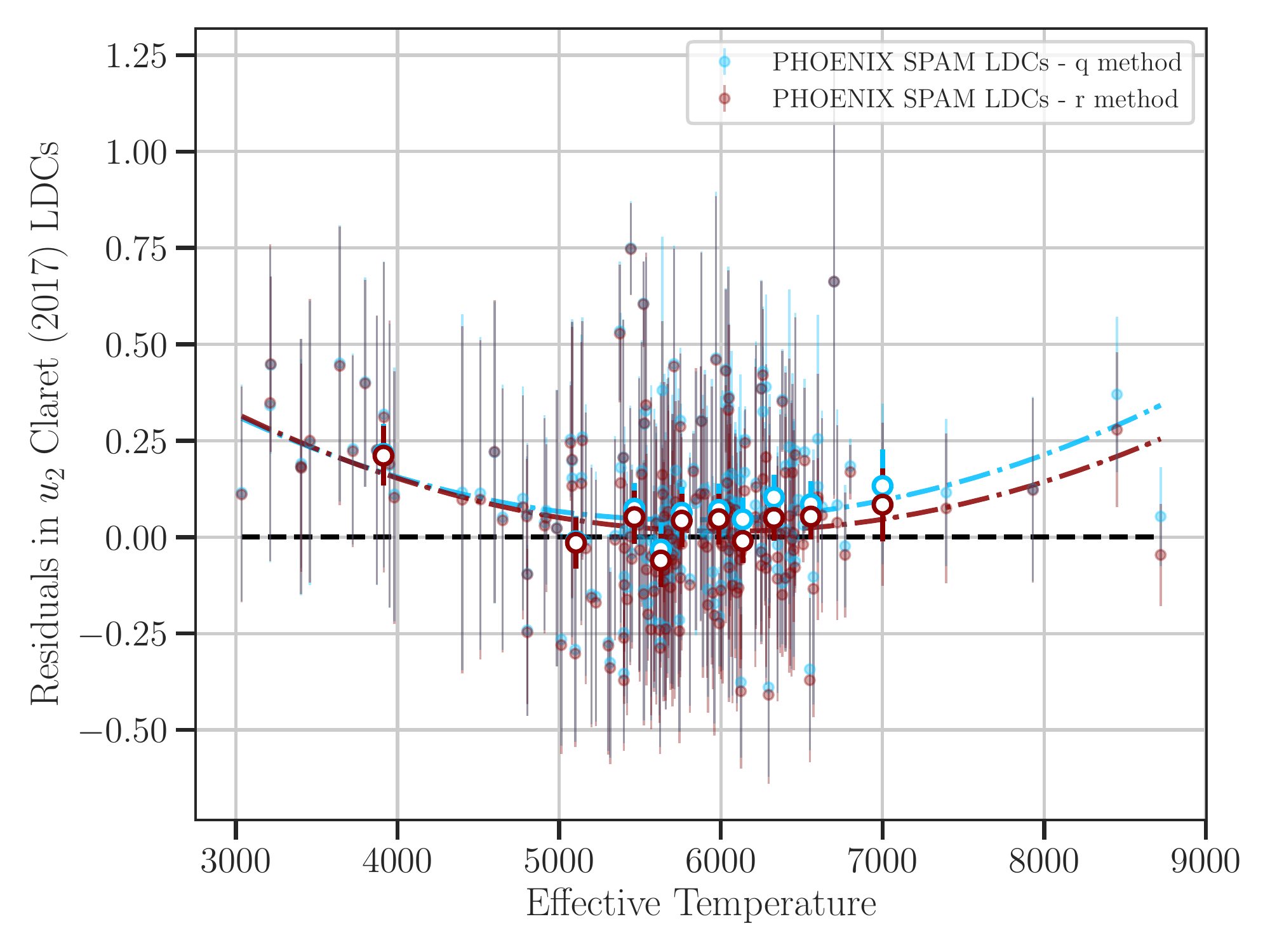}
    \caption{Same as Figure \ref{fig:u1_cla_te_r}, but now using SPAM LDCs.}
    \label{fig:u1_cla_te_r_SPAM}
\end{figure*}



\bibliographystyle{aasjournal}
\bibliography{ldbib} %



\end{document}